\documentclass{emulateapj}
\usepackage{graphicx}
\newcommand\gtsim{\mathrel{\lower0.6ex\hbox{$\buildrel {\textstyle >}\over {\scriptstyle \sim}$}}}
\newcommand\ltsim{\mathrel{\lower0.6ex\hbox{$\buildrel {\textstyle <}\over {\scriptstyle \sim}$}}}
\newcommand\sersic{S\'{e}rsic }
\usepackage{aas_macros}
\usepackage[footnotesize]{subfigure}
\shortauthors{Floyd et al.}

\slugcomment{To appear in ApJ}
\shorttitle{Radio galaxies in context}
\shortauthors{Floyd et al.}

\begin{document}
\title{Hubble Space Telescope Near Infrared Snapshot Survey of 3CR radio source counterparts\altaffilmark{1}\\
III: Radio galaxies and quasars in context}

\author{David J. E. Floyd\altaffilmark{2,3} \email{dfloyd@unimelb.edu.au} 
David Axon\altaffilmark{4,5}, 
Stefi Baum\altaffilmark{4}, 
Alessandro Capetti\altaffilmark{6}
Marco Chiaberge\altaffilmark{3,7}, 
Juan Madrid\altaffilmark{3,8}, 
Christopher P. O'Dea\altaffilmark{4}, 
Eric Perlman\altaffilmark{9}, 
and William Sparks\altaffilmark{3}}

\altaffiltext{1}{Based on observations with the NASA/ESA {\em Hubble Space Telescope}, obtained at the Space Telescope Science Institute, which is operated by the Association of Universities for Research in Astronomy, Inc. (AURA), under NASA contract NAS5-26555}
\altaffiltext{2}{AAO/OCIW Magellan Fellow. Current address: School of Physics, University of Melbourne, Victoria 3010, Australia}
\altaffiltext{3}{Space Telescope Science Institute, 3700 San Martin Drive, Baltimore, MD 21218, USA}
\altaffiltext{4}{Department of Physics, Rochester Institute of Technology, 84 Lomb Memorial Drive, Rochester, NY 14623, USA}
\altaffiltext{5}{School of Mathematical \& Physical Sciences, University of Sussex, Falmer, Brighton, BN2 9BH, UK}
\altaffiltext{6}{INAF--Osservatorio Astronomico di Torino, Via Osservatorio 20, I-10025 Pino Torinese, Italy}
\altaffiltext{7}{INAF--IRA, Via P. Gobetti 101, I-40129 Bologna, Italy}
\altaffiltext{8}{Centre for Astrophysics \& Supercomputing, Swinburne University of Technology P.O. Box 218, Hawthorn, VIC 3122, Australia}
\altaffiltext{9}{Florida Institute of Technology, Physics \& Space Sciences Department, 150 W. University Blvd., Melbourne, FL 32901 USA}

\begin{abstract}
We compare the near-infrared (NIR) $H$ band photometric and morphological properties of low-$z$ ($z<0.3$) 3CR radio galaxies with samples of BL~Lac object and quasar host galaxies, merger remnants, quiescent elliptical galaxies, and brightest cluster galaxies drawn from the literature. In general the 3CR host galaxies are consistent with luminous ($\sim L^\star$) elliptical galaxies. 
The vast majority of FR~II's ($\sim80$\%) occupy the most massive ellipticals and form a homogeneous population that is comparable to the population of radio-loud quasar (RLQ) host galaxies in the literature. 
However, a significant minority ($\sim20$\%) of the 3CR FR~II's appears under-luminous with respect to quasar host galaxies. 
All FR~II objects in this faint tail are either unusually red, or appear to be the brightest objects within a group.
We discuss the apparent differences between the radio galaxy and RLQ host galaxy populations.
RLQs appear to require $\gtsim10^{11}~M_\odot$ host galaxies (and  $\sim10^{9}~M_\odot$ black holes), whereas radio galaxies and RQQs can exist in galaxies down to $\sim 3 \times10^{10}~M_\odot$. 
This may be due to biases in the measured quasar host galaxy luminosities or populations studied, or due to a genuine difference in host galaxy. If due to a genuine difference, it 
%is difficult to reconcile with the simple unification picture, and 
would support the idea that radio and optical active galactic nucleii are two separate populations with a significant overlap.
\end{abstract}


%%%%%%%%%%%%%%%%%%%%%%%%%%%%%%%%%%%%%%%%%%%%%%%%%%
%The idea was is to use the 3CR as a blunt tool for collecting up all the most radio powerful sources in the northern sky, and then to simply weed out those that we have any reason to doubt the true radio power of.  We then lump all the genuinely radio-powerful (at a level consistent with ``radio-loud'' quasars) sources together, and confirm their consistency with the RQSO population. Then, we take this "RQSO-like" subsample of the 3CR and hold up various yardsticks to it. Essentially the 3CR subsample has been chosen in such a way as to provide a bridge, or stepping-stone between RQSO's (few; poorly studied; likely biased samples) to quiescent ellipticals, for which a far richer data set is available. The ideal would be to find a sample of ellipticals that is a perfect match to the RQSO-like 3CR sources in every available observational metric. That lacking, I have discussed which aspects of the individual samples characterize the RQSO-like 3CR and which diverge.

%\clearpage
\section{Introduction}
\label{sec-intro}
%Quasars and radio galaxies are understood to be essentially the same objects. But what is it that makes these particular galaxies active -- Are we simply observing a subset of the most massive elliptical galaxies that happen to be active at a given epoch due to a temporary abundance of fuel? 
Much effort has been invested over the last decade and a half to characterize the host galaxies of quasars using the {\it Hubble Space Telescope} ({\em HST}). Unfortunately by their very nature quasars remain difficult to study, and results remain ambiguous due to selection effects. Detailed isophotal analysis of quasar host galaxies is impossible due to the strongly anisotropic contamination by the nucleus and point-spread function (PSF) of the instrument, which also hinder accurate spectroscopy and thus measurement of host galaxy dynamics. 
In the absence of any forthcoming space-based coronographic mission (e.g., the Terrestrial Planet Finder Coronograph -- TPF-C~\footnote{http://planetquest.jpl.nasa.gov/TPF-C/tpf-C\_index.cfm}), researchers have begun to explore the nature of type 2 (heavily obscured) active galactic nucleii (AGN) to study their environments in greater detail (e.g.,~\citealt{zakamska+06}). Type 2 AGNs have been identified from the Sloan Digital Sky Survey (SDSS) out to $z=0.83$~\citep{zakamska+03}, but the  sample remains incomplete and biases difficult to measure. Radio galaxies offer us an alternative sample of type 2 radio-loud AGN, detectable to far higher redshifts in a statistically complete manner. 

In this paper, we compare the overall near-infrared (NIR; $H$ band) galaxy properties of the $z<0.3$ 3CR radio galaxies with those of quasars and BL~Lac objects at similar redshift, and to quiescent early-type galaxies and mergers available in the literature. For the reasons outlined above, we intend to use the radio galaxies as a link or bridge between the quasars 
%(for which detailed morphological analysis is impossible due to the strong anisotropic flux from the nucleus and PSF) 
and quiescent galaxies which can be studied in much greater structural, dynamical and population detail. 
Nearby radio galaxies have been statistically studied in recent papers~\citep{best+05b,mauchsadler07} and are well-known to occupy luminous galaxies in general. Indeed it is by now accepted that the primary requirement for radio-loud AGN activity is a supermassive black hole $\gtsim 10^{9}~M_\odot$, 
which generally requires a very luminous elliptical host galaxy.
%The principal aim is to determine what subset of galaxies, drawn from the nearby universe, are capable of hosting a radio-loud AGN.
The principal aim of this paper is to explore in detail the subset of galaxies drawn from the nearby universe that are capable of hosting a radio-loud AGN. We wish to determine what additional effects the host galaxy plays in determining the radio loudness of a source, and to identify any sources that are hosted by ``unusual'' galaxies.
%, beyond the well-known correlation between galaxy luminosity, black hole mass and radio power...???
A secondary aim is to determine whether there are any systematic biases introduced in the study of quasar host galaxies.

Our approach is to analyze the results of isophotal and parametric modeling of the host galaxies of 3CR radio sources from~\citet{madrid+06} -- hereafter Paper I, and~\citet{floyd+08} -- hereafter Paper~II, and to compare the results to those for similarly studied samples of galaxies and AGN.
In Paper~I we presented the imaging and photometry for the first part of our survey. In Paper~II we presented the modeling used here, and showed that two different techniques (widely adopted by the galaxy morphology and quasar host galaxy communities, respectively) reliably recovered the properties of the NIR host galaxies of the low-redshift ($z<0.3$) 3CR radio sources without strong optical nuclei. Furthermore, for synthetic quasars (created by taking a true galaxy and adding an artificial central point source), the quasar modeling technique was shown to recover the correct underlying morphological parameters to within 10\%, and galaxy luminosity within to 2\%.

The galaxies associated with the 3CR radio sources have long been known to be typically elliptical~\citep{matthews+64}, with a number of them associated with the brightest galaxies in a cluster or group~\citep{burns90,best+07}. Recent surveys with {\em HST} have revealed a wealth of information on the environments of these uniquely powerful sources at a range of wavelengths~\citep{dekoff+96,martel+99,dekoff+00,allen+02,madrid+06,privon+08,floyd+08}, and uncovered numerous new jets, dust disks, etc. 
But until recently~\citep{donzelli+07,floyd+08}, the 3CR lacked a detailed and systematic classification of their host galaxies in such a way that can be straightforwardly compared with existing samples of quiescent galaxies in the literature. Observing in the infrared offers important advantages in the study of both galaxies and AGNs providing a direct measure of dynamically-dominant old stellar population, an almost-constant mass-to-light ratio~\citep{zibetti+02}, and a peak in the ratio of galaxy to AGN continuum, giving the best chance of detecting the host galaxy in cases where AGN is unobscured. Thus, it seems a natural application of the NICMOS $H$ band data set to make a direct comparison to similar samples in the literature.
%We can perform identical analyses on most of the sample (excluding quasar-like objects exhibiting strong optical-IR nuclei) to that carried out on large samples of quiescent galaxies. We can also follow, for the entire sample, the somewhat simpler modeling techniques necessarily adopted for studying luminous quasar host galaxies. 
We need to be certain which radio galaxies correspond to which RLQs in order to be able to use radio galaxies as a proxy for quasars in detailed spectroscopic, dynamical and structural studies of the effect of AGN on environment and vice versa. 

This paper is structured as follows. In Section~\ref{sec-samp} we describe the samples used in this paper, and refer briefly to the observations and data reduction used. In Section~\ref{sec-res} we describe the results of sample-wide comparisons according to different observational properties of the galaxies. In Section~\ref{sec-disc} we discuss those comparisons, property-by-property across all the samples studied, and look at the $R_e-\mu_e$ Kormendy relation, host-to-nuclear luminosity distribution and host luminosity versus extended radio power distributions. We pick out several anomalously faint 3CR host galaxies for discussion in further detail in Section~\ref{sec-faint}. In Section~\ref{sec-cont} we attempt to draw this discussion together and look at the issues remaining in the study of quasar host galaxies. Finally we conclude in Section~\ref{sec-conc} with a summary of our findings and suggestions for future study.

%FRI's do not exhibit evidence of accretion activity exhibiting no strong emission lines, perhaps indicating a lack of accretion disk altogether  (e.g.,~\citealt{capetti+00}), or that accretion onto FR~I nuclei is advection-dominated with a small resulting radiative efficiency (Fabian \& Rees 1995, Reynolds et al. 1996). 
%FR~II's show clear accretion disc signatures when viewed within the putative torus cone angle. 
%However, they exhibit lower $L_{\mathrm MIR}/L_{\mathrm FIR}$ than quasars~\citep{haas+05, siebenmorgen+05} implying either that emission from the putative dust torus is optically thick even at MIR~\citep{pierkrolik93} -- hence anisotropic -- or that the intrinsic AGN luminosities are different. 

%%%%%%%%%%%%%%%%%%%%%%%%%%%%%%%%%%%%%%%%%%%%%%%%%%%
\section{Samples, observations and data reduction}
\label{sec-samp}
In this paper we take the revised 3C sample (3CR) as defined by~\citet{spinrad+85}, restricted to $z<0.3$, as a statistically complete sample of low-redshift radio galaxies, and compare it in various physical terms with samples of low-redshift quasars, quiescent galaxies and mergers that have deep imaging and structural modeling of the galaxies available in the literature. 
The 3CR covers sources in the northern hemisphere that are brighter than 9~Jy at 178~MHz.
\begin{figure}[htbf]
\centering
%{\includegraphics[width=8.0cm]{f1_noRQ.eps}}
\plotone{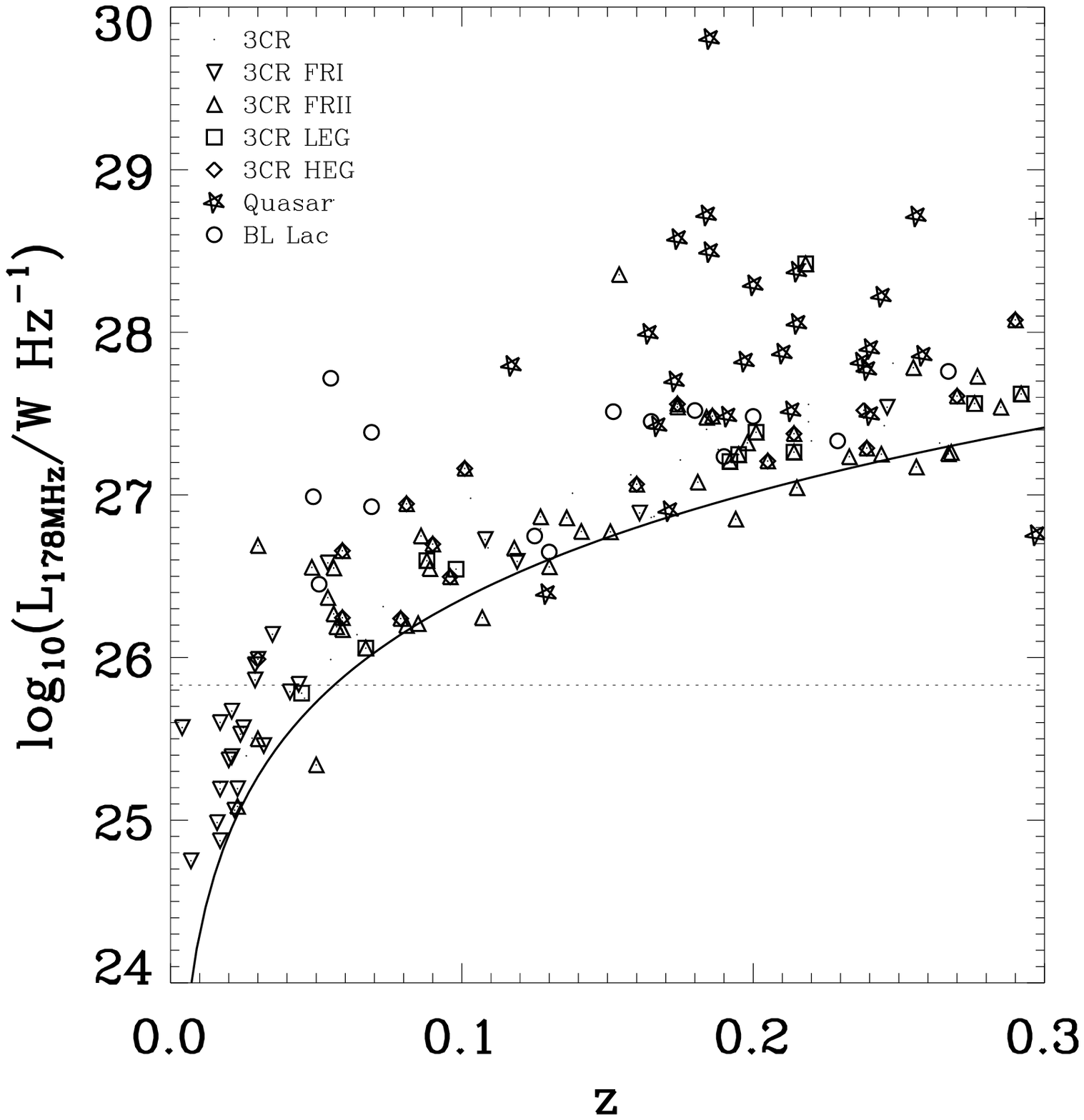}
\caption{\label{fig-radio} Rest-frame 178~MHz radio luminosity of the present sample (down-pointing triangles = FR~I's; up-pointing triangles = FR~II's; diamonds = HEGs; squares = LEGs), along with the quasars (stars) and BL~Lac objects (circles) with radio detections from ~\citet{urry+00,mcleod01,dunlop+03,floyd+04}. The dotted line shows the radio-quiet cutoff traditionally adopted in quasar studies ($\log_{10}(L_{178}) = 25.83$). The 3CR definition is of sources brighter than 9~Jy at 178~MHz.}
%Throughout this paper, radio-loud sources are shown by open symbols and radio-quiets are solid.}
\end{figure}
$H$ band snapshot observations of the low-$z$ 3CR sample (as defined by~\citealt{spinrad+85}) began in {\em HST} Cycle 13, using NICMOS 2, and we presented the first year's imaging in paper I. Observations continued through Cycle 14 yielding a total of 101 sources (an 87\% completeness),  including 11 drawn from the NICMOS 2 archive. Modeling and classification of all 101 galaxies were presented in Paper~II. In the present paper we revisit those modeling results, drawing comparisons with other galaxy samples drawn from the literature. See Paper~I for full details of the 3CR NIC2 observations and notes on half the sample. See Paper~II for details of the 3CR NIC2 data reduction, modeling results and notes on the remainder of the sample. 
In particular, the machine-readable Tables 3, 4 and 5 of Paper~II contain the basic radio properties, one-dimensional and two-dimensional \sersic profile fits, respectively, for the entire present sample of 101 3CR sources. Also see~\citealt{jenkins+77,giovannini+88} for 5~GHz radio properties and~\citet{lillylongair84} for $K$ band fluxes.
All reduced, science ready data is available for public download from the STScI archive Web site\footnote{http://archive.stsci.edu/prepds/3cr/}. 

Note that from the original sample modeled in Paper~II, we omit 3C~258 from study in this paper because, as discussed in Paper~II, it is found to be a higher redshift ($z\gtsim1$) RLQ located behind a $z=0.165$ irregular galaxy. 
%Note that we have slightly refined the original 3CR sample studied in papers I and II to the ``3CRR'' sample that is complete to a 178-MHz flux limit of 10.9 Jy over an area of 4.2 sr. This selection omits the following $z<0.3$ 3CR sources: 3C~133 (too low in galactic plane -- $|b|<10^\circ$); 277.3, 303.1, 460 (have too low fluxes). Furthermore we follow~\citet{willott+03, mclure+06} in omitting the nearby starburst M82 (3C~231). Lastly we omit 3C~258 because, as discussed in Paper~II, it is found to be a higher redshift ($z\gtsim1$) RLQ located behind a $z=0.165$ irregular galaxy. Our final sample thus consists of 6 fewer objects than that of papers I and II, of which we presented observations of 4 (3C~133, 258, 277.3, 460). We thus include 97 sources in this study -- an 88\% completeness for the $z<0.3$ 3CRR.

%%%%%%%%%%%%%%%%%%%%%%%%%%%%%%%%%%%%%%%%%%%%%%%%%%%
\begin{deluxetable*}{llrllll}[ht]
%\begin{table*}{llrllll}[ht]
  \tabletypesize{\tiny}
  \tablecolumns{7}
  \tablewidth{0pc}
  \tablecaption{\label{tab-samp} Details of the comparison samples used in this paper.}
  \tablehead{
    \colhead{Paper} & \colhead{Sample Type} & \colhead{$N_s^1$} & 
    \colhead{Telescope/Instrument} & \colhead{Filter} &
    \colhead{Sample Distance} & \colhead{Selection Notes}}
\startdata
U+00$^{a}$ & BL~Lac object & 30 & {\em HST}/WFPC2 & $R$ & $0.0<z<1.3$ & Composite sample; 30 sources at $z<0.3$\\
MM01$^{b}$ & RQQ & 16 & {\em HST}/NICMOS2 & $H$ & $z\approx0.3$ & Luminous QSO's at $z<0.3$\\
D+03$^{c}$ & RLQ, RQQ, RG & 33 & {\em HST}/WFPC2 & $R$ & $z=0.2$ & Optically matched samples of RLQ and RQQ\\
F+04$^{d}$ & RLQ, RQQ   & 17 & {\em HST}/WFPC2 & $I$ & $z=0.4$ & Optically luminous QSO's\\
T+96$^{e}$ & RLQ, RQQ, RG   & 33 & UKIRT/IRCAM & $K$ & $z=0.2$ & We use D+03$^{c}$ subsample\\
RJ04$^{f}$ & Merger     & 51 & U. Hawaii 2.2~m/QUIRC & $K$ & $z<0.045$ & Single-nucleus candidate disk-merger remnants\\
BBF92$^{g,h}$& Elliptical & 80 & Calar-Alto 1.23~m & $V, R, I$ & $D<140$~Mpc & Nearby $B$ band luminous ellipticals\\
P99$^{i}$ & Elliptical &341 & Palomar 60 inch; LCO 1~m, 2.5~m & $K$ & $z<0.03$ & Various nearby galaxies, groups and clusters\\
M+99$^{j}$ & Elliptical & 48 & UKIRT/IRCAM3 & $K$ & $\bar{D}=99$~Mpc & Coma cluster ellipticals\\
F+06$^{k}$ & Early type &100 & {\em HST}/ACS-WFC & $g, z$ & $\bar{D}=18$~Mpc & Virgo cluster early types\\
G+96$^{l,m}$ & BCG        &119 & CTIO 1.5m; KPNO 2.1m, 4m & $R$ & $0.01<z<0.054$ & Abell brightest cluster galaxies\\
I, II$^{n,o}$ & RG & 101 & {\em HST}/NICMOS & $H$ & $z<0.3$& Present study\\
\hline
\enddata
\tablecomments{$^1$ Sample size used}
  \tablerefs{$^a$ \citet{urry+00}; 
    $^b$\citet{mcleod01}; 
    $^c$\citet{dunlop+03}; 
    $^d$\citet{floyd+04}; 
    $^e$\citet{taylor+96}; 
    $^f$\citet{rothberg+04}; 
    $^g$\citet{BBF92}; 
    $^h$\citet{bendermoellenhoff87}; 
    $^i$\citet{pahre99}; 
    $^j$\citet{mobasher+99}; 
    $^k$\citet{ferrarese+06}; 
    $^l$\citet{graham+96}; 
    $^m$\citet{postmanlauer95};
    $^n$\citet{madrid+06};
    $^o$\citet{floyd+08}}
\end{deluxetable*}
%\end{table*}
%%%%%%%%%%%%%%%%%%%%%%%%%%%%%%%%%%%%%%%%%%%%%%%%%%

\subsection{Cosmology, Photometry, and Evolution}
\label{sec-cosm}
We throughout assume a flat, $\Lambda$-dominated cosmology in which $\Omega_{\Lambda}=0.7$, and $h_0=0.7$.
For each source we have $k$-corrected the apparent magnitudes to rest-frame H band assuming a spectral index of $\alpha=1.5$ for the host galaxy and $\alpha=0.2$ for the nucleus where necessary ($f_\nu\propto\nu^{-\alpha}$). Galactic extinction corrections have been applied following~\citet{schlegel+98}. Comparison samples have been converted to the rest-frame wavelength (and correct angular diameter distance) of our study following the same cosmological assumptions and using a throughput  template for the NICMOS2 F160W ($H$ band) filter and the appropriate filter/instrument configuration for the original observations.

Note that few samples exist to precisely match our redshift distribution. Out to $z=0.3$ we are probing a total look-back time of 3.4~Gyr. While evolution is not negligible across such a time frame, on the basis of our hypothesis that the 3CR represent a subsample of the most massive elliptical galaxies at low redshift, it is safe to assume that most of the mass has formed much earlier and that now the galaxies are evolving more-or-less passively. 
The radio sources themselves will have evolved significantly, but again, following our null hypothesis, all early-type galaxies should have hosted a radio source at some point over this epoch.
For quasars we can construct samples that span a similar redshift range, while for quiescent galaxies existing morphological studies tend to concentrate on local samples, but it is informative to compare with quiescent galaxy samples where they are available.

\subsection{Comparison Samples}
We explored the literature for suitable samples with comparable published data -- morphological information from one- or two-dimensional fitting.
We take BL~Lac objects and quasars with deep imaging and well-studied host galaxies from the samples of~\citet{taylor+96,urry+00,mcleod01,dunlop+03,floyd+04}. 
Quiescent early-type galaxies are taken from~\citet{BBF92,pahre99,mobasher+99,ferrarese+06}. 
Merger remnants we take from~\citet{rothberg+04}. 
Finally, we use the brightest cluster galaxy (BCG) sample studied by~\citet{graham+96}. 
The samples in each of these papers are briefly discussed below. See Table~\ref{tab-samp} for basic details of the comparison samples at a glance. 

In each case, we have taken care to compare published properties with equivalent properties within our sample. A wide range of conventions are used in the literature for describing scale length, surface brightness and diskiness. We have made use of the conversions discussed by~\citet{milvangjensenjorgensen99}. In cases where the total magnitude of the (modeled) host galaxy was unavailable, we have calculated it from the basic model parameters using the prescriptions in the same paper.

Note that, at present, there are insufficient quasars and BL~Lac objects with observed host galaxies to produce a complete or volume-limited sample. The radio luminosity--redshift distributions are compared in Figure~\ref{fig-radio}. The 178~MHz luminosity distribution of the combined RLQ$+$BL~Lac object sample forms a good match to that of the 3CR, with a Kolmogorov--Smirnov (K--S) test returning $D=0.20$, with $p=0.76$, indicating no significant differences between the radio power distributions of the two samples. 

%The redshift distributions differ somewhat however ($D=0.25$,~$p=6\%$)., and a 2-D K-S test returns a probability of just 1\% that the samples are drawn from the same population. This is an unavoidable effect of the small number of known quasars and BL~Lac objects and AGN. There is also a difference in the observation wavelengths of each sample, and so precise correlation may not be possible. However, we attempt to provide the best comparison possible of the different samples available, and to explore whether any differences are likely due to observational or$H$ band selection effects, or to inherent properties of each sample.

%\begin{figure}[htf]
%\centering
%{\includegraphics[height=4.0cm]{/Users/floyd/3C_Work/plots/radio_hist.eps}}
%{\includegraphics[height=4.0cm]{/Users/floyd/3C_Work/plots/radio_hist_qso.eps}}
%\caption{\label{fig-5GHz} 5~GHz radio luminosity distributions.}
%\end{figure}

\subsubsection{Quasars and BL~Lac objects}
\begin{figure*}[t]
\centering
{\includegraphics[height=4.0cm]{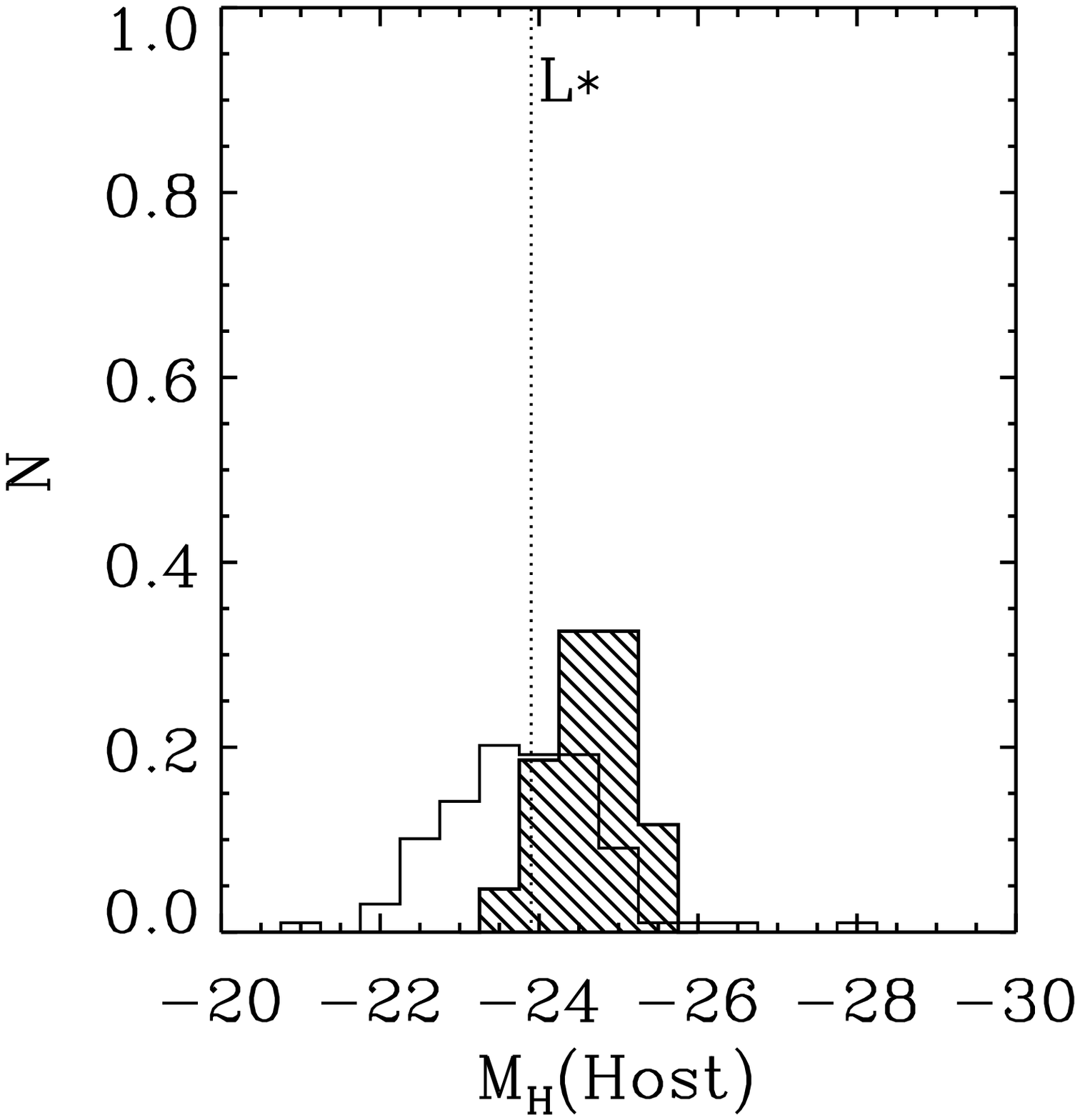}}
{\includegraphics[height=4.0cm]{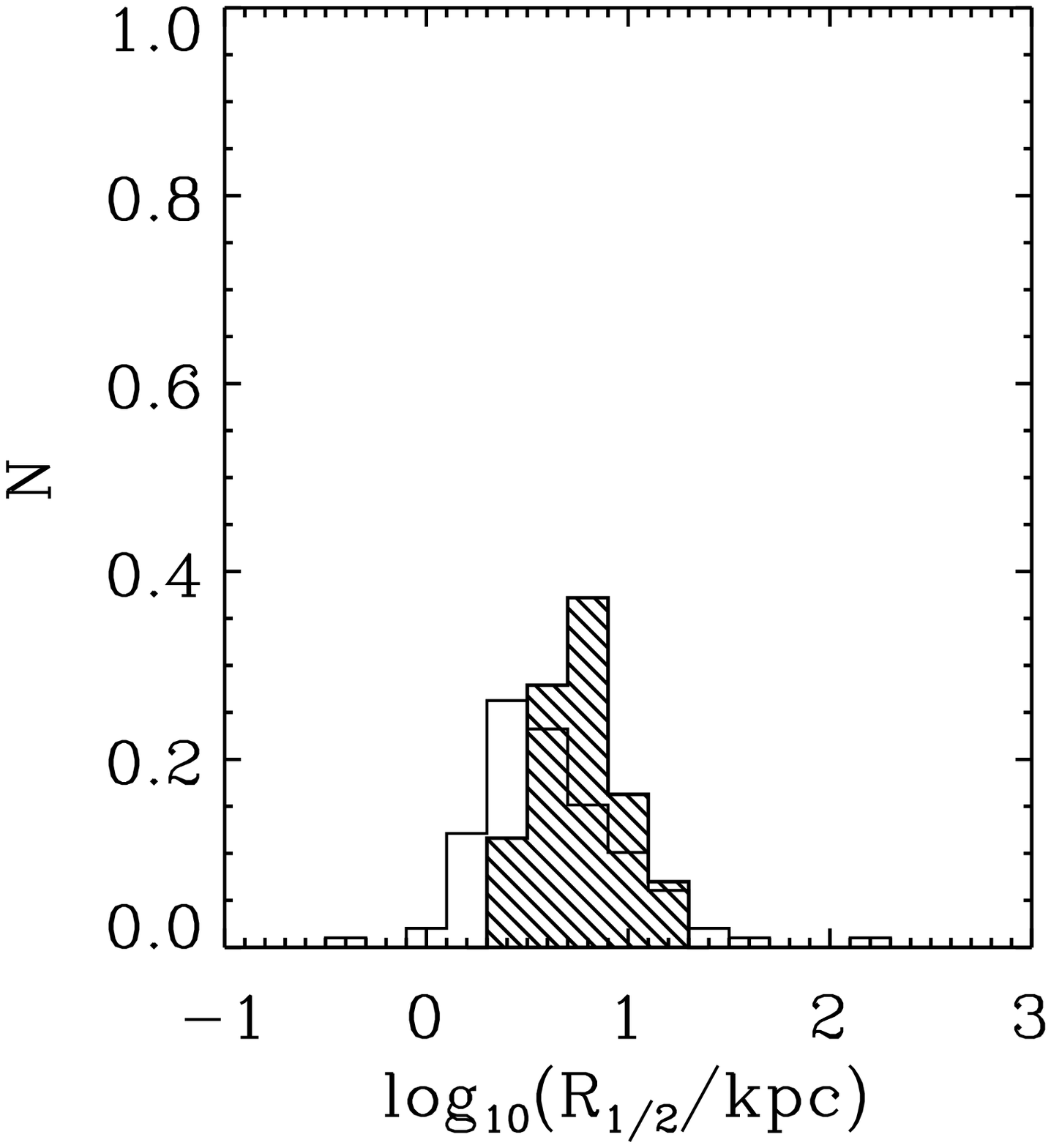}}
{\includegraphics[height=4.0cm]{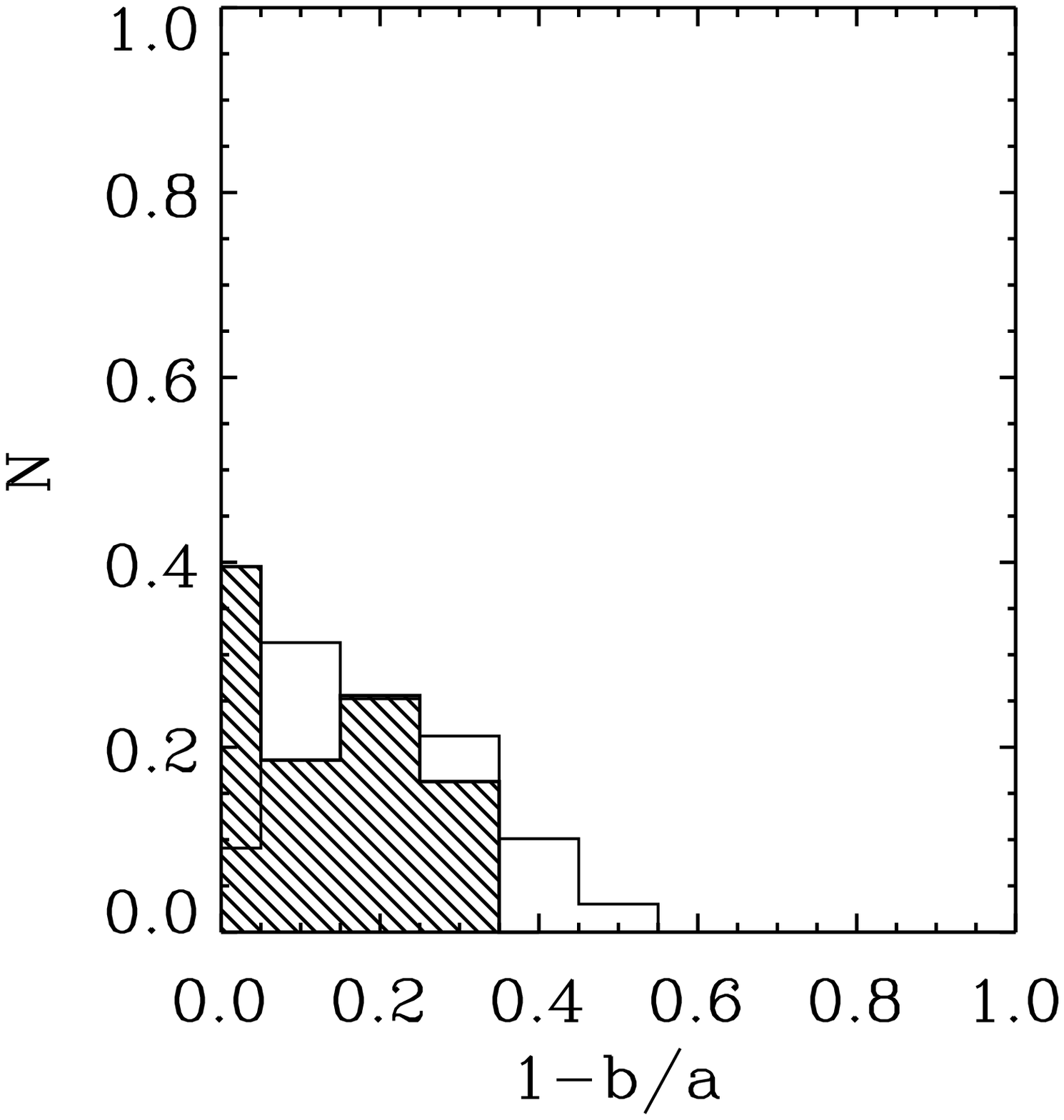}}
{\includegraphics[height=4.0cm]{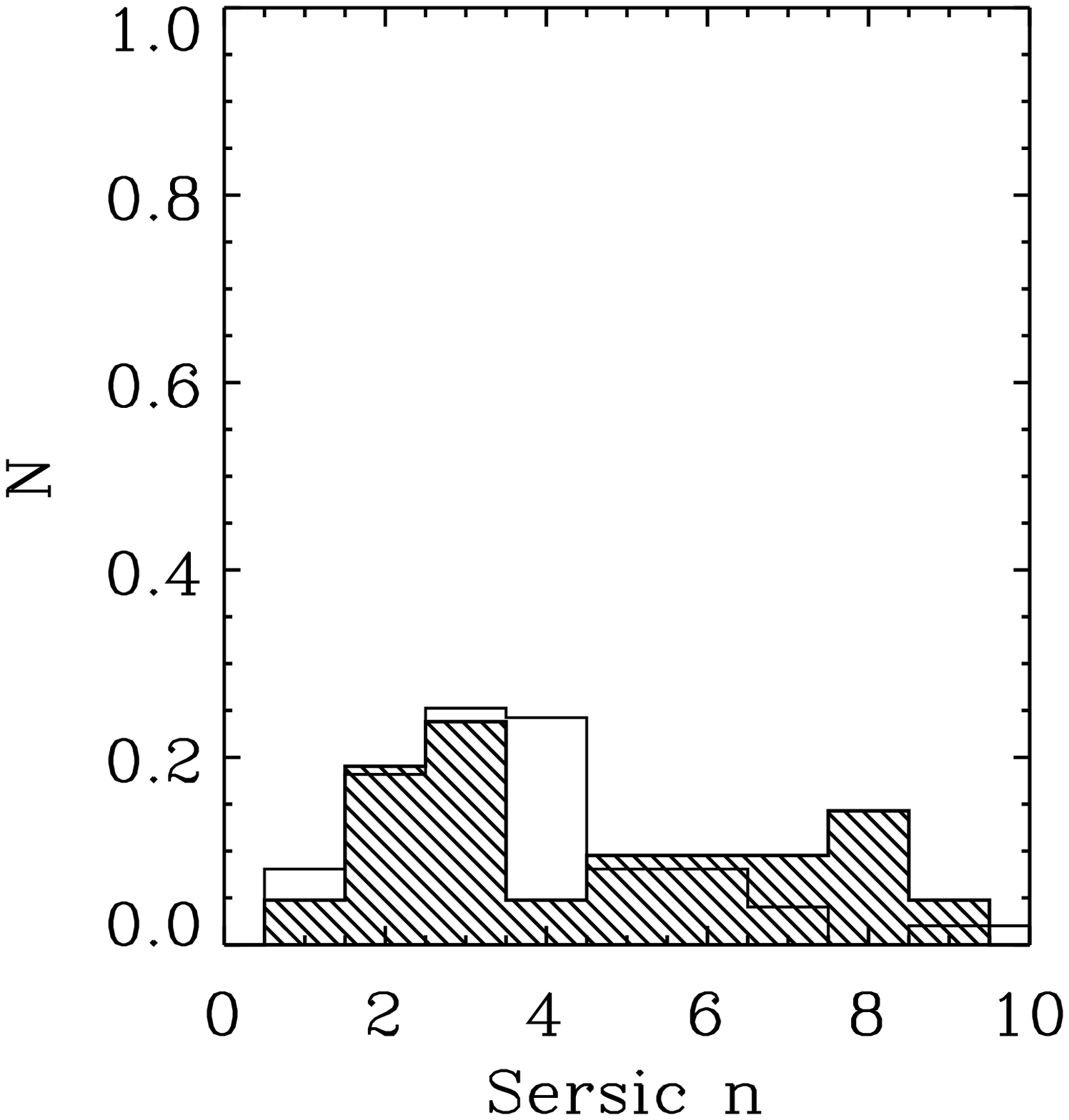}}
\caption{\label{fig-urry} Histograms for the present 3CR sample (open) and the~\citet{urry+00} BL~Lac object sample (shaded) for the following properties (from left to right): host galaxy luminosity (converted $H$ band); scale length; ellipticity; \sersic index. $L^{\star}$ is indicated by a dotted line on the left-hand figure. The results of a two-sided K-S test for each distribution are presented in Table~\ref{tab-stats}.}
\end{figure*}

{\bf Urry et al. (2000): }
\citet{urry+00} obtained {\em HST}/WFPC2 F702W ($R$ band) images of 110 BL~Lac objects drawn from a 132-strong sample at $0\ltsim z \ltsim 1.3$. 
The sample is a composite of seven complete catalogs selected in the radio, X-ray, and optical.
%, and offers an unbiased view of these objects. 
We use only those 30 sources that fall into the redshift range of our sample ($0.0<z<0.3$). Urry et al. apply a one-dimensional fit to an azimuthally averaged radial profile of the source, simultaneously fitting both host and nucleus. They adopt an exponential disk and a de Vaucouleurs $R^{1/4}$ law, but do not fit \sersic models, nor ellipticities. 
%Two-dimensional fits to the $z<0.3$ sample were produced by~\citet{falomo+00}, including ellipticities that we use to compare with ellipticities from our sample. 
As part of a survey of QSO hosts in the {\em HST} archive (D. J. E. Floyd et al. in preparation), we have refit the Urry et al. host galaxies with a \sersic model, using the same technique as described in~\citet{floyd+04} and Paper~II. We use these results (which are consistent in terms of host luminosity with the original data) to compare \sersic parameters.
See Figure~\ref{fig-urry} for a comparison of the Urry sample with the 3CR.

\begin{figure*}
\centering
{\includegraphics[height=4.0cm]{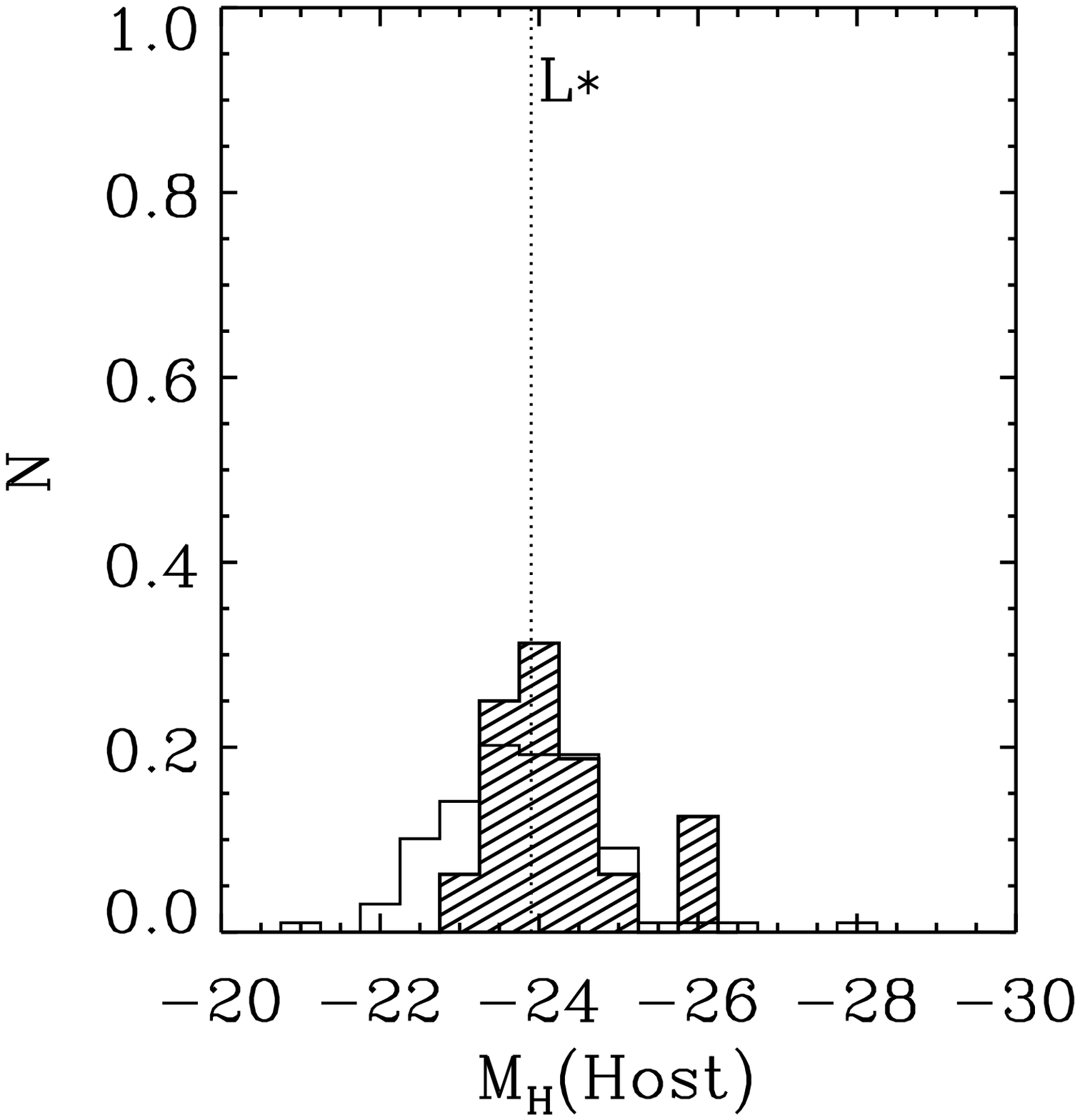}}
{\includegraphics[height=4.0cm]{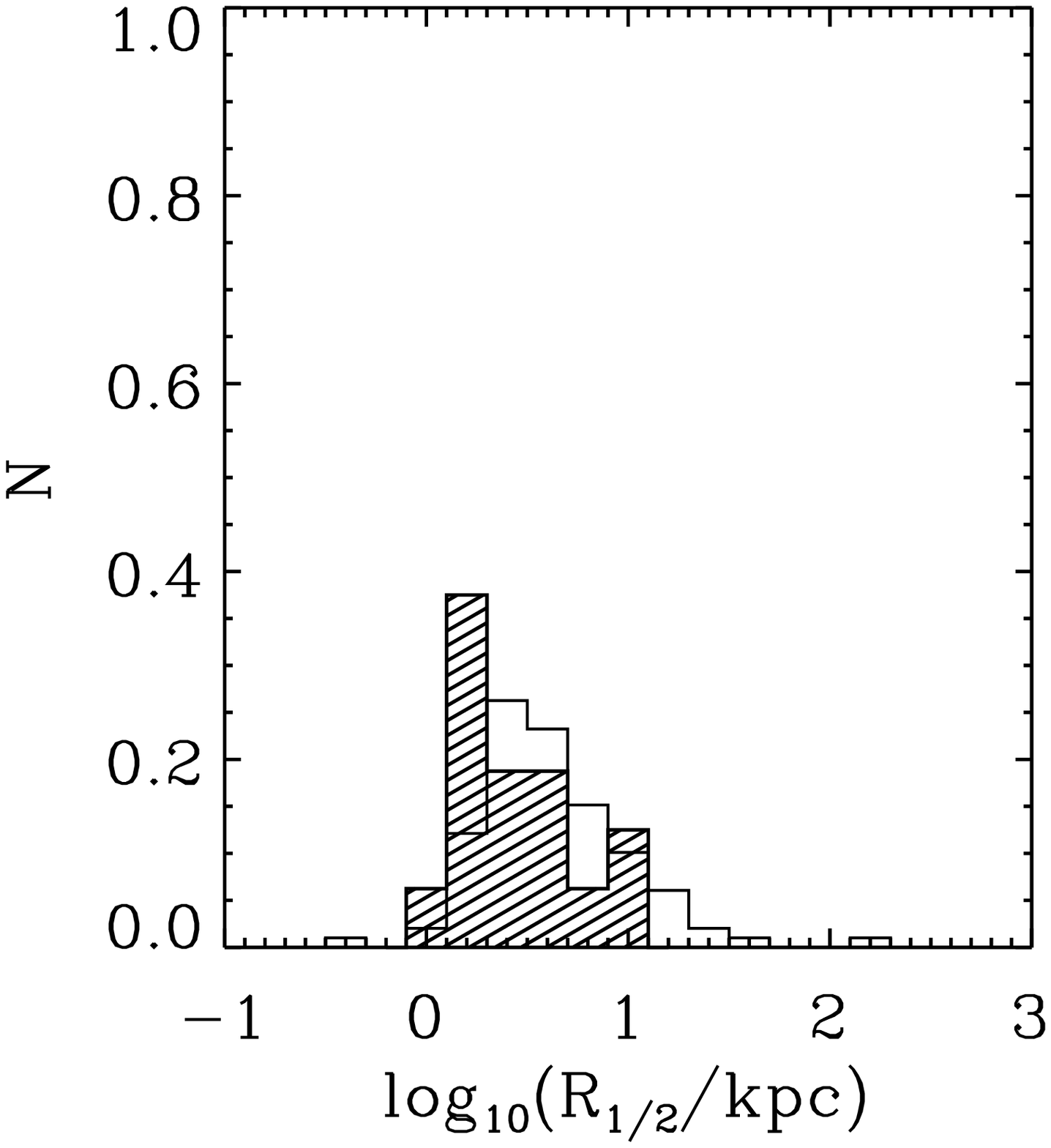}}
{\includegraphics[height=4.0cm]{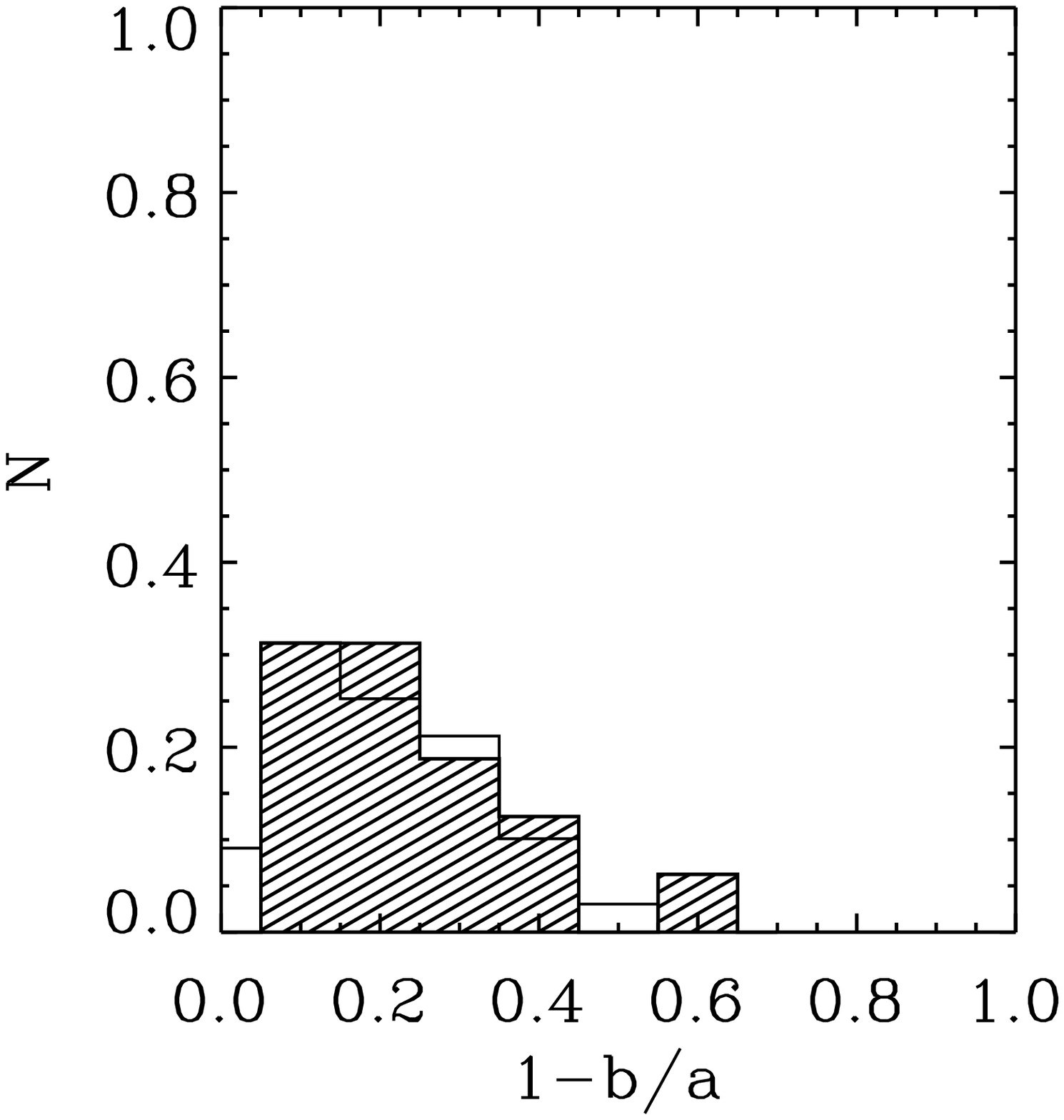}}
\caption{\label{fig-mcleod} Histograms for the present 3CR sample (open) and the~\citet{mcleod01} RQQ sample (shaded) for the following properties (from left to right): host galaxy luminosity (converted to $H$ band); scale length; ellipticity. $L^{\star}$ is indicated by a dotted line on the left-hand figure. The results of a two-sided K-S test for each distribution are presented in Table~\ref{tab-stats}.}
\end{figure*}
\begin{figure*}
\centering
{\includegraphics[height=4.0cm]{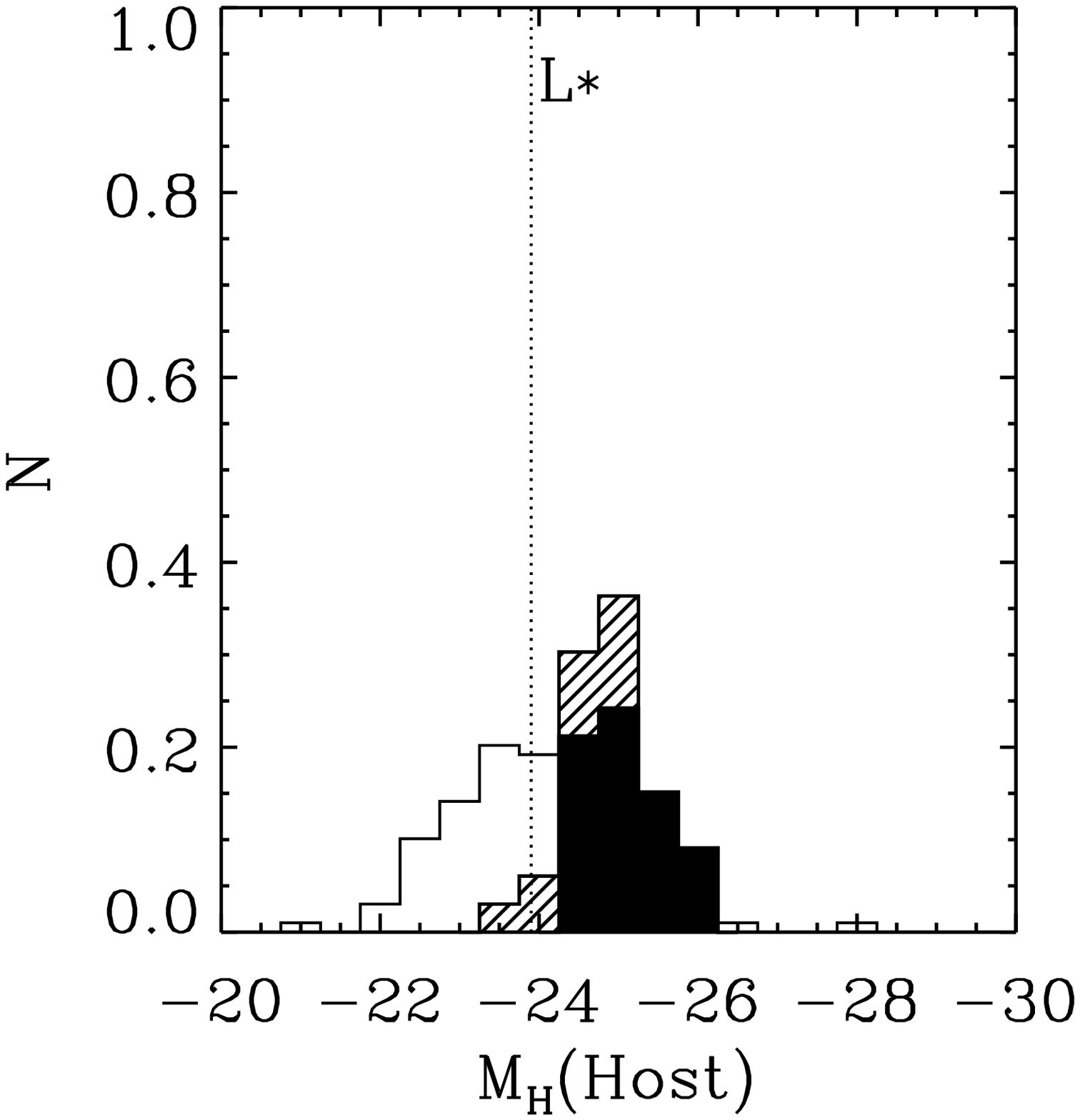}}
{\includegraphics[height=4.0cm]{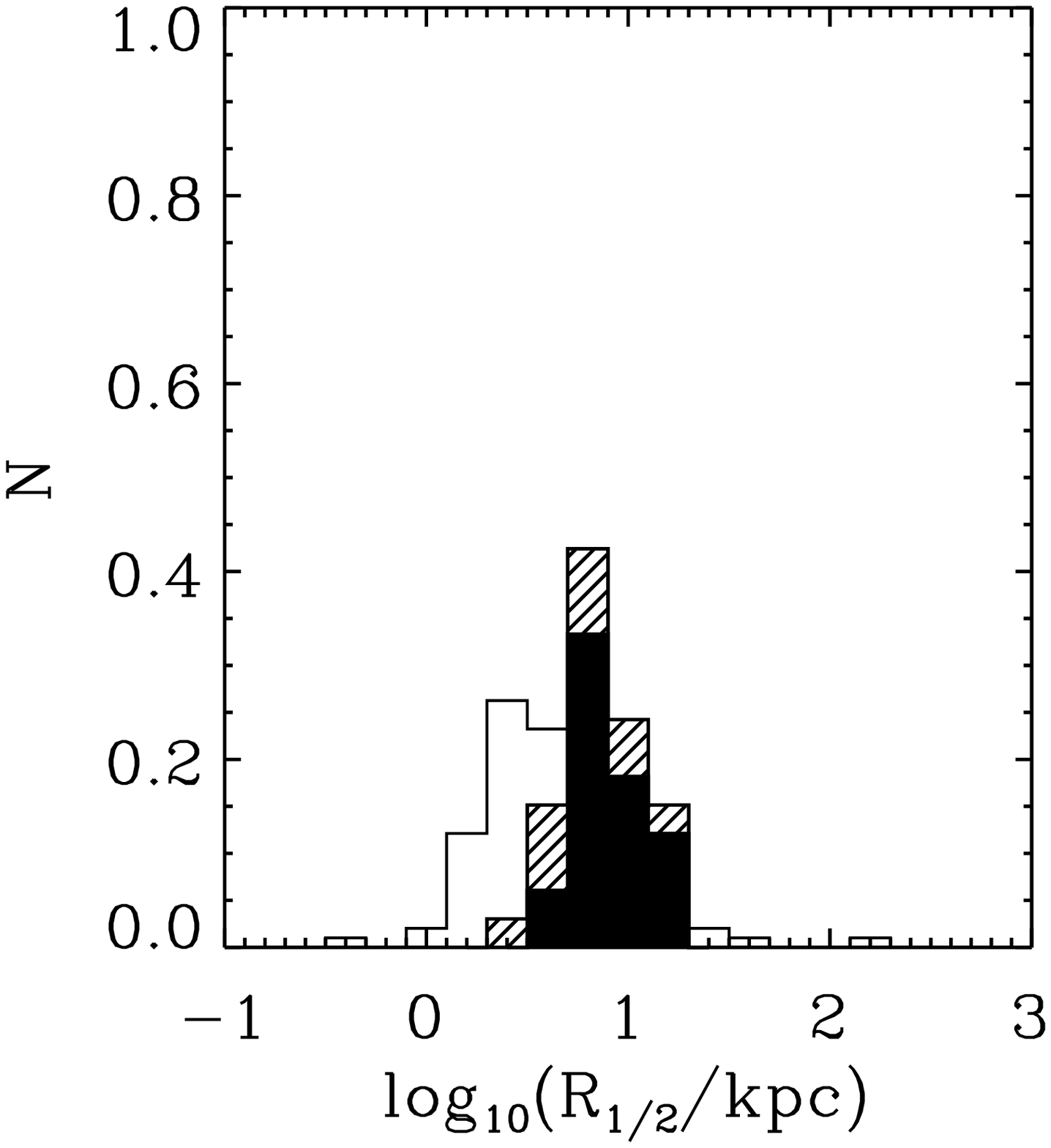}}
{\includegraphics[height=4.0cm]{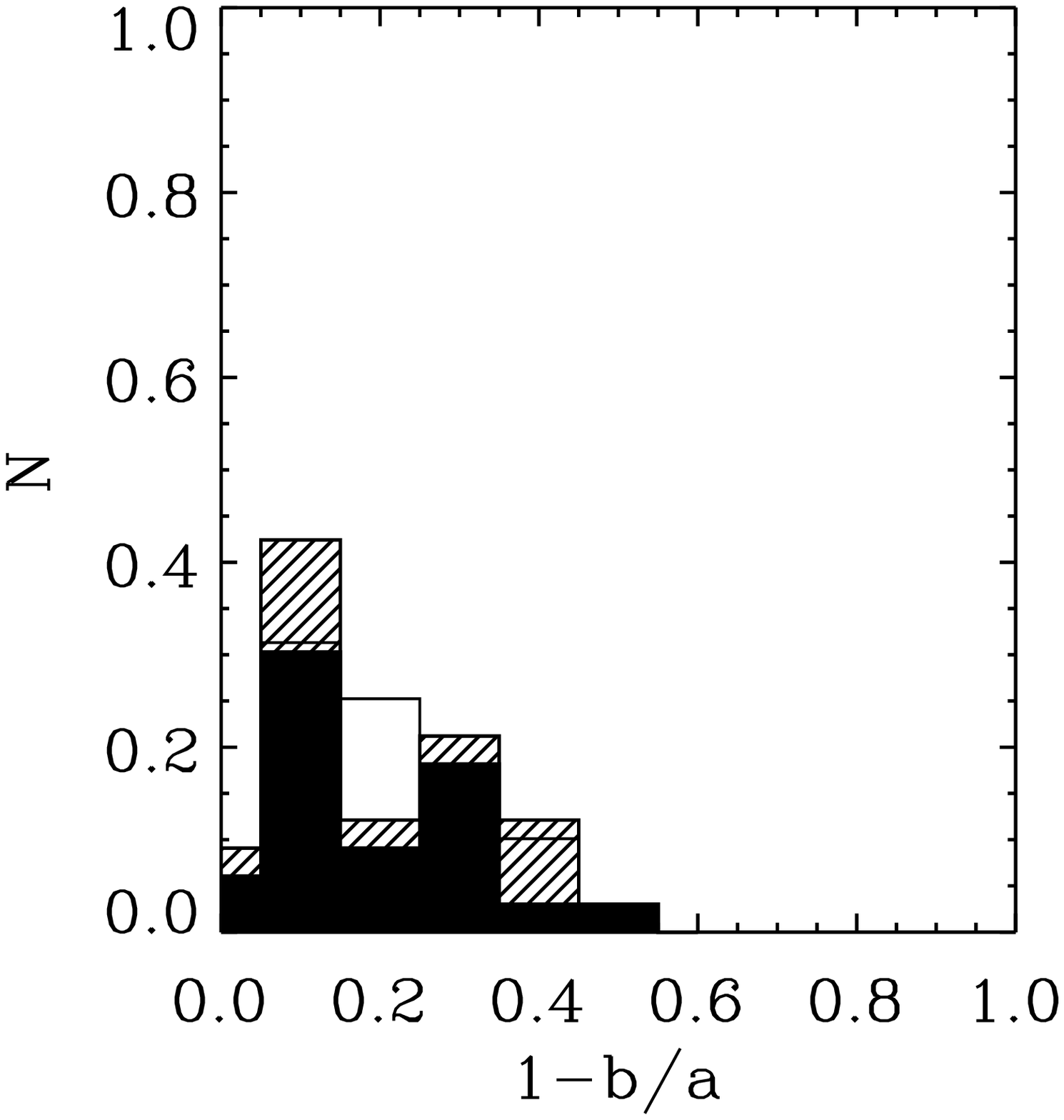}}
{\includegraphics[height=4.0cm]{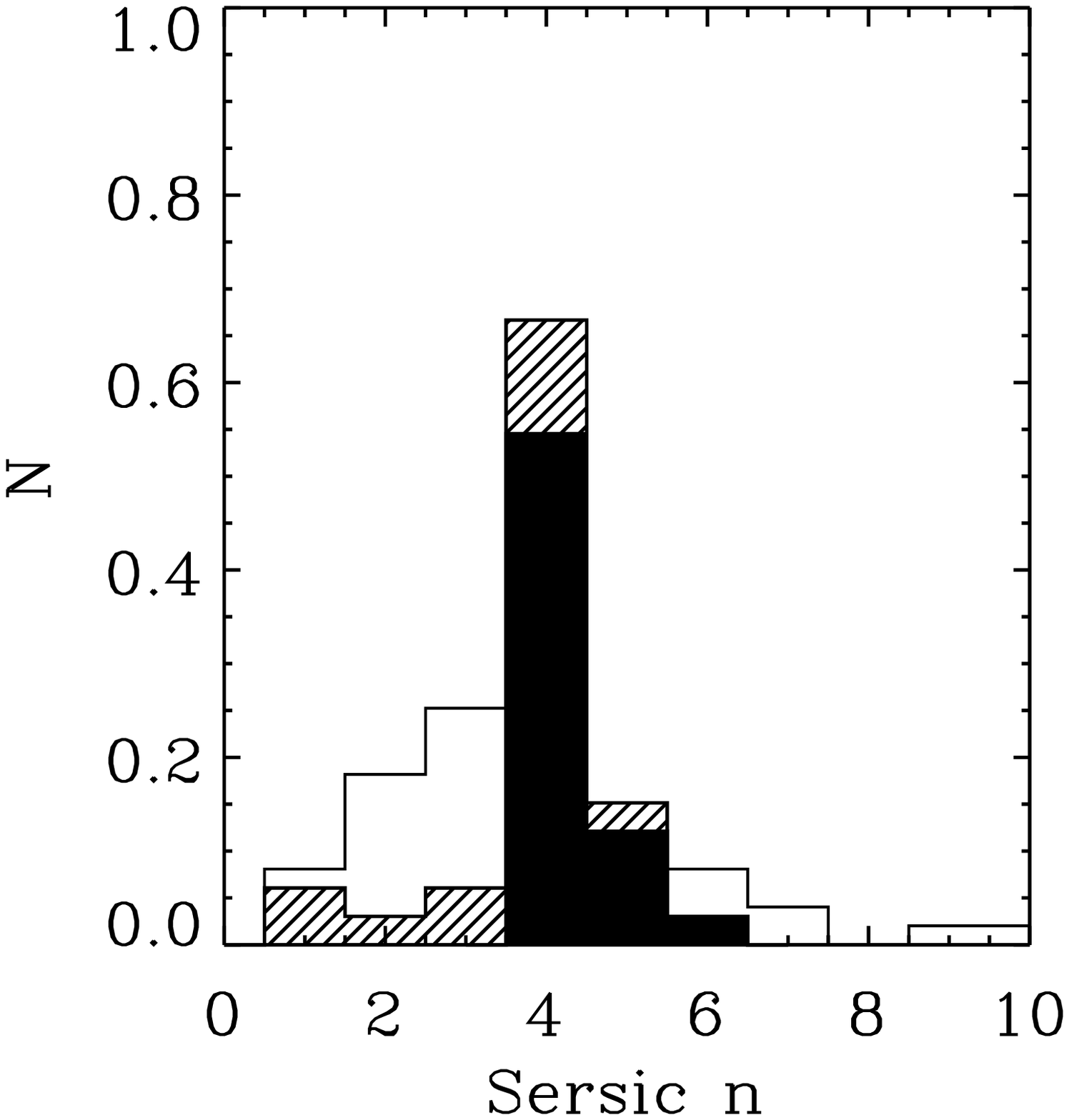}}
\caption{\label{fig-dunlop} Histograms for the present 3CR sample (open) and the~\citet{dunlop+03} quasar sample (full sample -- shaded; RLQs -- black) for the following properties (from left to right): host galaxy luminosity (converted $H$ band); scale length; ellipticity; \sersic index. $L^{\star}$ is indicated by a dotted line on the left-hand figure. The results of a two-sided K-S test for each distribution are presented in Table~\ref{tab-stats}.}
\end{figure*}

{\bf McLeod \& McLeod (2001): }
\citet{mcleod01} studied 16 $z\approx0.3$ optically-luminous radio-quiet quasar (RQQ) host galaxies in the NIR F160W filter ($H$ band) using NICMOS. The authors adopted a PSF-subtraction technique, followed by modeling of the residual host galaxy using both one-dimensional and two-dimensional analysis. We compare our two-dimensional results to their two-dimensional results which fit exponential disk and de Vaucouleurs elliptical models, but no \sersic profile. Due to the presence of saturated cores in almost all of the images, we were unable to remodel this data using our technique.
Figure~\ref{fig-mcleod} shows the various properties of the sample compared with those of the 3CR.

\begin{figure*}[htbf]
\centering
{\includegraphics[height=4.0cm]{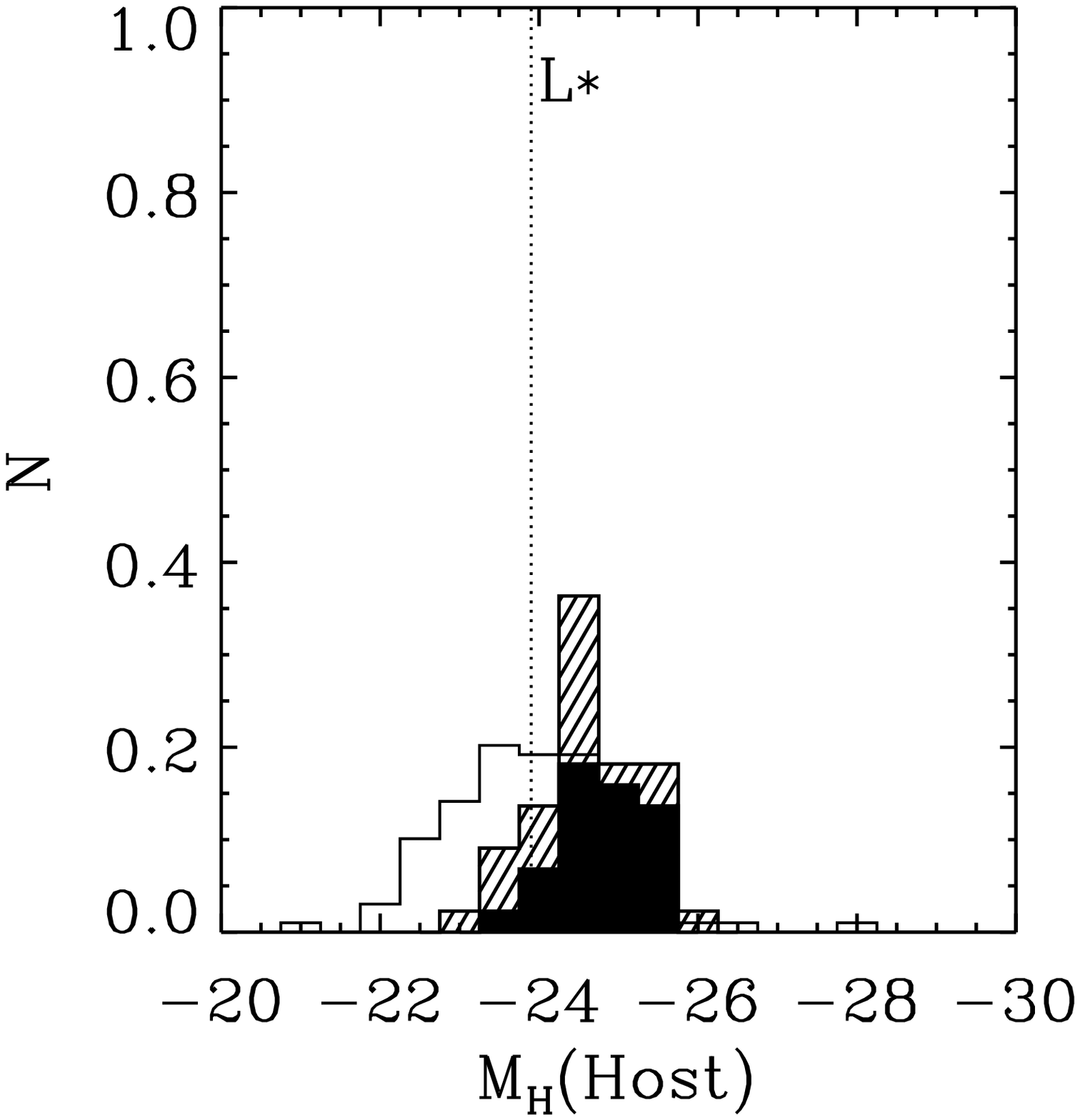}}
{\includegraphics[height=4.0cm]{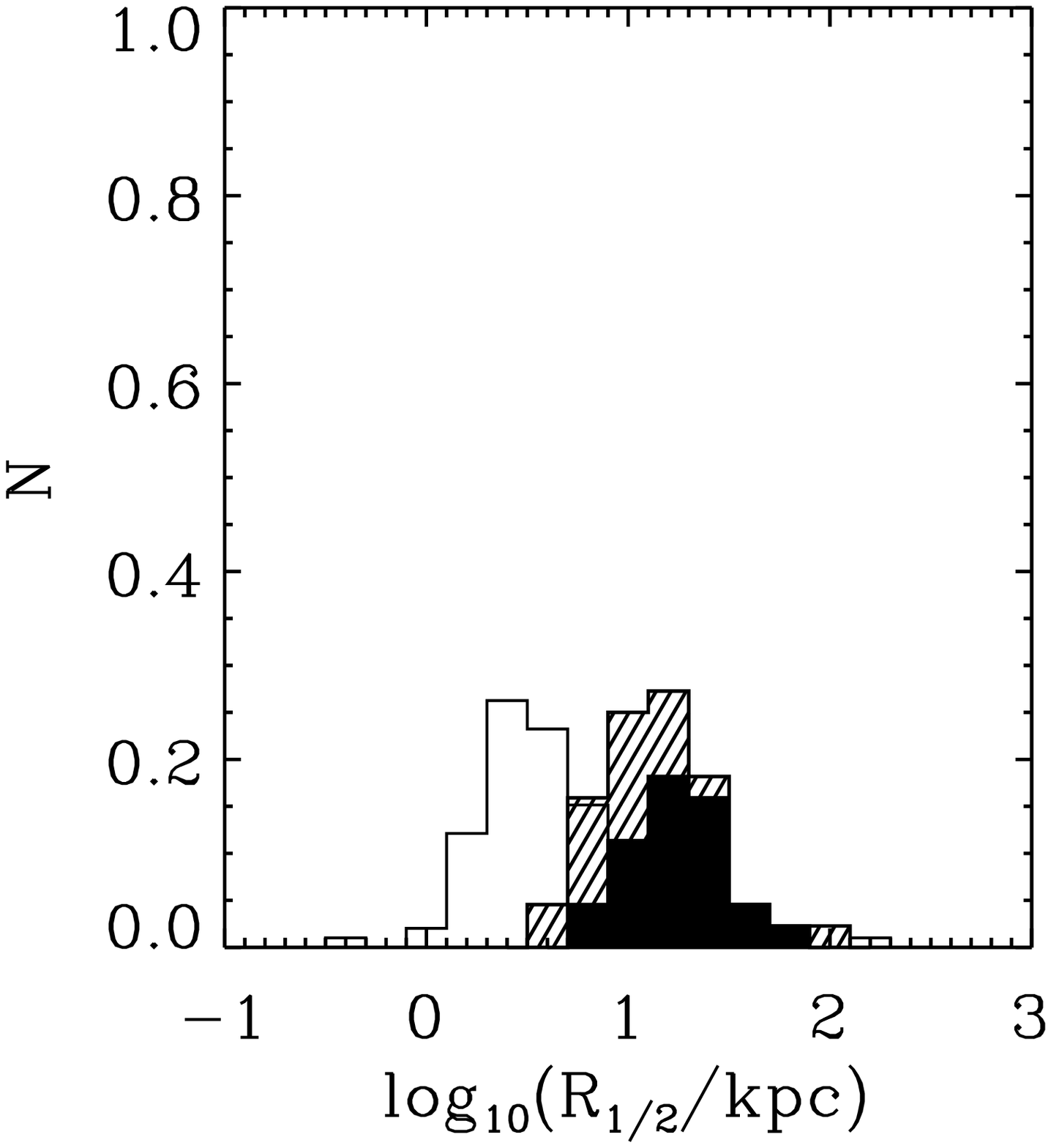}}
{\includegraphics[height=4.0cm]{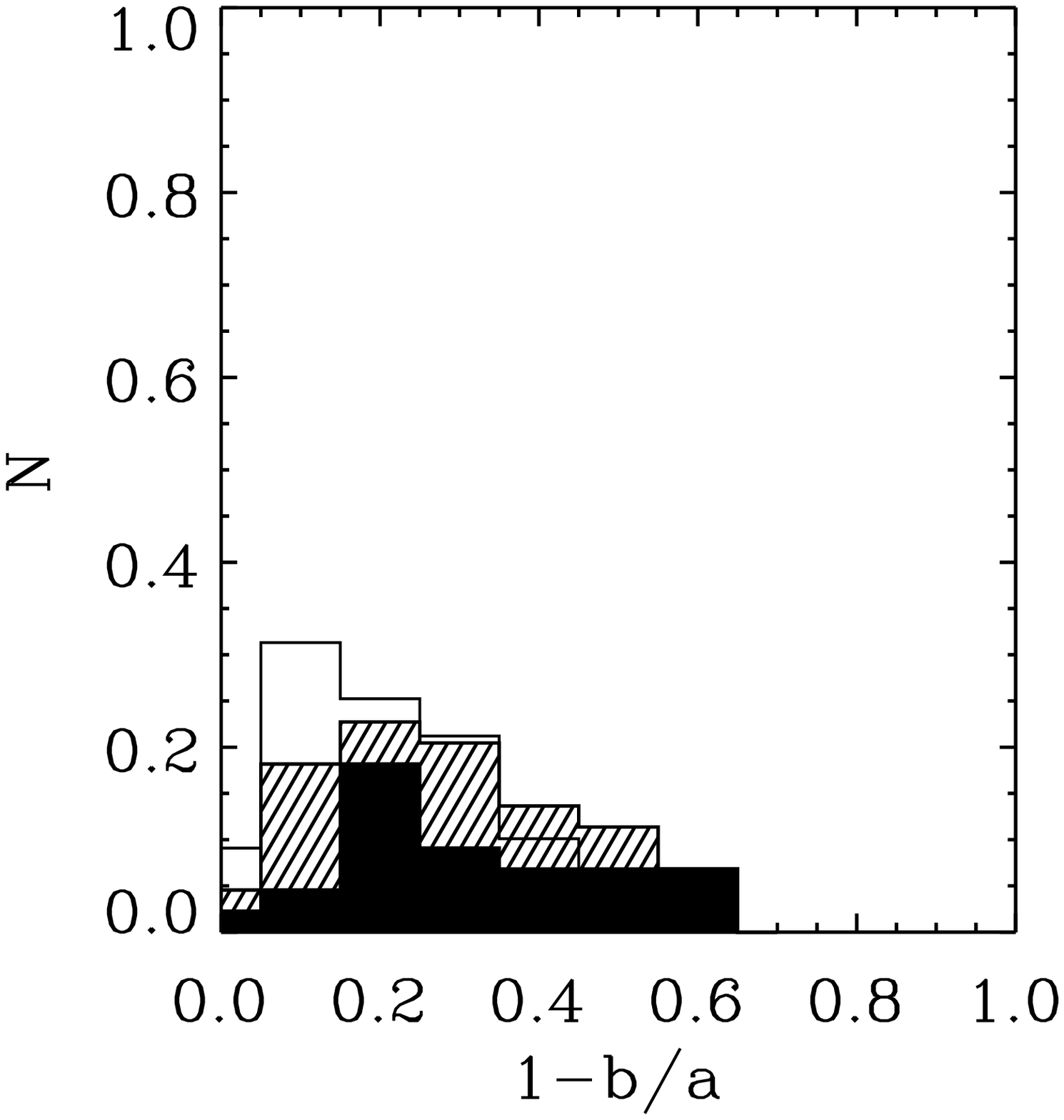}}
\caption{\label{fig-taylor} Histograms for the present 3CR sample (open) and the~\citet{taylor+96} quasar sample (full sample -- shaded; RLQs -- black) for the following properties (from left to right): host galaxy luminosity (converted $H$ band); scale length; ellipticity. $L^{\star}$ is indicated by a dotted line on the left-hand figure. The results of a two-sided K-S test for each distribution are presented in Table~\ref{tab-stats}. The Taylor et al. sample is the same as the Dunlop et al. one, but studied at $K$ band from the ground (see the text).}
\end{figure*}

{\bf Dunlop et al. (2003), Taylor et al. (1996): }
\citet{dunlop+03} studied the largest and most stringently selected sample of low-redshift quasars: 10 RLQs, 13 RQQs, and 10 radio galaxies at $z\approx0.2$. The radio-loud and RQQ subsamples were optically matched, and the RLQs and radio galaxies were matched in terms of their radio luminosities. The sample selection avoided core-dominated radio sources in order to minimize the chances of beaming producing artificially boosted apparent radio luminosities. Targets were observed using {\em HST}/WFPC2 with the F675W ($R$ band) filter. Figure~\ref{fig-dunlop} illustrates the various properties of the sample compared with those of the 3CR. The same objects were previously observed from the ground using UKIRT/IRCAM in $K$ band by~\citet{taylor+96} -- see Figure~\ref{fig-taylor}.

\begin{figure*}
\centering
{\includegraphics[height=4.0cm]{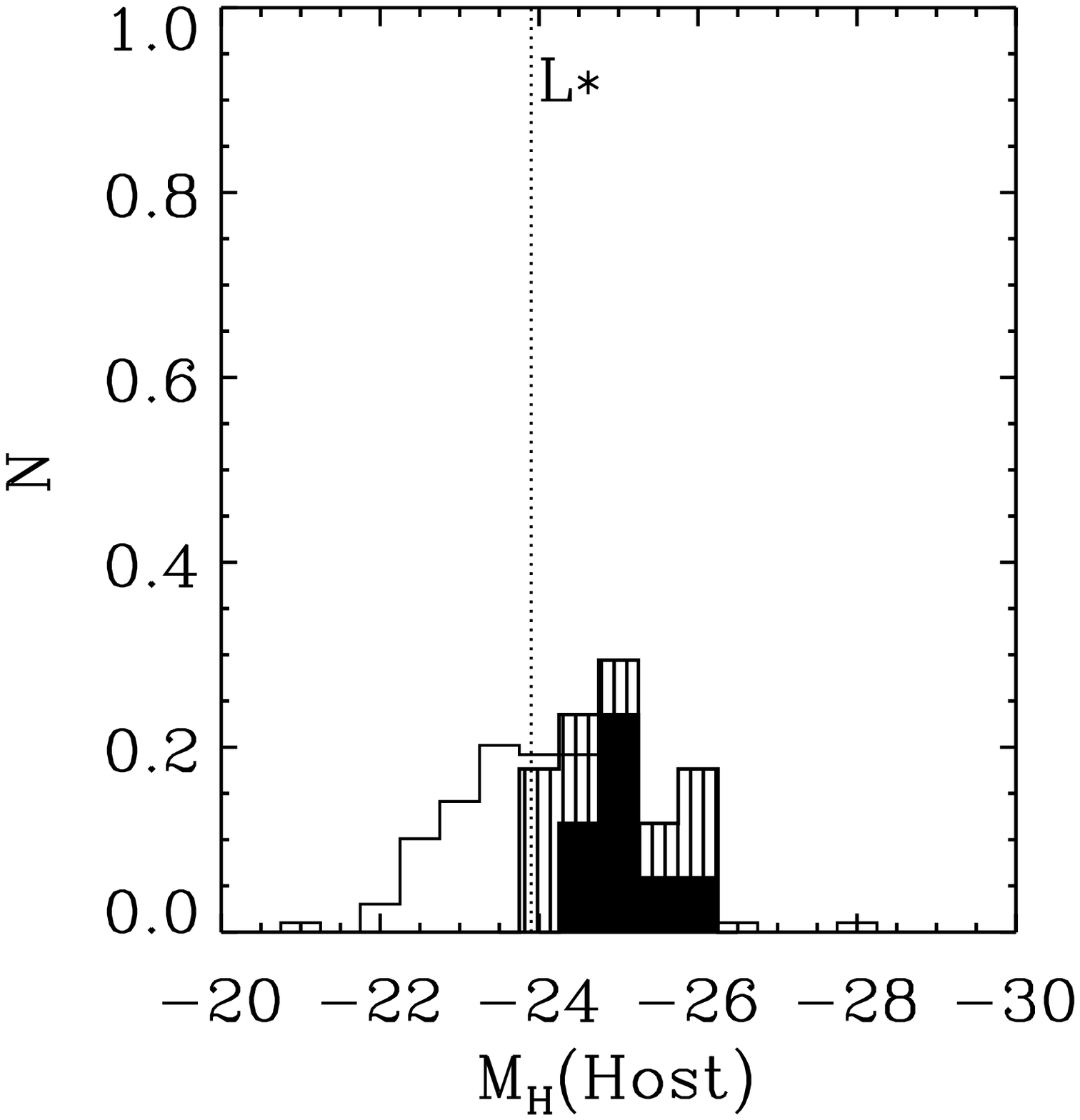}}
{\includegraphics[height=4.0cm]{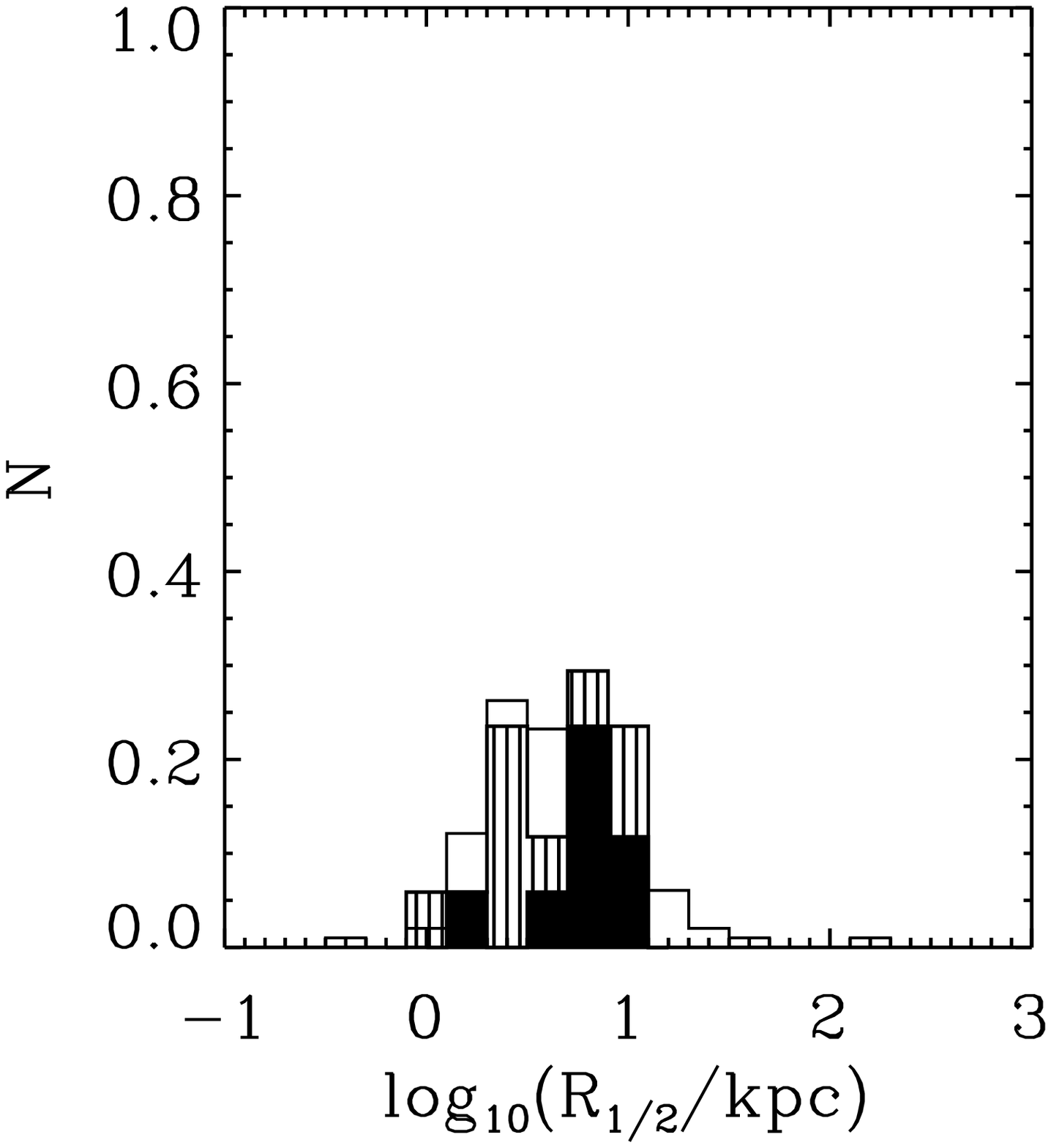}}
{\includegraphics[height=4.0cm]{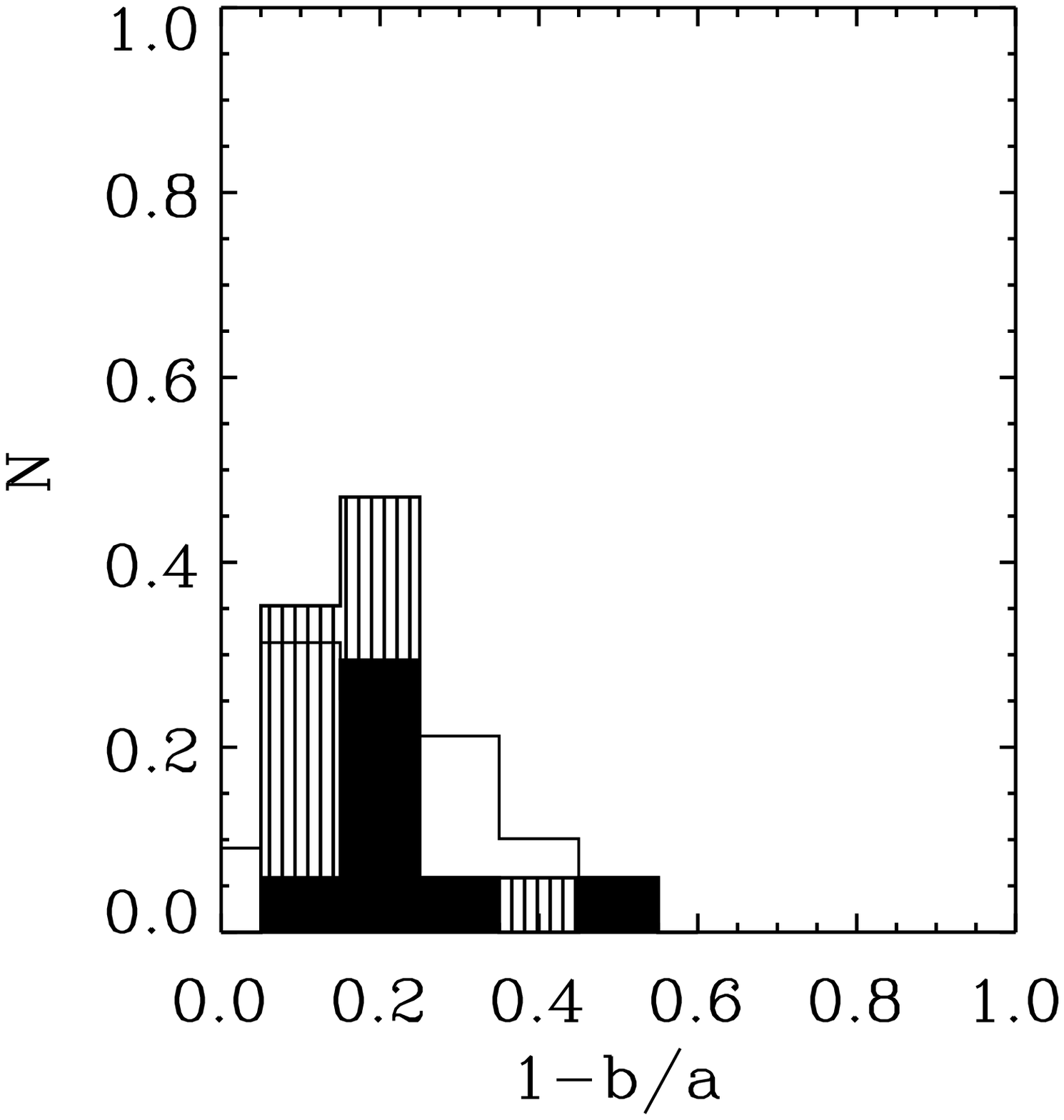}}
{\includegraphics[height=4.0cm]{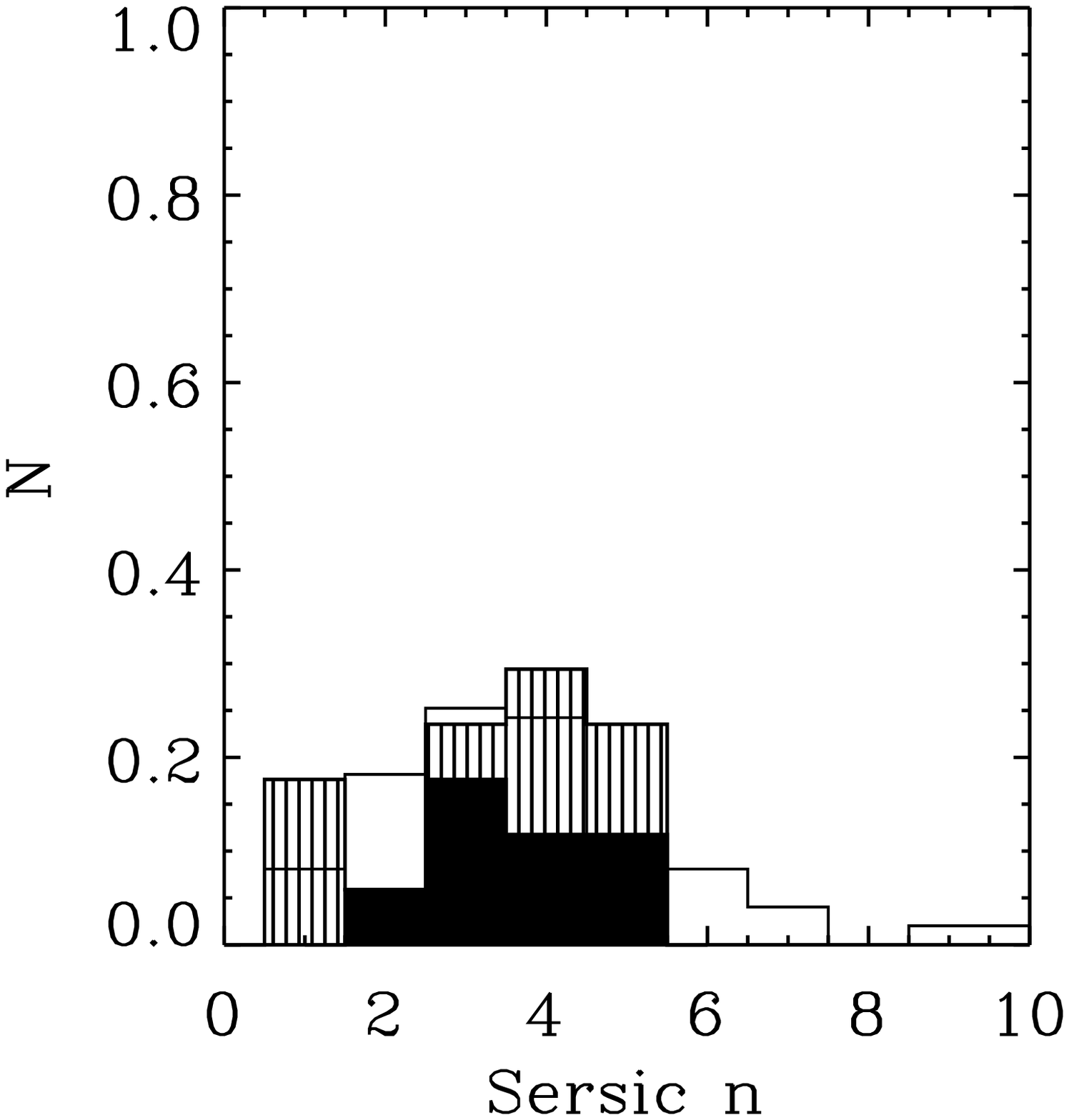}}
\caption{\label{fig-floyd} Histograms for the present 3CR sample (open) and the~\citet{floyd+04} quasar sample (full sample -- shaded; RLQs -- black) for the following properties (from left to right): host galaxy luminosity (converted $H$ band); scale length; ellipticity; \sersic index. $L^{\star}$ is indicated by a dotted line on the left-hand figure. The results of a two-sided K-S test for each distribution are presented in Table~\ref{tab-stats}.}
\end{figure*}
\begin{figure*}
\centering
{\includegraphics[height=4.0cm]{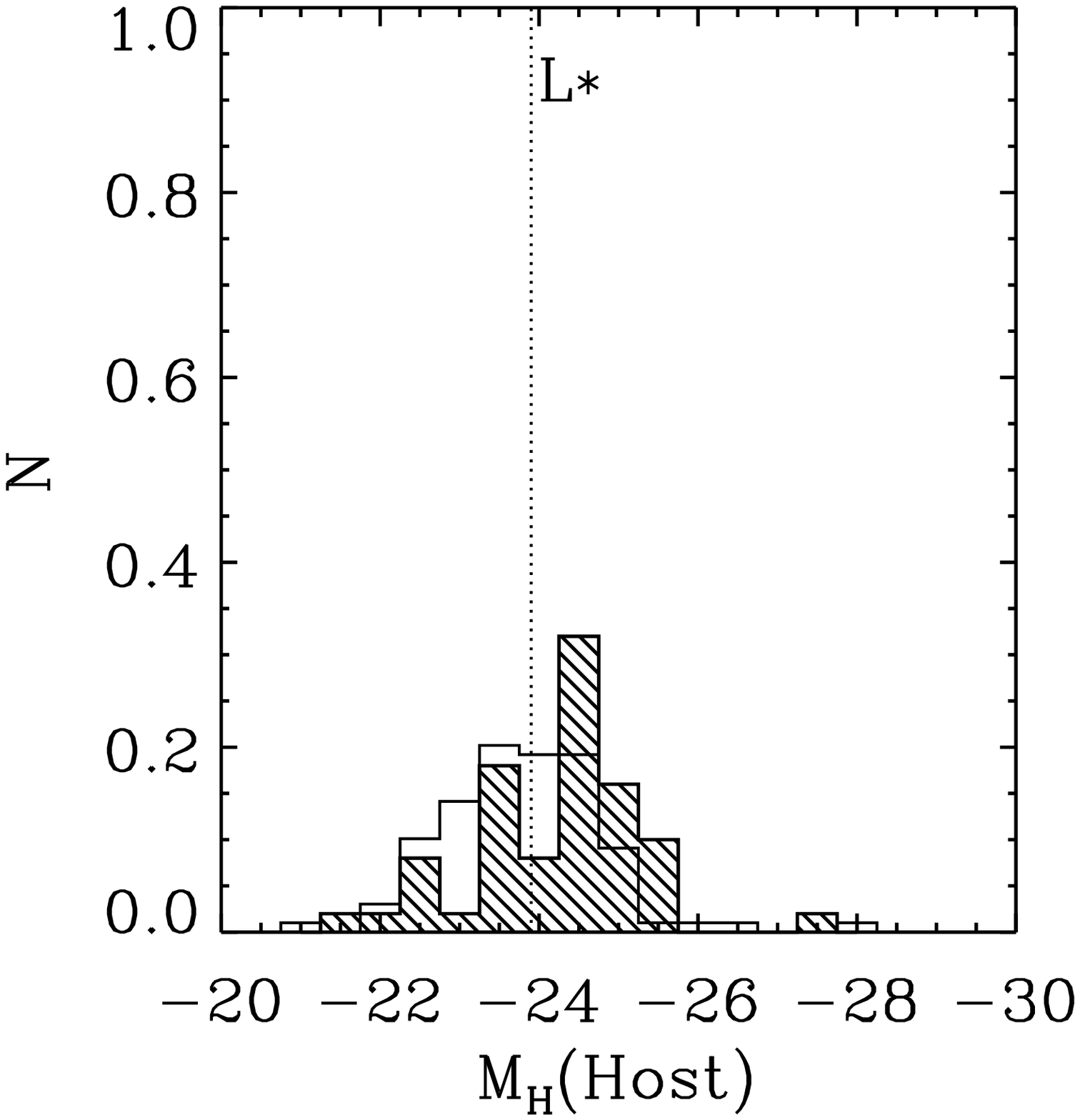}}
{\includegraphics[height=4.0cm]{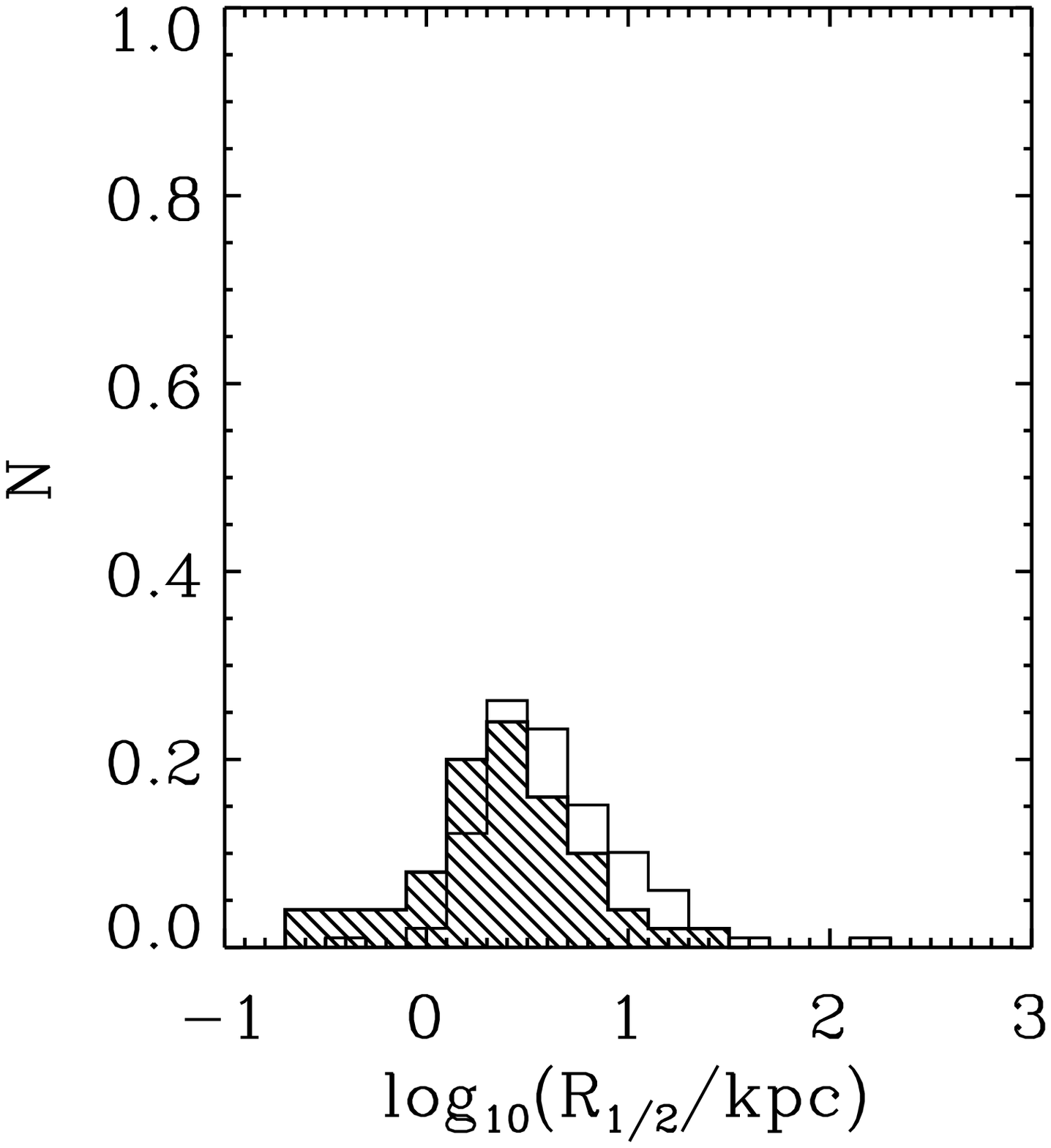}}
{\includegraphics[height=4.0cm]{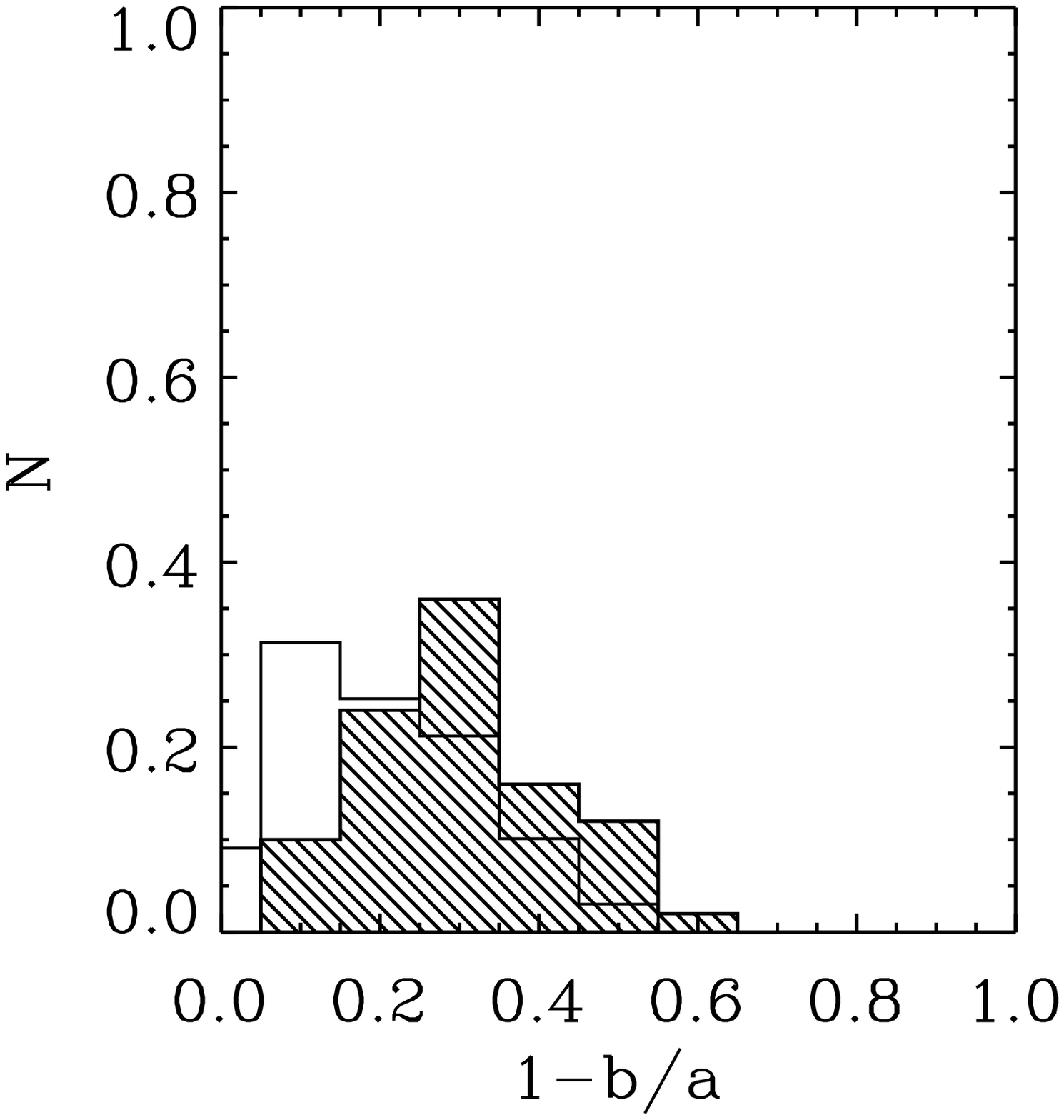}}
{\includegraphics[height=4.0cm]{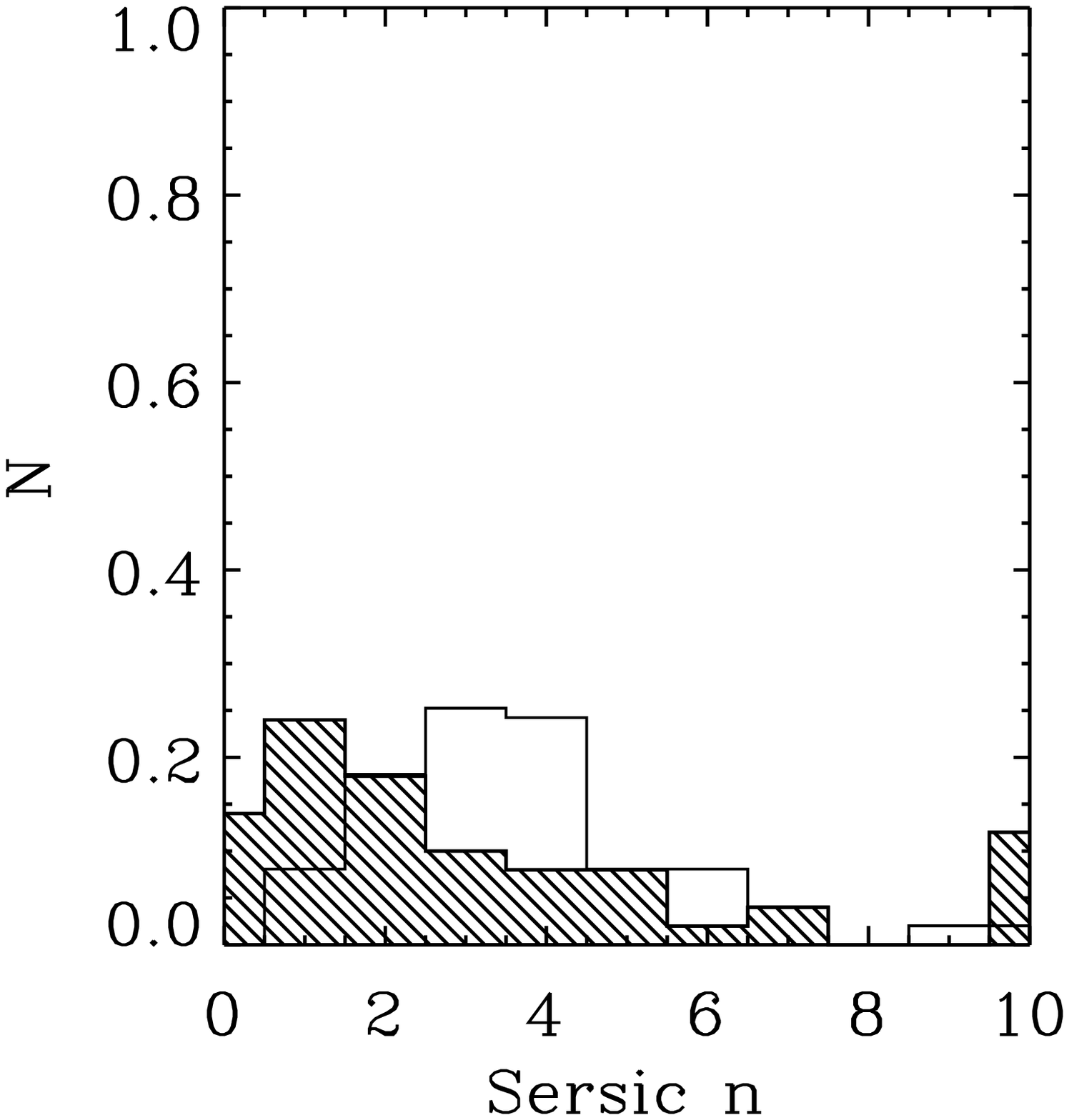}}
\caption{\label{fig-roth} Histograms for the present 3CR sample (open) and the~\citet{rothberg+04} merger sample (shaded) for the following properties (from left to right): host galaxy luminosity (converted $H$ band); scale length; ellipticity; \sersic index. $L^{\star}$ is indicated by a dotted line on the left-hand figure. The results of a two-sided K-S test for each distribution are presented in Table~\ref{tab-stats}.}
\end{figure*}

{\bf Floyd et al. (2004): }
\citet{floyd+04} studied the host galaxies of 17 quasars at $z\approx0.4$ (including those of the seven optically brightest QSO's at $z<0.4$) using {\em HST} / WFPC2 F814W ($I$ band) imaging. The lower luminosity subsample (10 objects) was divided into equal-sized subsamples of radio-loud and RQQs that are optically matched both with each other and with the sample of Dunlop et al. (2003). Similar care was taken to avoid beamed radio sources as in the Dunlop sample. Figure~\ref{fig-floyd} shows the various properties of the sample compared with those of the 3CR.

\subsubsection{Mergers}
\begin{figure*}[t]
\centering
{\includegraphics[height=4.0cm]{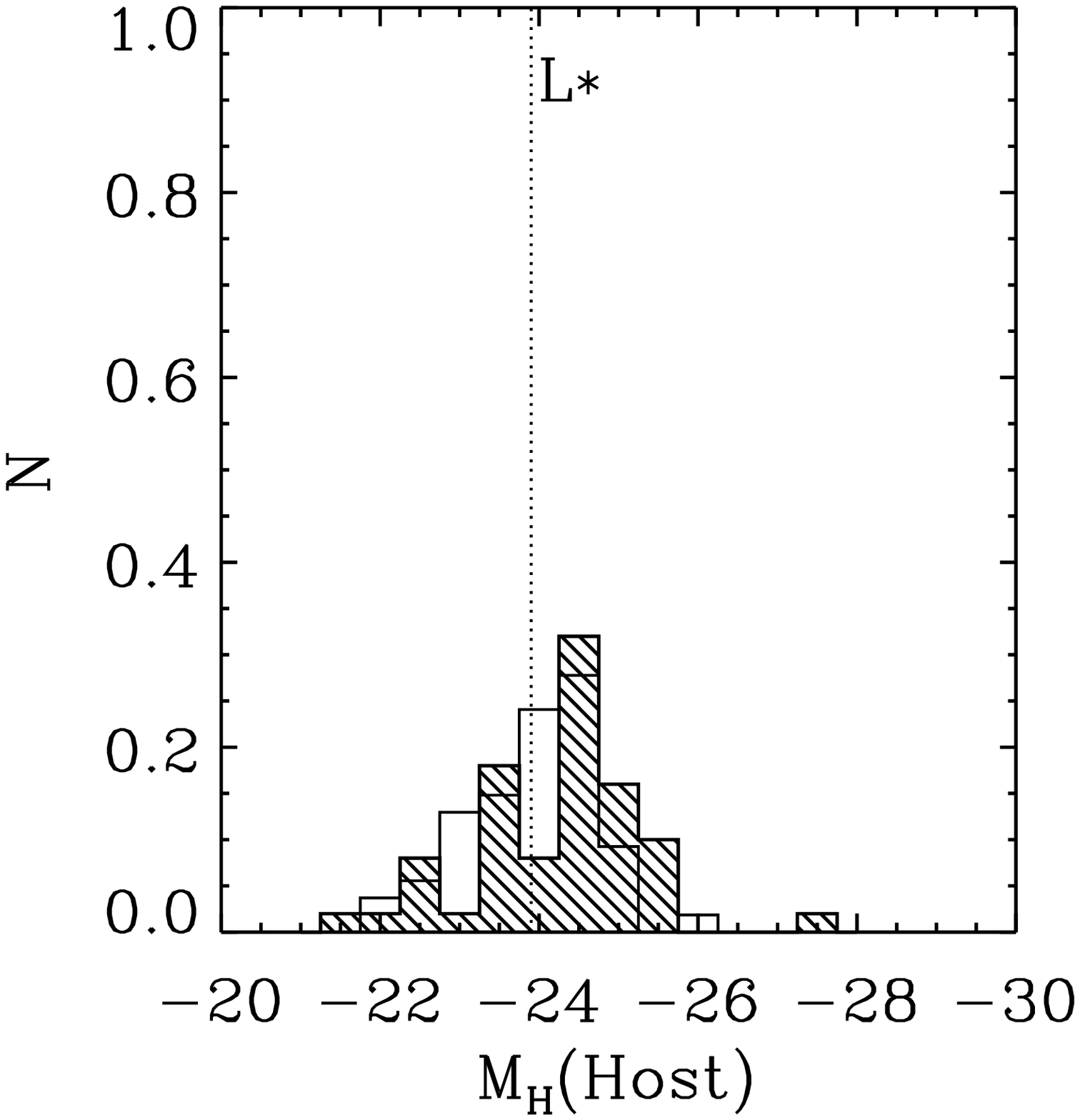}}
{\includegraphics[height=4.0cm]{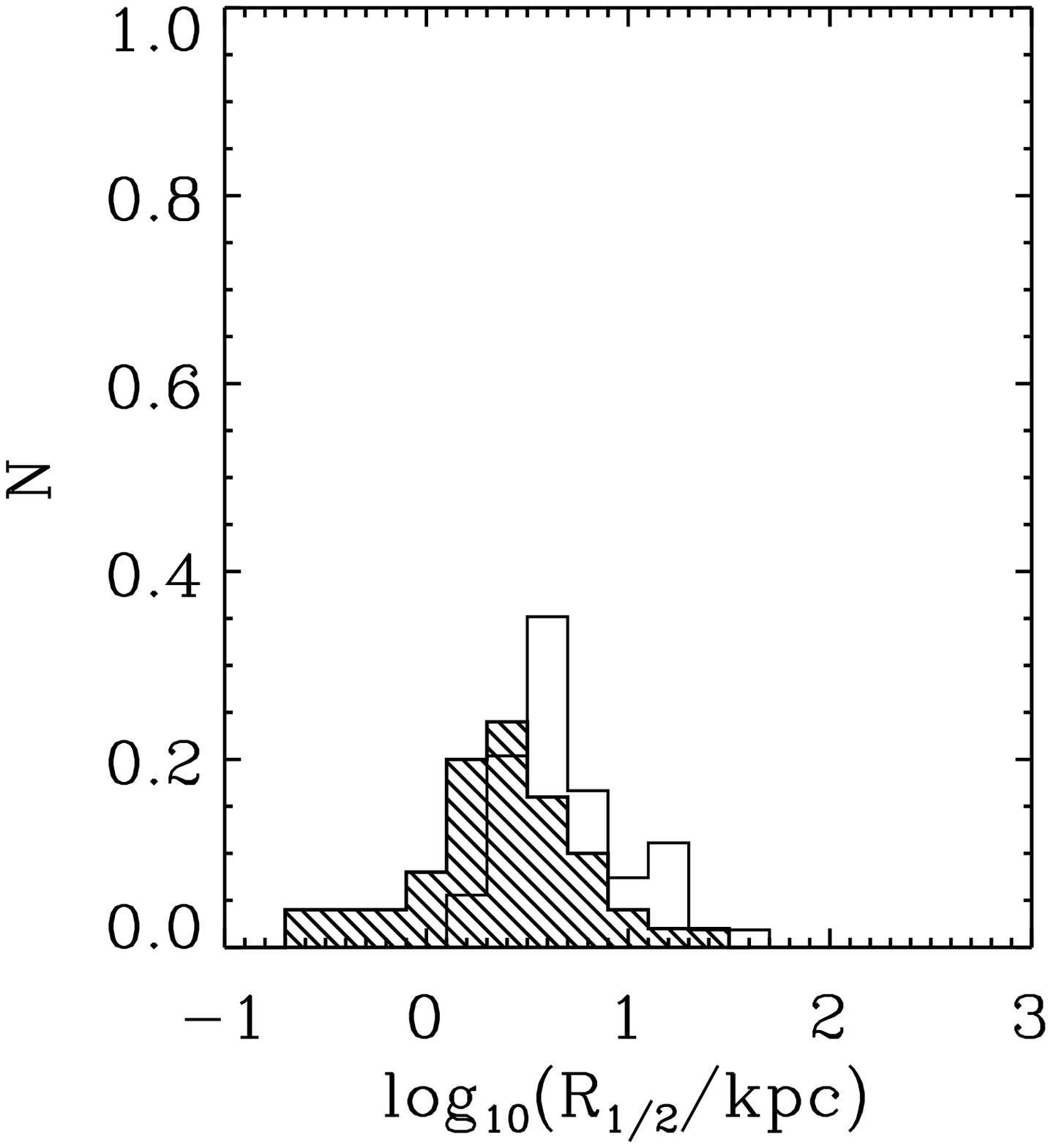}}
{\includegraphics[height=4.0cm]{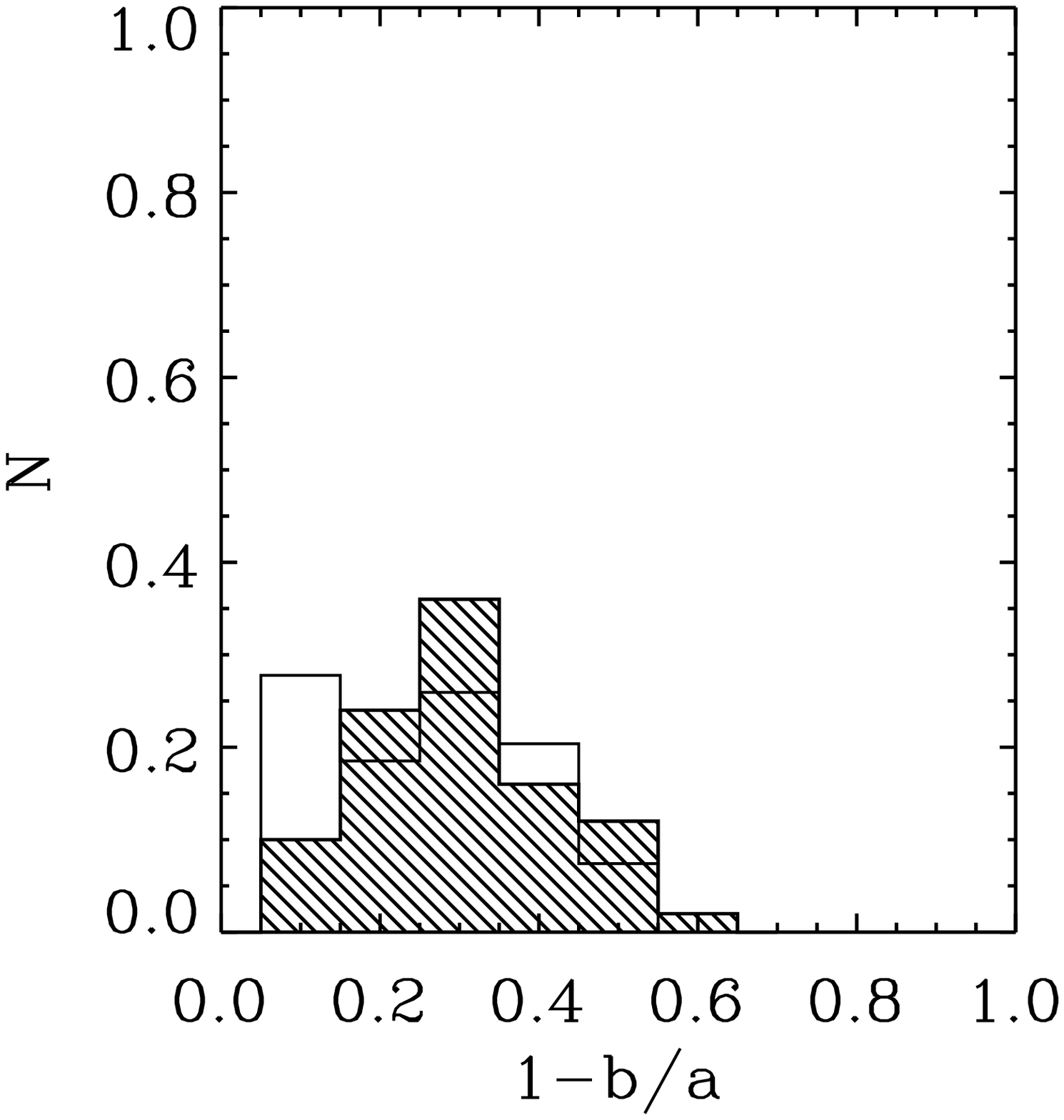}}
{\includegraphics[height=4.0cm]{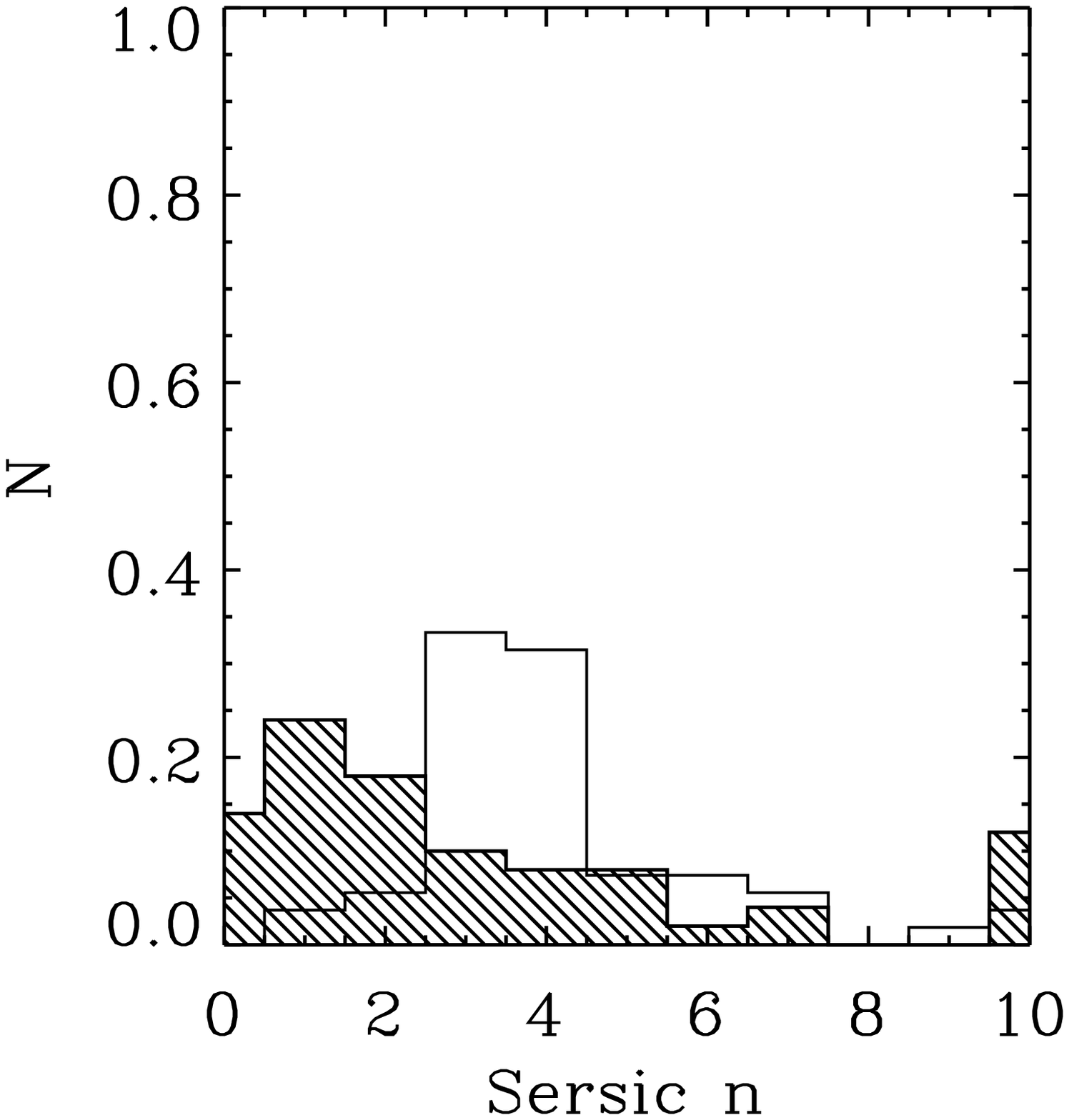}}
\caption{\label{fig-roth-merg} As for Figure~\ref{fig-roth}, but for Rothberg \& Joseph (2004 -- shaded) compared to only those 3CR sources that are flagged as {\it merger remnants} (open).}
\end{figure*}

{\bf Rothberg \& Joseph (2004): }
\citet{rothberg+04} studied a sample of 51 nearby ($z<0.045$) candidate disk-merger remnants, all with a single nucleus, and based on optical morphologies to include objects with tidal loops, tails and shells. The study used ground-based $K$ band data and one-dimensional radial profile fitting to elliptical isophotal values in the same way as our one-dimensional approach. Observations were taken using QUIRC at {\em f}/10 on the University of Hawaii 2.2~m telescope. This is the only large sample or mergers that has been studied in such morphological detail to date. 
Figure~\ref{fig-roth} shows the Rothberg sample compared to the 3CR. Figure~\ref{fig-roth-merg} shows the Rothberg sample compared to only those 3CR sources that were flagged as mergers in Paper~II.

\begin{figure*}
\centering
{\includegraphics[height=4.0cm]{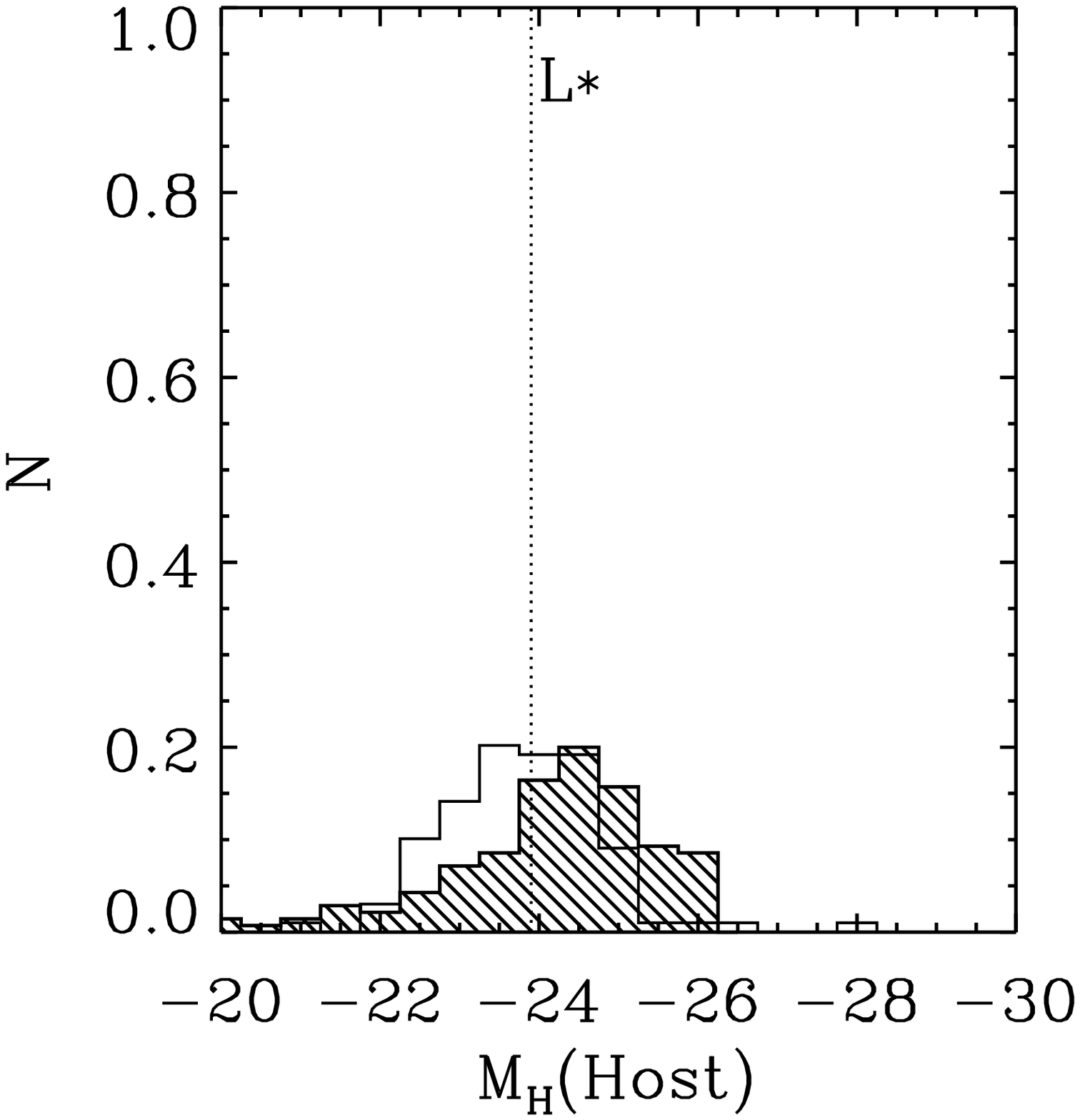}}
{\includegraphics[height=4.0cm]{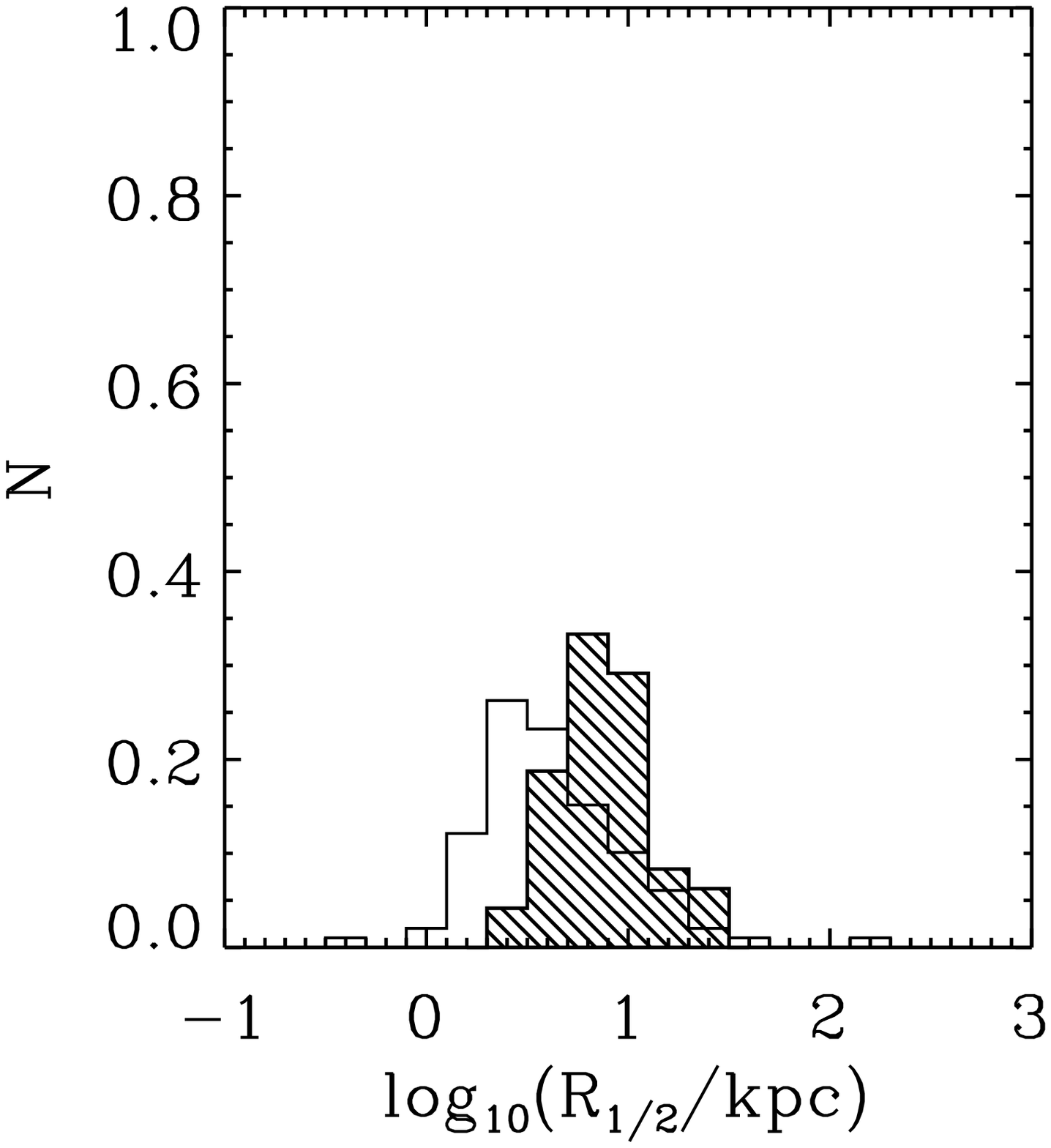}}
{\includegraphics[height=4.0cm]{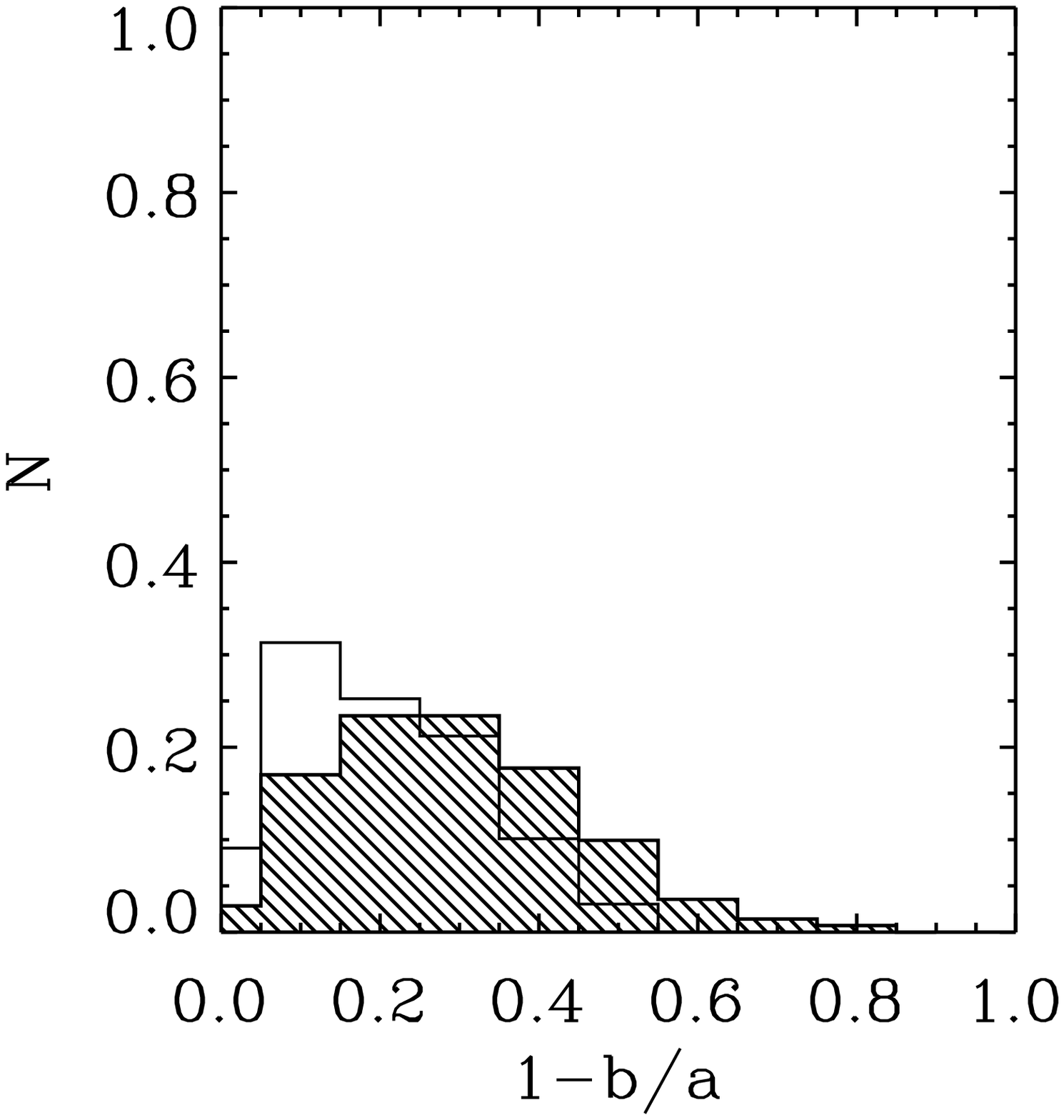}}
\caption{\label{fig-BBF92} Histograms for the present 3CR sample (open) and~\citet{BBF92} early-type galaxy sample (shaded) for the following properties (from left to right): host galaxy luminosity (converted $H$ band); scale length; ellipticity. $L^{\star}$ is indicated by a dotted line on the left-hand figure. The results of a two-sided K-S test for each distribution are presented in Table~\ref{tab-stats}.}
\end{figure*}
\begin{figure*}
\centering
{\includegraphics[height=4.0cm]{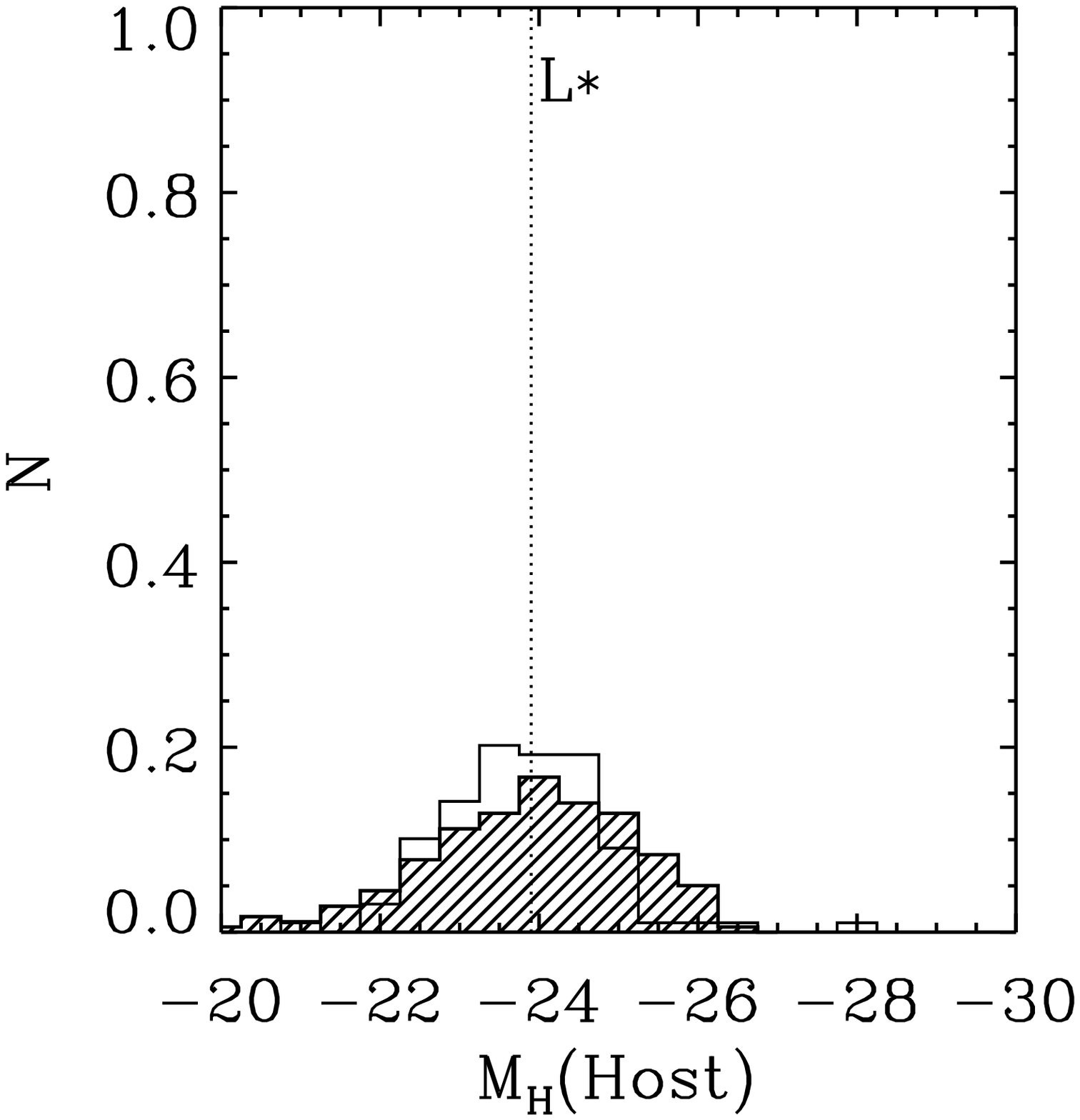}}
{\includegraphics[height=4.0cm]{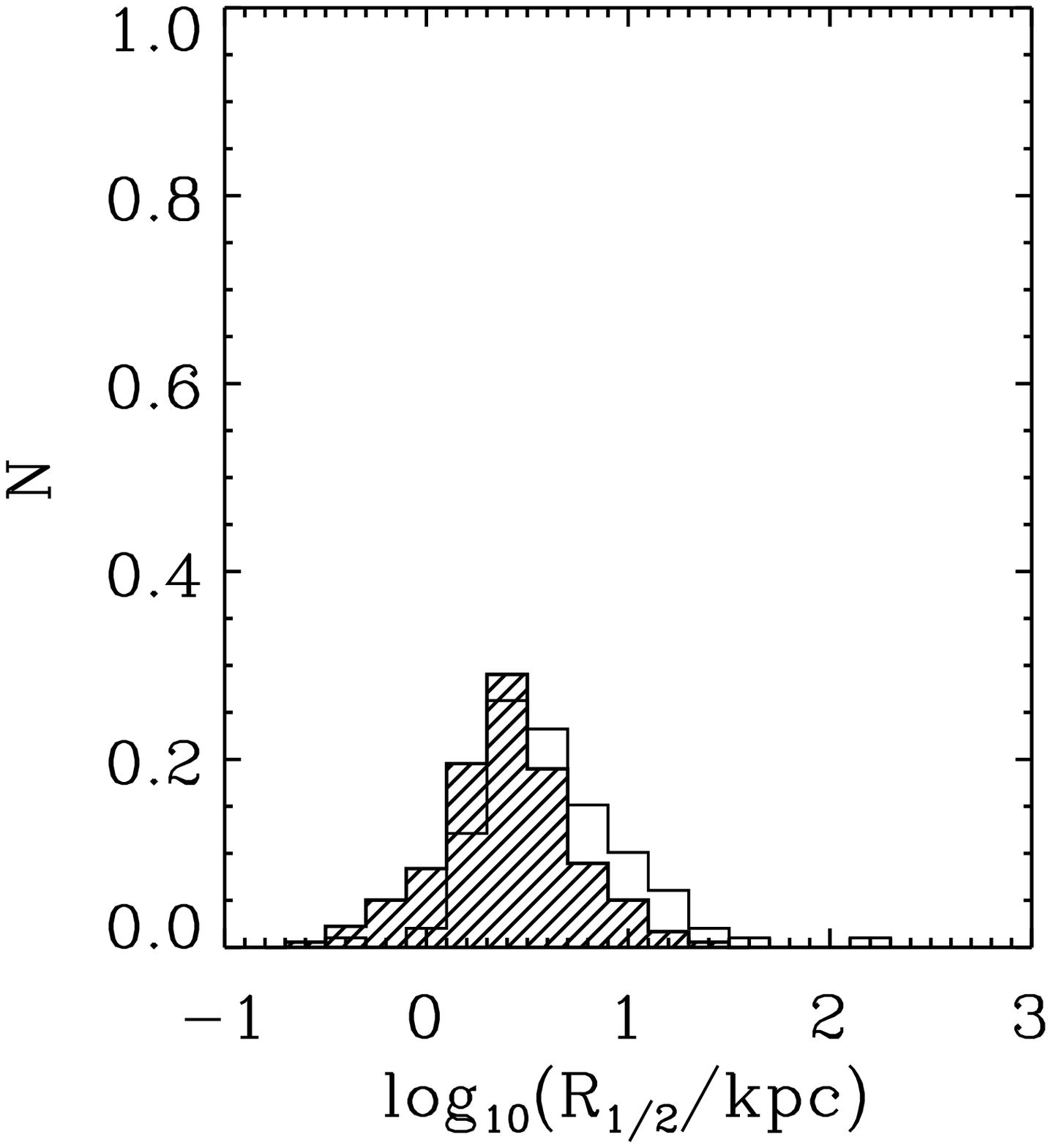}}
{\includegraphics[height=4.0cm]{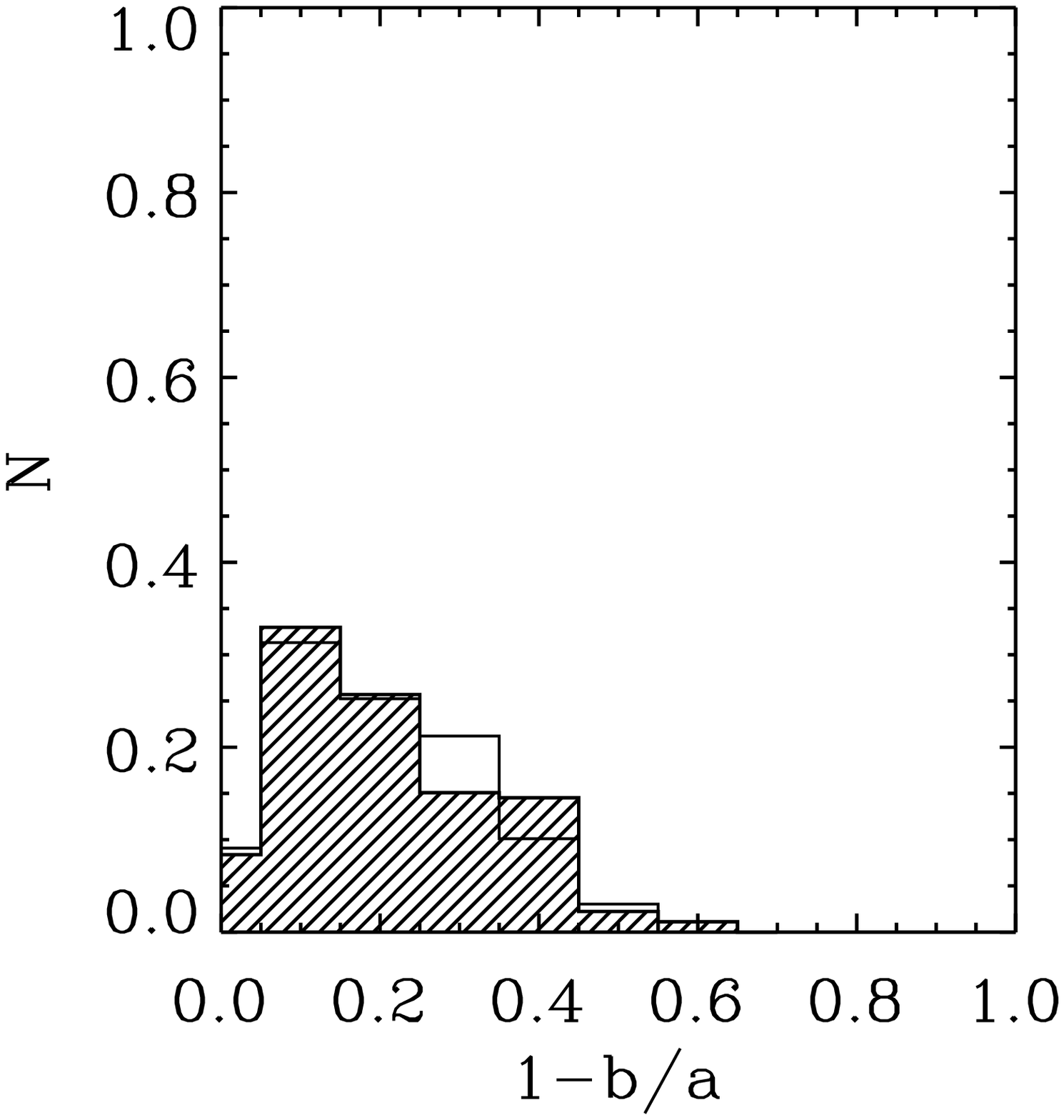}}
\caption{\label{fig-pahre} Histograms for the present 3CR sample (open) and the~\citet{pahre99} early-type galaxy sample (shaded) for the following properties (from left to right): host galaxy luminosity (converted $H$ band); scale length; ellipticity. $L^{\star}$ is indicated by a dotted line on the left-hand figure. The results of a two-sided K-S test for each distribution are presented in Table~\ref{tab-stats}.}
\end{figure*}

\subsubsection{Elliptical and Early-Type Galaxies}
{\bf Bender, Burnstein \& Faber (1992): }
\citet{BBF92} studied a sample of local ($D<140$~Mpc) ``dynamically hot galaxies'', that are in practice a range of spheroids and bulges selected in terms of their $B$ band luminosity. The sample includes 48 giant ellipticals (defined as having $M_B\leq-20.5$), 20 intermediate ellipticals ($-20.5 \leq M_B \leq -18.5$), 12 bright dwarf ellipticals ($M_B>-18.5$) as well as 4 compact ellipticals, 19 bulges and 5 lower luminosity dwarf spheroids. Observations and modeling were presented by~\citet{bendermoellenhoff87}. Observations were made using the Calar-Alto 1.23~m telescope and a NEC P8603/A detector. modeling followed a similar method to our one-dimensional fits to elliptical isophotal values, presented in Paper~II. The sample is compared with the 3CR in Figure~\ref{fig-BBF92}.
We adopt this sample as a useful cross-section of the elliptical population.

{\bf Pahre (1999): }
\citet{pahre99} observed 341 nearby ($z\ltsim0.03$) early-type galaxies in $K$ band using the Palomar 60~inch and the Las Campanas 1~m (Swope) and 2.5~m (Du Pont) telescopes and Rockwell NICMOS3 HgCdTe detectors. Most are from 13 rich clusters (85\%), but also with 12\% from loose groups and 3\% from the general field. These span the full range of early types, forming a useful reference for the 3CR. Modeling was performed in a similar way to our one-dimensional fits to elliptical isophotes, as described in Paper~II, although all masking was performed automatically, instead of some removal by hand of obvious companion sources as in our modeling. No \sersic fit is given. Figure~\ref{fig-pahre} illustrates the various properties of the sample compared with those of the 3CR.

\begin{figure*}[t]
\centering
{\includegraphics[height=4.0cm]{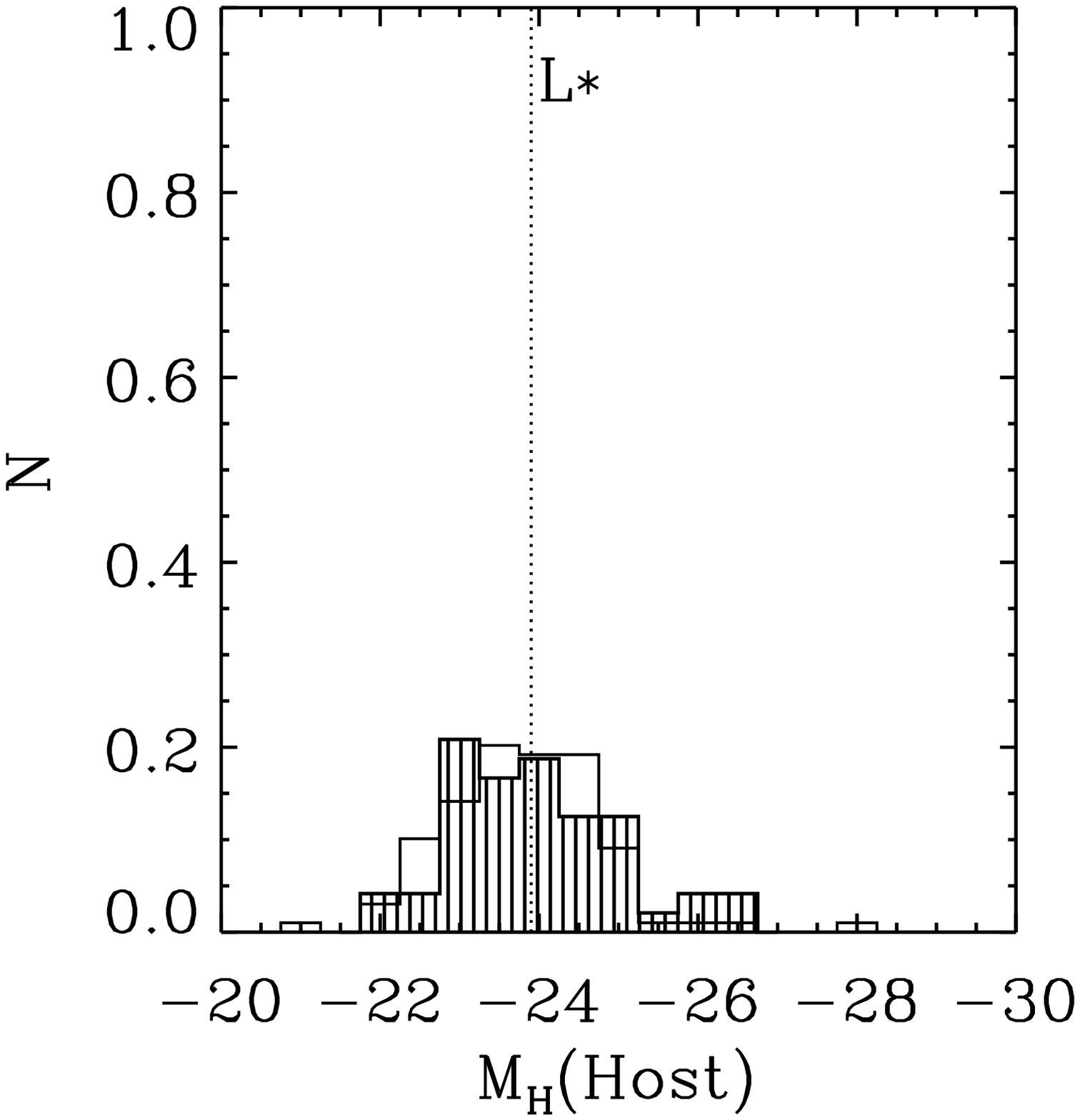}}
{\includegraphics[height=4.0cm]{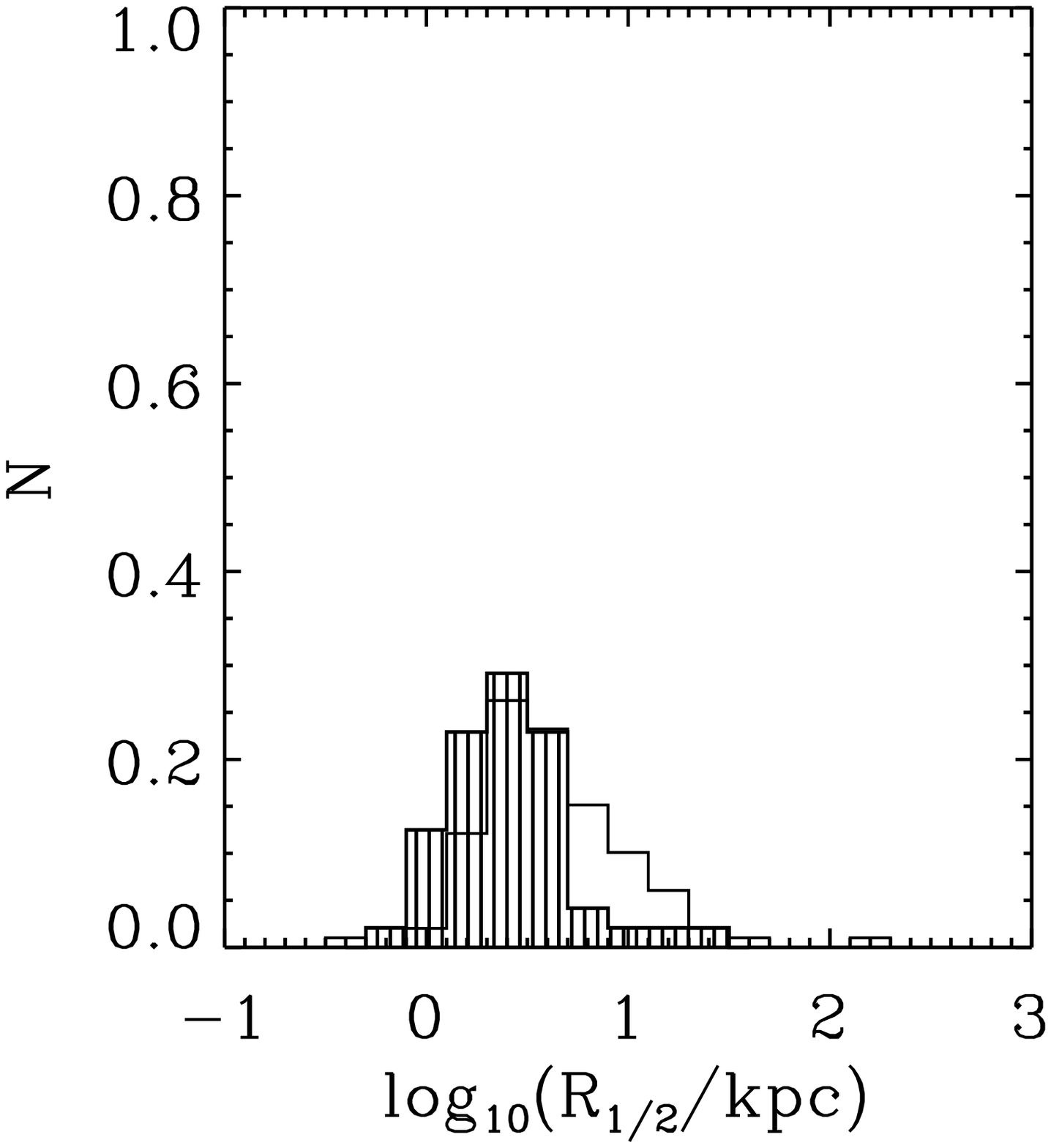}}
\caption{\label{fig-mob} Histograms for the present 3CR sample (open) and the~\citet{mobasher+99} early-type galaxy sample (shaded) for the following properties (from left to right): host galaxy luminosity (converted $H$ band); scale length. $L^{\star}$ is indicated by a dotted line on the left-hand figure. The results of a two-sided K-S test for each distribution are presented in Table~\ref{tab-stats}.}
\end{figure*}

{\bf Mobasher et al. (1999)}
\citet{mobasher+99} studied the Fundamental Plane in a sample of 48 elliptical galaxies (no lenticulars) from the Coma cluster ($\bar{D}=99$~Mpc) in $K$ band. They were selected to have optical half-light diameters of less than 1~arcmin in order to fit into the field of view of IRCAM3 on UKIRT. Curves of growth were produced for each target, rather than a functional form being fitted. Effective diameter and effective mean surface brightness were calculated directly from the curve of growth. Figure~\ref{fig-mob} illustrates the various properties of the sample compared with those of the 3CR.

\begin{figure*}
\centering
{\includegraphics[height=4.0cm]{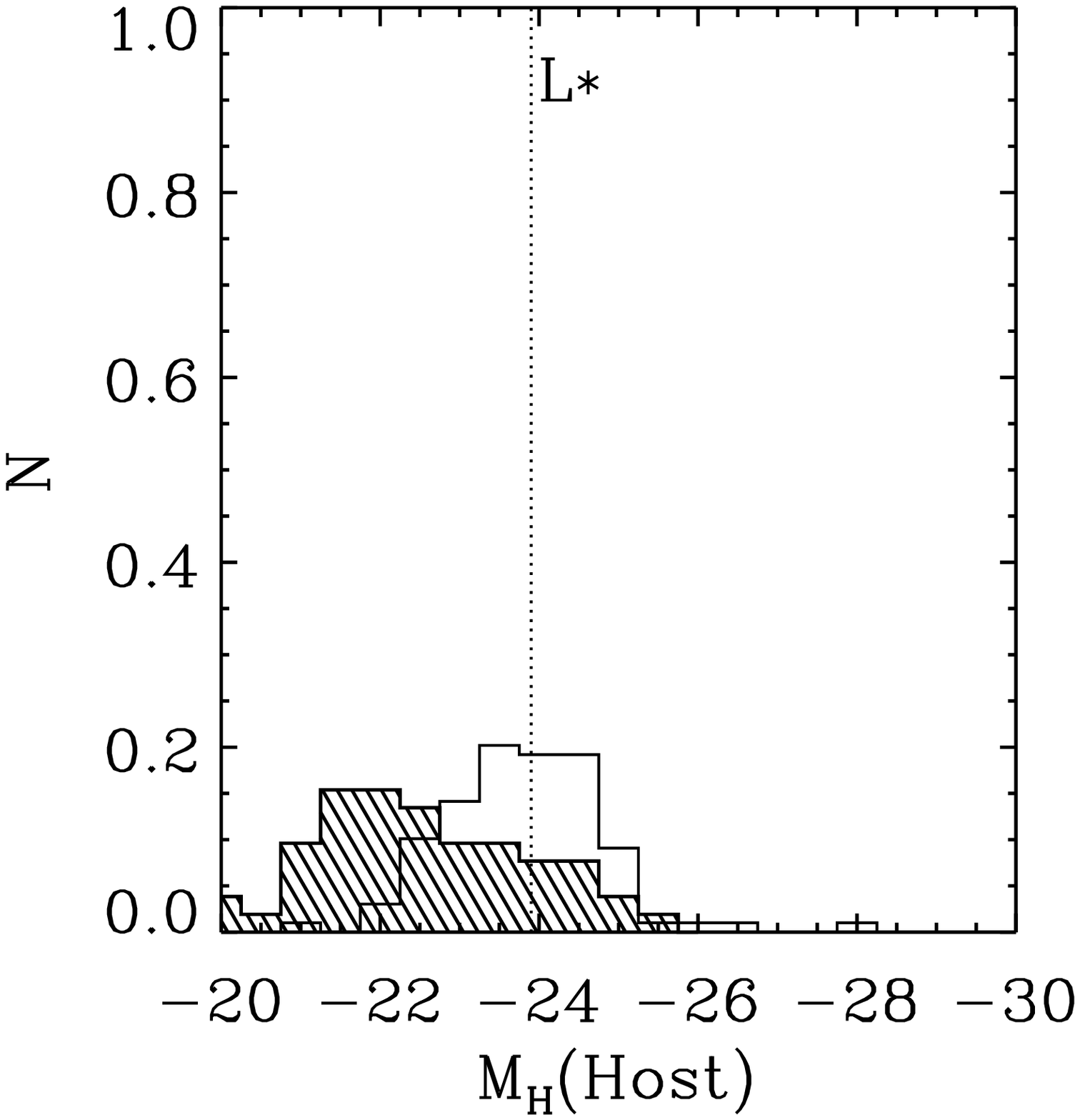}}
{\includegraphics[height=4.0cm]{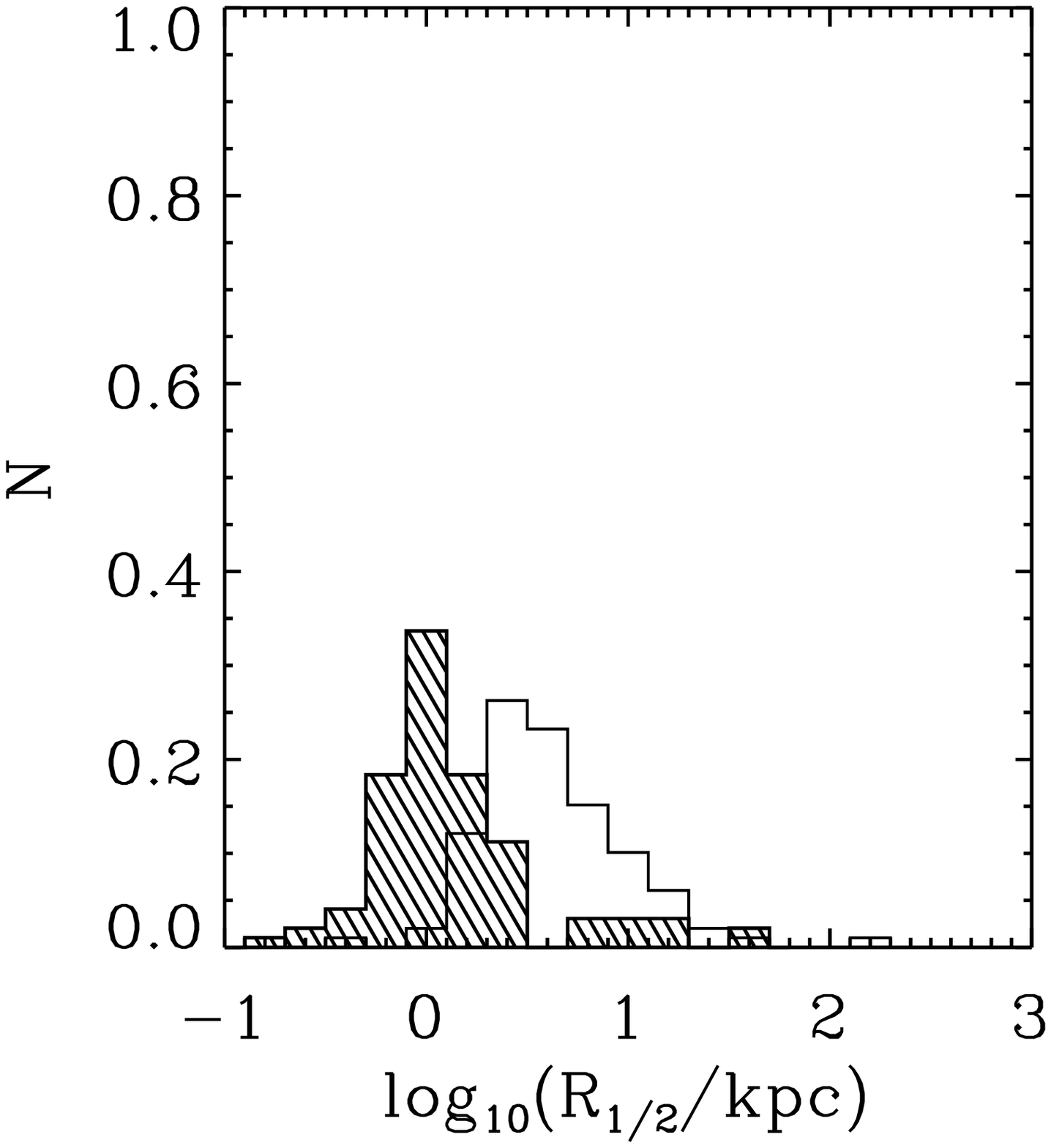}}
{\includegraphics[height=4.0cm]{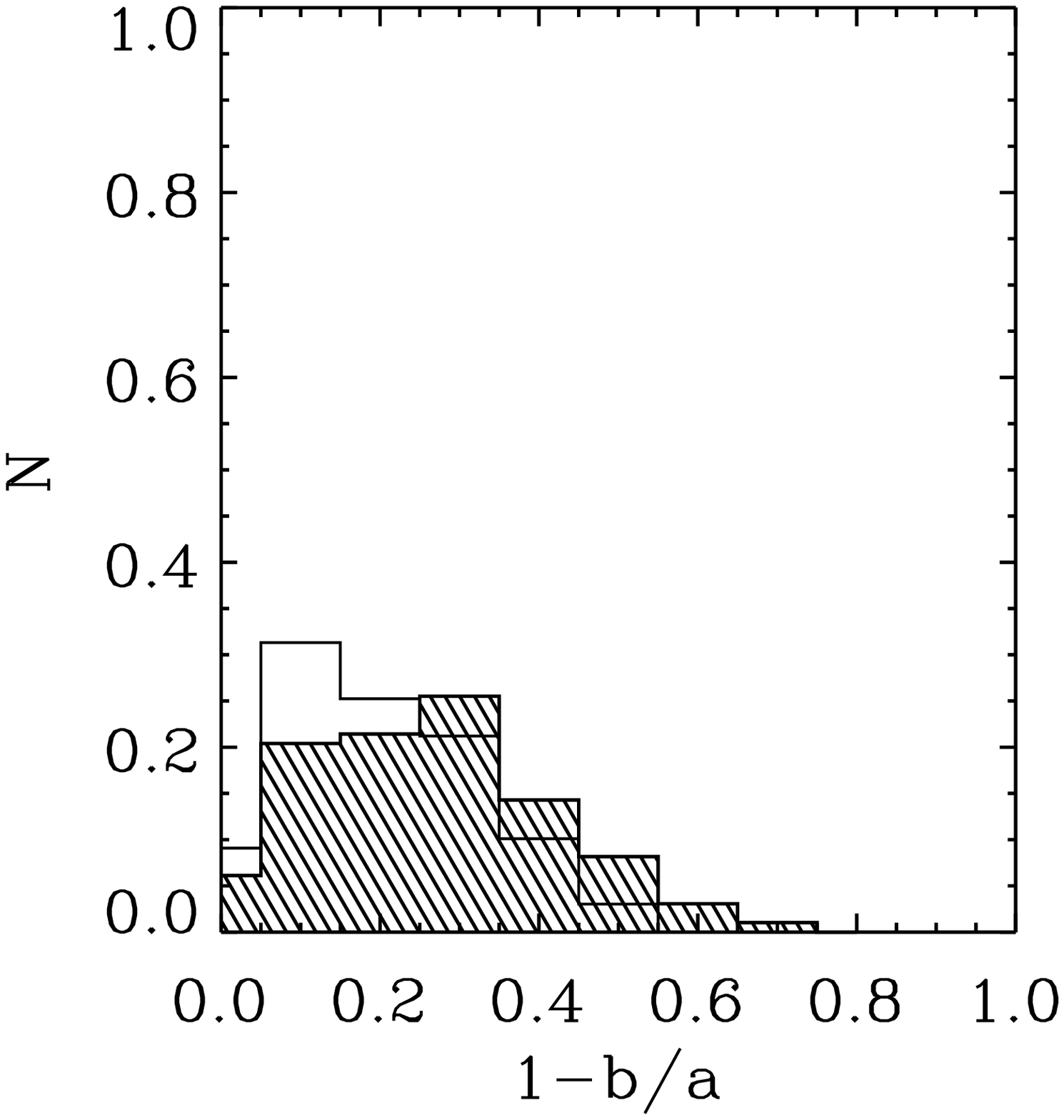}}
{\includegraphics[height=4.0cm]{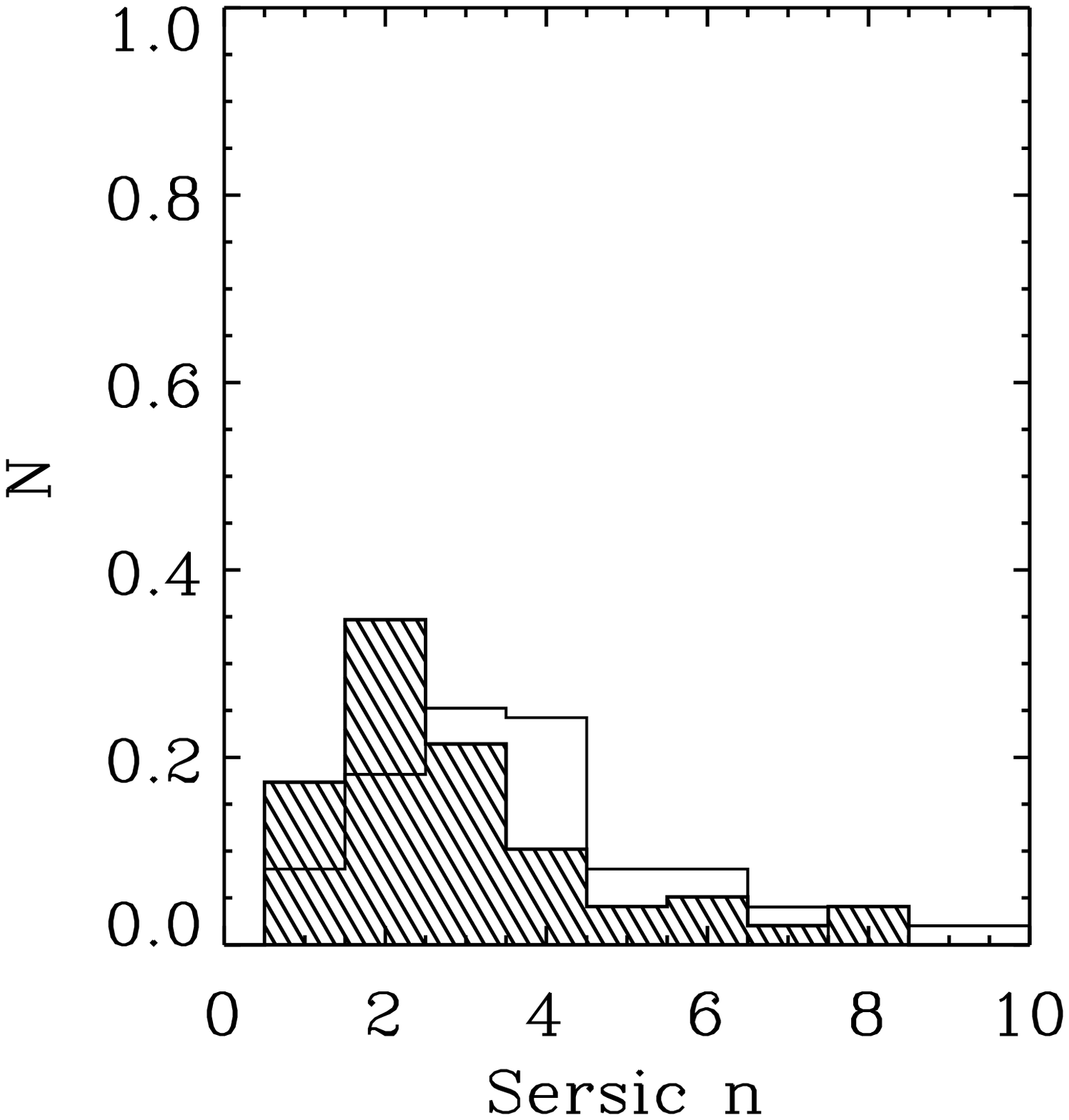}}
\caption{\label{fig-ferr} Histograms for the present 3CR sample (open) and the~\citet{ferrarese+06} galaxy Virgo cluster galaxy sample (shaded) for the following properties (from left to right): host galaxy luminosity (converted $H$ band); scale length; ellipticity; \sersic index. 
$L^{\star}$ is indicated by a dotted line on the left-hand figure. The results of a two-sided K-S test for each distribution are presented in Table~\ref{tab-stats}.}
\end{figure*}

{\bf Ferrarese et al. (2006): }
As part of the Advanced Camera for Surveys (ACS) Virgo Cluster Survey,~\citet{ferrarese+06} studied a sample of 100 early-type members of the Virgo cluster ($\bar{D}=18$~Mpc) from dwarfs ($M_B=-15.1$~mag) to giants ($M_B=-21.8$~mag) imaged in F475W and F850LP ($g$ and $z$ bands). This forms a useful local early-type galaxy baseline with which to compare our the luminosity of our sample, allowing us to rank the host galaxies of the 3CR against the entire contents of a nearby cluster. Modeling performed by Ferrarese et al. was identical to our ellipse fitting followed by radial profile fits to the elliptical isophote values in Paper~II, although Ferrarese et al. also fit core--\sersic~\citep{graham+03} models to objects that are sufficiently close to detect a ``Nuker''--like~\citep{lauer+95} turnover or ``core'' in some of the galaxy centers. We compare results to their $z$ band data, converted to $H$ as described above. Figure~\ref{fig-ferr} illustrates the various properties of the sample compared with those of the 3CR.
We note that many of these objects have noticeable disks in addition to their bulges (S0 types), reflected in their intermediate \sersic index values. This use of a single \sersic profile in such cases is questionable.

\begin{figure*}[htbf]
\centering
{\includegraphics[height=4.0cm]{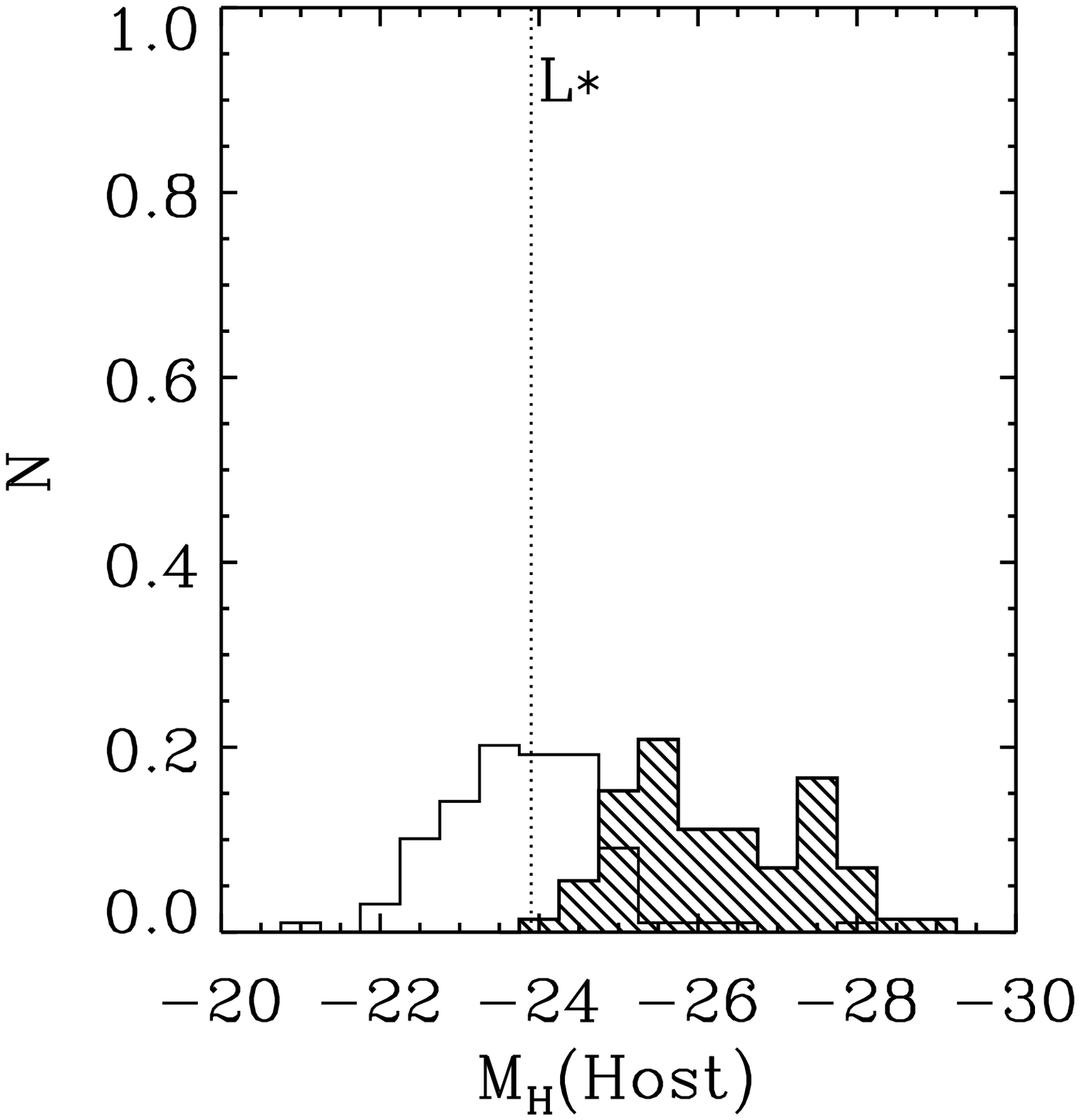}}
{\includegraphics[height=4.0cm]{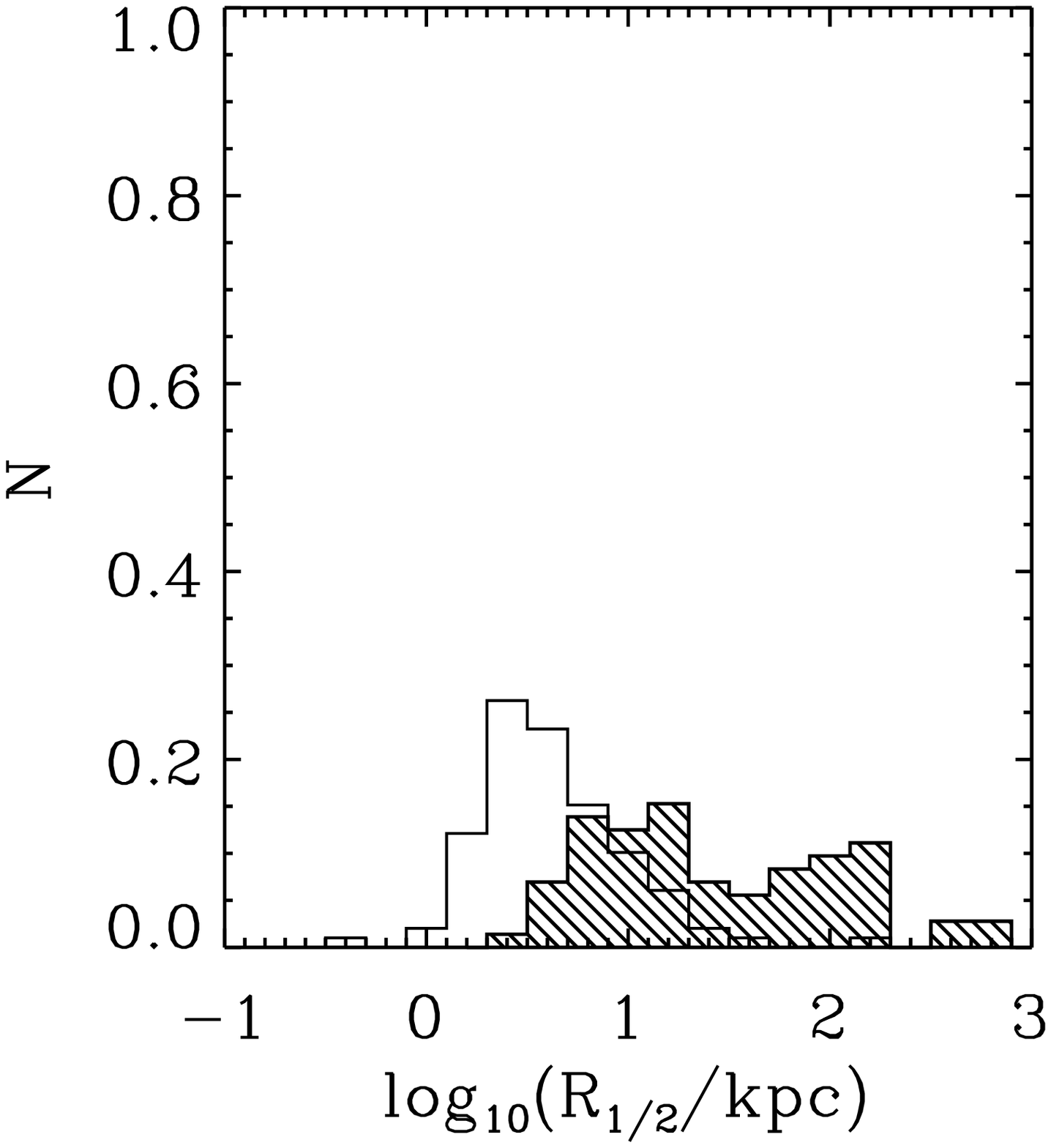}}
{\includegraphics[height=4.0cm]{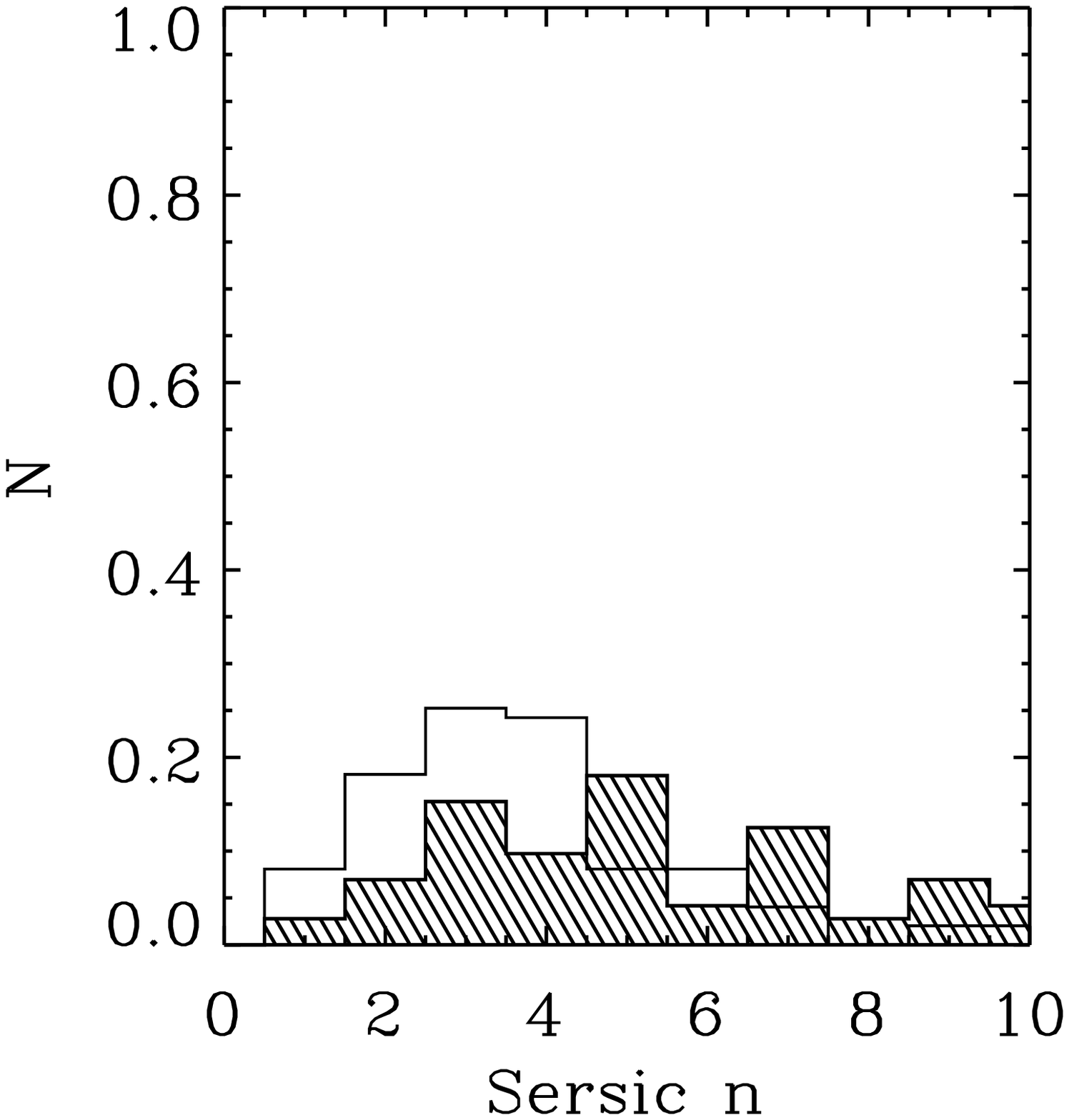}}
\caption{\label{fig-graham} Histograms for the present 3CR sample (open) and the~\citet{graham+96} BCG sample (shaded) for the following properties (from left to right): host galaxy luminosity (converted $H$ band); scale length; \sersic index. $L^{\star}$ is indicated by a dotted line on the left-hand figure. The results of a two-sided K-S test for each distribution are presented in Table~\ref{tab-stats}.}
\end{figure*}

\subsubsection{BCGs}
{\bf Graham et al. (1996): }
\citet{graham+96} took the sample of 119 Abell BCGs ($0.01<z<0.054$) originally observed in Cousins $R$ band by~\citet{postmanlauer95} on the CTIO 1.5~m and the KPNO 2.1~m and 4~m  telescopes. \sersic profiles were fit to the semi-major axis surface brightness profiles of~\citet{postmanlauer95}, similar to our one-dimensional fitting of profiles to the elliptical isophotal levels in Paper~II. Figure~\ref{fig-graham} illustrates the various properties of the sample compared with those of the 3CR. This remains a state-of-the-art data set for the study of BCG's in spite of the age of the observations.

%???\citet{grahamguzman03}: \sersic parameters 

%%%%%%%%%%%%%%%%%%%%%%%%%%%%%%%%%%%%%%%%%%%%%%%%%%%

\section{Results}
\label{sec-res}
For the full tabulated results of the morphological modeling of each individual source, see Paper~II, Tables 4 and 5. Here we only discuss those results in a sample-wide comparison. Histograms comparing the distributions of various properties of these samples with those of our 3CR sample are presented in Figures~\ref{fig-urry}--\ref{fig-graham}. We performed a K-S test to compare each property of each comparison sample to the equivalent property of our 3CR sample. This returns the K-S statistic, $D$, reflecting the cumulative difference between two samples, and a likelihood, $p$ that the two samples are drawn from the same population.
Results of the K-S test, showing the goodness of fit between each sample and the 3CR are presented in Table~\ref{tab-stats}. 
Table~\ref{tab-stats_FRi} shows each sample compared to just the FR~I sources in our 3CR sample. 
Table~\ref{tab-stats_FRii} shows each sample compared to just the FR~II sources in our 3CR sample.
In the following we present a broad sample-by-sample comparison, followed by more detailed comparisons in each observational metric (morphology, ellipticity, luminosity, scale length and the $R_e-\mu_e$ Kormendy relation). In the discussion Section~\ref{sec-disc} below, we engage in a more detailed discussion of each property.

\subsection{Sample-by-sample Comparisons}
The BL~Lac object host galaxies of the~\citet{urry+00} sample span the upper luminosity and scale length range of the 3CR host galaxies -- Figure~\ref{fig-urry}. Galaxies as dim (small) as the very bottom end of the 3CR galaxy luminosity (scale length) distribution are conspicuously missing, and this is also found with the other active galaxy samples discussed below. We return to discuss this point in some detail in Sections~\ref{sec-disc-lum} and~\ref{sec-faint}. Interestingly, the BL~Lac object ellipticities overlap nicely with the round majority of the 3CR, but lack the eccentric tail representing merging sources among the radio galaxies. Overall the BL~Lac objects represent a poor fit to the 3CR sample. Even considering the FR~I and FR~II subsamples separately produces little improvement in the match.
%Unexpectedly, the FR~II's form a better match to the BL~Lac object sample in terms of \sersic index, scale length and host luminosity, while the FR~I's exhibit the closer match in terms of ellipticity.

The~\citet{mcleod01} RQQ host galaxy luminosities and ellipticities exhibit a good match with the 3CR -- Figure~\ref{fig-mcleod}. The quasar host galaxies appear to be somewhat smaller on average, in contrast to the other samples of quasars discussed below that include both radio-loud and radio-quiet sources. Redshifts range from 0.15 to 0.4 with an average at $\bar{z}=0.25$. 

The~\citet{dunlop+03} quasar host galaxies appear to be consistent with the bright end of the 3CR host galaxy distribution, assuming a simple conversion from $R$ to $H$ band -- see Figure~\ref{fig-dunlop}. The more certain conversion from $K$~\citep{taylor+96} to $H$ provides a similar picture (Figure~\ref{fig-taylor}) though based on less certain host galaxy luminosities due to the less stable ground-based PSF. The higher ellipticities and scale lengths in the Dunlop / Taylor quasars are again possibly due to PSF correction problems. Note that the Dunlop et al.\ sample exhibits a strong cutoff in \sersic index in their radio-loud subsample that is {\em not} mirrored in the 3CR population.

The quasar host galaxy sample of~\citet{floyd+04} again matches the upper reaches of the 3CR host galaxy luminosity distribution, while the scale length, ellipticity and \sersic indices all show a similar spread in values -- Figure~\ref{fig-floyd}. Note that a key finding of both Floyd et al. and Dunlop et al. was the relative similarity of RLQ and RQQ properties, although the smallest and dimmest of the quasar host galaxies in each sample are radio quiet in each case. This is reflected in the histograms presented in Figures~\ref{fig-dunlop} and~\ref{fig-floyd}. 

In general, and unsurprisingly, the Rothberg \& Joseph (2004) sample forms a poor match to the 3CR host galaxies (Figure~\ref{fig-roth}). 
However, if we examine only the subset of 3CR host galaxies that are classed as exhibiting merger signatures (as defined in Paper~II), we see some improvement (Figure~\ref{fig-roth-merg}). 
In particular, there is a noticeable increase in the likelihood that the two samples are drawn from the same ellipticity distribution. 
But even among these merging 3CR sources, the host galaxies are significantly less disky than the general merger population explored by Rothberg \& Joseph and the disk-merger population contains many sources much smaller than the 3CR mergers, while the luminosity distributions span a similar range. 

The \citet{BBF92} sample forms a poor match to the 3CR host galaxies in all observational metrics (see Figure~\ref{fig-BBF92}), being generally larger and more luminous, as well as slightly more eccentric in shape.
% perhaps due to the $B$ band selection and imaging, which is a poor tracer of stellar mass and makes the conversion to $H$ band highly uncertain. 

Generally the~\citet{pahre99} sample forms a good match to the 3CR host galaxies, although the Pahre sample does exhibit a long low-luminosity / compact galaxy wing, beyond the cutoff in the 3CR host luminosity and scale length distributions -- Figure~\ref{fig-pahre}. 
In terms of ellipticity, the match is excellent.
%It is clear that a more luminous elliptical host galaxy (2 mag below $L^\star$) is a prerequisite for membership of the 3CR %See also the discussion on faint host galaxies in Section~\ref{sec-faint}.

The~\citet{mobasher+99} sample forms a very good match in terms of host galaxy luminosity to the 3CR -- Figure~\ref{fig-mob}. It is noticeable that the 3CR galaxies tend to be slightly larger, with a poorer match between the scale length distributions, due to the field-of-view imposed size cutoff in the Mobasher sample.

Unsurprisingly, the 3CR host galaxies are much larger and more luminous than the typical cluster galaxy represented by the~\citet{ferrarese+06} sample -- Figure~\ref{fig-ferr}. However, they also exhibit larger \sersic indices in general, in spite of the fact that Ferrarese et al. deliberately study only morphologically-selected early-type galaxies. 
Many of the Ferrarese objects are S0's with noticeable disks in addition to their bulges, yielding intermediate \sersic values.
S0's are not accurately represented with a single \sersic profile.

The 3CR host galaxies span a far lower range of luminosity than the BCG's of~\citet{graham+96}, are significantly smaller and have lower \sersic indices -- Figure~\ref{fig-graham}. The distributions have dissimilar shapes but there is significant overlap in terms of \sersic index.

\subsection{Host Galaxy Morphology}
\label{sec-morph}
We find a broad distribution of \sersic indices in our sample, spanning the full range from $n=1$ (exponential disks) to $n=4$ (de Vaucouleurs ellipticals), and higher. In Paper~II we discussed the FR~II radio sources that are hosted by disk-like ($n<2$) galaxies, all of which have close companion objects and are either merging, or are likely to be post-merger systems.
In comparison with normal cluster early-types (Figures~\ref{fig-BBF92} --~\ref{fig-ferr}), this range of \sersic index is not unusual. 
For example, by comparison to the local sample of~\citet{ferrarese+06}, the {\em range} of \sersic index is identical, but the peak in \sersic index in our sample falls significantly higher: in the Virgo cluster the peak \sersic index is at $n=2$, with a significant lower-$n$ tail, whereas in the 3CR, the peak is at $n=3$, and the tail exhibits a much more rapid falloff to low-$n$ (Figure~\ref{fig-ferr}). When we compare the Ferrarese sample to only the FR~I's in the 3CR there is a dramatic improvement ($D=0.10,~p=0.99$), with both populations exhibiting similar peaks in the \sersic index distribution. However, whereas the Ferrarese objects are typically S0's with distinct bulge and disk components, the FR~I's are typically found in diffuse giant (D, cD) elliptical galaxies~\citep{owenlaing89}. Both populations are better described using two \sersic profiles.

In comparison with the BCGs of~\citet{graham+96} there is again a clear distinction, with the BCG's exhibiting a far larger range of $n$ to high values (Figure~\ref{fig-graham}). The peak in the BCG \sersic index distribution is close to $n=5$, with a number of objects having $n>10$. However,  at the low-$n$ end, the 3CR closely follows the cutoff in \sersic index seen in these most massive cluster-dominating galaxies. These sources are also be better fit by a two-\sersic component model~\citep{seigar+07}.

Morphology also distinguishes the 3CR host galaxies from typical merger systems that have \sersic indices even lower than the cluster ellipticals, peaking at $n=1$ (Figure~\ref{fig-roth}). These sources are posited as post-merger systems on their way to forming massive ellipticals, but not generally the first-ranked elliptical galaxies in clusters~\citep{rothberg+04,rothberg+10}.
%nor even the average luminosity $L^\star$ objects~\citep{rothberg+04}. 
Even when comparing to only the 3CR sources flagged as mergers in Paper~II, we see a far stronger tendency for the 3CR mergers to be identified as bulge dominated ($n>2$) than the general merger population (Figure~\ref{fig-roth-merg}).

The morphologies of the sample do fit in well with the (far smaller) QSO samples of~\citet{dunlop+03} and~\citet{floyd+04} -- see Figures~\ref{fig-dunlop},~\ref{fig-floyd}. However, these QSO hosts appear to show a bimodal distribution, with a smaller disky ($n=1$) peak, as well as the elliptical peak close to $n=4$. As noted by~\citet{floyd+04}, however, the three disky objects are all {\em radio-quiet} quasars, at low-optical luminosity similar to a classical Seyfert galaxy. The same is true for the Dunlop sample. Thus, while the 3CR are morphologically similar to the RLQs studied in each of these papers, they are distinct from the general host galaxy population of RQQs. 
Using our own \sersic fits to the \citet{urry+00} BL~Lac object sample we find a similar spread in \sersic indices, although there is a long tail to high values among the BL~Lac objects.

%It is immediately clear from the table that a range of morphological types (quantified by the \sersic $n$ value) are present, in contrast to evidence for quasar host galaxies in the literature.
%Figure~\ref{fig-morph} shows the distributions of the ellipticity, $1-b/a$, and the S\'{e}rsic index, $n$, (and diskiness?) compared to those of samples of quasars. While the distributions of ellipticity are quite consistent with those of quasar host galaxies of Floyd et al. (2004) and Dunlop et al. (2003) the S\'{e}rsic $n$ distributions appear somewhat different, and this is confirmed by a two-sided Kolmogorov-Smirnov test, returning  a maximum difference of $D=0.334$, with a significance $p=0.118$.

\subsection{Ellipticity} 
The ellipticity distribution of the 3CR matches very well with that of the~\citet{pahre99} sample of $K$ band studied elliptical galaxies drawn from a range of environments (K-S test: $D=0.06,~ p=0.99$). However the samples of~\citet{BBF92,ferrarese+06}, which focus on entire populations from a cluster, show (typically E2-E3) ellipticity distributions that are closer to that of the mergers than that of the 3CR.  ~\citet{disney+84} also found that radio-loud ellipticals are inherently rounder than radio-quiet ones. Unfortunately, no ellipticity data is available for the BCG's, but they are quite round in general (from visual inspection of the original data of~\citealt{postmanlauer95}).

The quasar data seems to support a similar ellipticity distribution for both the Floyd and the Dunlop samples ($D=0.22,~p=0.42$ and $D=0.14,~p=0.67$, respectively). In particular, the radio-loud objects exhibit similar peaks at around E1 in both samples, but there is an additional highly-elliptical (E5) peak in both quasar samples that may be due to small number statistics, or indicate objects in which asymmetries from the PSF have interfered with proper characterization of the quasar host galaxy.

The most striking contrast in terms of ellipticity is with the merger remnants of~\citet{rothberg+04} -- see Figure~\ref{fig-roth} ($D=0.34,~p=0.001$). The merger remnants typically have much higher median ellipticities (E3) than those of the present 3CR sample (E1). This is perhaps unsurprising given that the Rothberg \& Joseph sample is targeted at objects that have recently undergone a major merger and have not had time to fully relax to their (presumably elliptical galaxy) endpoints. But it does demonstrate quite graphically that the powerful radio sources are not, in general, consistent with recent major-merger systems. 
%(in contrast with, e.g.,~\citealt{heckman+86,wu+98,canalizostockton01})
However, if we turn our attention to those objects in the 3CR that exhibit any kind of tidal or merger-like disturbances (flagged with mergers, shells or tidal tails in Paper~II), we find a somewhat different situation -- see Figure~\ref{fig-roth-merg}: a bimodal ellipticity distribution that follows the ellipticity distribution of Rothberg \& Joseph's mergers somewhat better ($D=0.20,~p=0.22$).

\subsection{Galaxy Luminosity}
\label{sec-MH}
The low-$z$ 3CR host galaxy population peaks at close to $L^\star$ ($M_H=-23.9$), but with significant spread.
The high end is dominated by the FR~II sources, but there are several FR~II sources in the low-luminosity wing as well.
%generally containing the lower redshift and lower-radio-luminosity (FR~I) sources 
%and a longer cutoff to the high end, dominated by FR~II sources. 
%Both FR~I and FR~II sources are significantly more luminous than the typical cluster galaxies of the~\citet{ferrarese+06} sample.
The merger systems studied by Rothberg \& Joseph are generally as luminous as the 3CR host galaxies, if not more so. 
%This is likely due to ongoing star formation, as 
Rothberg \& Joseph conclude that these systems end up largely as $L^\star$ (or brighter) elliptical galaxies after the starburst fades.

Comparing with elliptical galaxies in general, we see a good match in host luminosity distribution with the sample of~\citet{mobasher+99} ($D=0.11$; $p=0.82$) -- particularly to the FR~I's ($D=0.12$; $p=0.83$). The sample of~\citet{pahre99} gives a somewhat poorer match ($D=0.15$; $p=0.11$) which is improved if we look only at the FR~II's ($D=0.15$; $p=0.26$).
The~\citet{ferrarese+06} sample covers the entire Virgo cluster, and understandably peaks at far lower luminosities, but it is instructive to see the crossover between the 3CR and the entire contents of such a cluster (Figure~\ref{fig-ferr}). The 3CR hosts are representative of the top one third, in terms of luminosity, of the galaxies in Virgo. While there is some overlap with the BCG population, in general when compared the 3CR are smaller and less luminous. While the FR~I's were found to have a similar \sersic index distribution, any apparent similarity with the Ferrarese sample is eliminated by comparing their luminosities.

The median luminosity of the 3CR host galaxies is somewhat dimmer than that of the typical quasar host galaxy, or BL~Lac object, based on simple transformations from the fluxes of~\citet{urry+00,dunlop+03} and~\citet{floyd+04} assuming an old stellar population (see Section~\ref{sec-cosm}). 
We acknowledge that there are  difficulties in comparing to luminosities measured in different bands in this manner, but with the existing samples of low-redshift quasar host galaxies, a single band is generally all that can be compared, and so there is no constraint on the mass-to-light ratios of these objects. 
However, for the Dunlop et al. sample, ground-based $K$ band imaging was published in an earlier paper by~\citet{taylor+96}, offering us a convenient test-case: although the $K$ band morphologies are less reliable due to the problem of disentangling the host and the nucleus from the ground, luminosity in most cases is well constrained, and they can be much more reliably converted into an $H$ band luminosity. 
We see an essentially identical luminosity distribution to that exhibited by the Dunlop/Taylor data -- Figures~\ref{fig-dunlop} and~\ref{fig-taylor}. Once again, we see a significant population of sub-$L^\star$ 3CR hosts that are not mirrored in the quasar population. 
As was stated earlier and in Paper~II, the low-luminosity 3CR objects tend to be low-redshift, low-radio-power FR~I sources. We note that the~\citet{urry+00} sample exhibits a slightly greater crossover in luminosity with the 3CR at this low end, in accord with the unification of these objects with the BL~Lac objects, but there remains a deficit of the faintest objects seen in the 3CR.

\subsection{Galaxy Scale Length}
\label{sec-R}
The 3CR host galaxy scale lengths, in keeping with the host galaxy luminosities, are typical of large ($\sim L^\star$) elliptical galaxies, with a median of 3.6~kpc, and a few sources larger than 10~kpc in size. They are somewhat smaller than reported in many parts of the quasar literature, but are consistent with those of~\citet{floyd+04} ($D=0.19$, with $p=0.65$) and somewhat consistent with~\citet{mcleod01} ($D=0.32$, with $p=0.09$). 
They are inconsistent  with the scale length distributions reported by~\citet{urry+00} ($D=0.33,~p=0.002$),~\citet{dunlop+03} ($D=0.52,~p\sim10^{-4}$) and ~\citet{taylor+96} ($D=0.63,~p\sim10^{-9}$). The latter data set in particular is known to overestimate the scale lengths, due to atmospheric seeing problems: compare results with~\citet{dunlop+03}, who study the same sample in $R$ band using {\em HST} data -- Figures~\ref{fig-dunlop} and~\ref{fig-taylor}. The Taylor et al. ground-based work generally recovers accurate host galaxy luminosities (see~\citet{dunlop+03}), but poorly constrained morphological parameters, such as scale length and surface brightness. That the 3CR scale lengths agree at least with the Floyd et al.\ quasar sample is reassuring, and it is possible that earlier studies overestimated the quasar host galaxy scale lengths somewhat. Considering the 3CR FR~II sources alone, we get a considerable improvement to the comparison with the Floyd et al.\ quasars ($D=0.18,~p=0.78$).
Considering 3CR FR~I sources alone we find a significant improvement in the comparison to the~\citet{mcleod01} RQQ host galaxy scale lengths ($D=0.13$, with $p=0.99$).

While the Merger sample of Rothberg \& Joseph (2004) is generally as luminous as the 3CR, we find that in general, the 3CR host galaxies are somewhat larger, consistent with the merger's fluxes being dominated by active star formation in their inner regions.

Comparing to normal elliptical galaxies there are no statistically good matches in terms of scale length. The 3CR are slightly larger on average than the Pahre sample and Mobasher et al. sample of ellipticals, but smaller than the BBF92 sample. Interestingly, the 3CR span the noticeable gap in the Ferrarese sample scale length distribution (Figure~\ref{fig-ferr}). The 3CR are smaller in general than the Graham et al.\ BCG's which lack objects smaller than 2~kpc.

%\subsection{Diskiness}
%Diskiness is quantified in different ways by different authors. In particular,~\citet{bender87} re-examined their~\citet{BBF92} sample to obtain diskiness profiles, and calculated the {\em maximum} $a_4$ value across the profile. This data has been restudied by~\citet{rothberg+06,naab+06}. By comparison,~\citet{rothberg+04} publish median $a_4$ values, along with values at the effective radius of each galaxy. We have attempted to compare like-with-like throughout this work, and so different characterizations were obtained from the profiles presented in Paper~II. Against the data of~\citet{BBF92} we compare the extreme value of the $a_4/a$ parameter across the profile range between a radius of 0\farcs5 and where the flux falls to less than $3\sigma_{sky}$. However, maximum diskiness is a highly biased measure of a source's diskiness.  The range of our maximum diskiness is significantly smaller than the range in the elliptical galaxies of Bender et al.

\subsection{Kormendy Relation (R$_e$-$\mu_e$)}
\label{sec-korm}
\begin{figure*}%[htbp]
\centering
%{\includegraphics[width=8.0cm]{f14.eps}}
\plotone{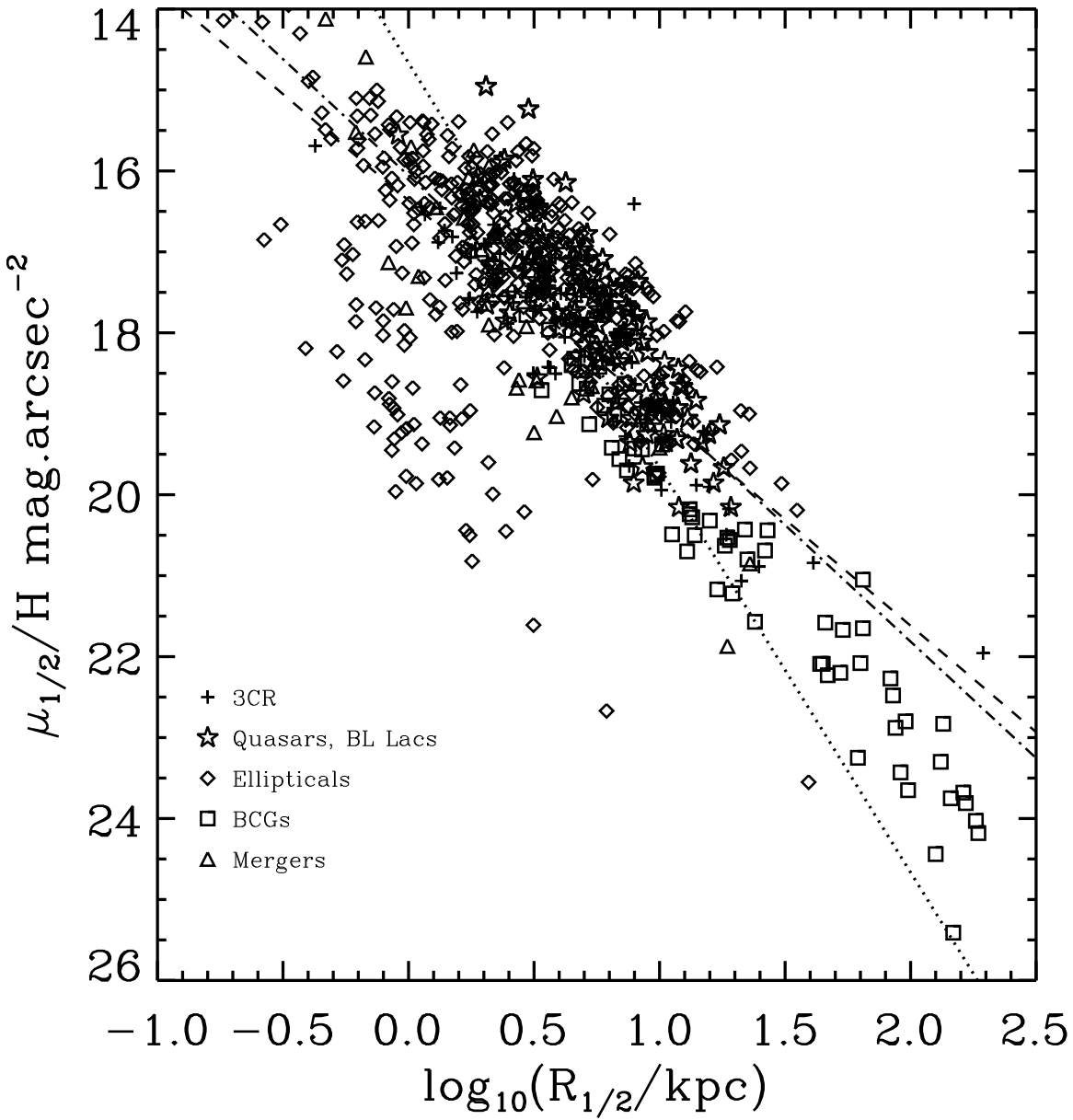}
\caption{\label{fig-korm} scale length--surface brightness for the 3CR along with comparison samples (after conversion to H band) of quasar and BL~Lac object host galaxies, quiescent ellipticals, BCG's and mergers as presented in the text. Best fitting scale length--surface brightness relations are shown for the FR~I's (dashed) and the FR~II's (dot-dash). The locus of an $L^\star$ galaxy is shown by the dotted line. The 3CR fits in well with the structural properties of the giant ellipticals, and is distinct from the dwarf elliptical subsample within the~\citet{ferrarese+06} sample from the Virgo cluster.}
\end{figure*}

The strong correlation between scale length and surface brightness is a well-known feature of bright elliptical galaxies~\citep{kormendy77}, and is a projection of the more physically significant ``fundamental plane''. The relation followed by the present low-redshift 3CR sample is close to that of quiescent massive elliptical galaxies, following a slope of $2.97\pm0.1$ (Paper~II).
The BCG's follow a similar trend but extend to far larger radii. 
The same relation is also a good fit to the quasar host galaxy data of~\citet{dunlop+03} and~\citet{floyd+04}. 
This provides strong evidence that all of these objects are structurally similar, dynamically relaxed systems, although proof of this requires spectroscopy to study the dynamics. 
Mergers too fit remarkably well with the overall trend in the AGN and quiescent elliptical populations.
Putting all of the samples together, including the Virgo cluster galaxies of~\citet{ferrarese+06} results in the plot shown in Figure~\ref{fig-korm}. It is clear here that the giant ellipticals, and powerful AGNs (quasars, radio galaxies, BL~Lac objects) form a single population in the $R_e-\mu_e$ space. 
Dwarf ellipticals do not obey the same trend~\citep{capacciolicaon91,grahamguzman03}): for a given scale length these objects are far dimmer than their cousins on the Kormendy relation, and these objects are clearly not represented by the 3CR. 
The~\citet{rothberg+04} mergers appear to be on the low side of the typical elliptical galaxy Kormendy relation.

\section{Discussion}
\label{sec-disc}
\subsection{Host Galaxy Morphology}
\label{sec-disc-morph}
The modeling used in this paper is deliberately simple, aimed at fitting the bulk of each galaxies flux with the minimum of parameters. Many if not all luminous galaxies have core regions that deviate from simple \sersic models, and which may be detected in many of the closest objects presented here, but in the interests of comparing like with like we have adopted a \sersic model and compared to the large literature on fitting similar models to control samples of galaxies and quasars.
Regarding morphology of quasar host galaxies it is difficult to distinguish an $n=6$ bulge, say, from an $n=4$ one underneath the glare of a quasar nucleus ($M_V\leq-23$). The spread in $n$ present in the 3CR host galaxies does in fact connect them more firmly with the quiescent elliptical galaxy population, which does not in general follow a perfect de Vaucouleurs type $r^{1/4}$ law which tends to be seen in quasar host galaxies. It seems likely that when applying such morphological models to the host galaxies of quasars, there is insufficient signal under the photon shot noise in the central region to accurately determine $n$ in many cases: although it is clear that in most cases the difference between an elliptical or bulge-dominated and disky or disk-dominated host is measurable~\citep{mclure+99,mcleod01,dunlop+03,floyd+04}, it is difficult to determine the precise morphology as characterized by a single continuous parameter given current observational technology. 

%The 3CR FR~I's fit well into the morphology distributions observed for the cluster ellipticals of Ferrarese et al., suggesting that these objects are hosted by galaxies that are representative of the entire early type galaxy spectrum. 
%In Paper~II we identified eight powerful FR~II radio sources that are hosted by disk-like ($n<2$) galaxies, five of which are at $z>0.1$: 3C~173.1, 3C~234, 3C~288, 3C~323.1 and 3C~332. The remaining low-$n$ (``disky'') galaxies all host FR~I sources at low redshift. All five of the low-$n$ FR~II sources have close companion objects, and are either merging, or are likely to be post-merger systems (see Paper~II for details). They generally show evidence of dusty environments from their optical and IR images, except for 3C 323.1 which has a quasar-like nucleus in both the optical and IR. 

The 3CR host galaxies follow the same Kormendy relation to the host galaxies of quasars, and giant ellipticals (Figure~\ref{fig-korm}): even the very lowest luminosity ones are clearly distinguished from the dwarf ellipticals. 
However, what is clearly missing is dynamical information on these sources that would provide a real physical insight into the conditions inside the host galaxies. 
%Are they dynamically relaxed, lying on the standard Fundamental Plane like normal ellipticals? 
A study of southern radio galaxies by~\citet{bettoni+01} has shown that those sources are indeed hosted by dynamically relaxed elliptical galaxies.%, and there is no reason to suppose that the 3CR should differ. 
The closest comparison sample is that of Mobasher et al. (1999) which already has suitable spectroscopic data. 
It would be a valuable study to investigate the relative dynamical states of these two samples. 
Spectroscopic data for the 3CR host galaxies would enable us to determine the dynamics and stellar populations of these objects in order to explore their ages, merger histories, and the differences between sources lying in rich environments and those lying in the field.

\subsection{Host Galaxy Luminosity}
\label{sec-disc-lum}
The most surprising finding from this survey has been the relative faintness in $H$ band of a small number of the 3CR host galaxies relative to the luminosities expected from studies of quasar host galaxies. Even when we consider only FR~II sources, several host galaxies exist that are rather fainter than the cutoff of $\sim L^\star$ inferred from the RLQ host galaxy luminosity functions of Dunlop et al. (2003) and Floyd et al. (2004). 
Assuming a mass-to-light ratio for early-type galaxies in general, $M/L_H\sim1.2$ (with a scatter of 0.2~dex~\citealt{zibetti+02}) we have just three sources with masses $>5\times10^{11}~M_\odot$: 3C~130, 3C~196.1 and 3C~338. 
%Is it possible that radio-loud AGN activity (in terms of high radio luminosity) is {\em not} the sole preserve of the most massive elliptical galaxies as we have come to expect from studies of RLQs? ???
%This possibility has been discussed before by~\citep{best+05b}.
% M<3e10: 3c61.1 3c76.1 3c84 3c264 3c272.1 3c442 3c445 3c449
Firstly we note that the median mass of the sample estimated in this simplistic way is $10^{11}~M_\odot$ and the mean mass $2\times10^{11}~M_\odot$. This is significantly higher than the $\sim3\times10^{10}~M_\odot$ mass cutoff inferred for a transition in galaxy formation mode~\citep{kauffmann+03a,khochfarsilk06,khochfarsilk09}.
Secondly, the vast majority of the anomalously faint tail ($M_H>-23$) in the 3CR host galaxy distribution is made up of very nearby (11 at $z<0.05$, 19 at  $z<0.1$) radio sources that are relatively radio-dim, some even classically ``radio-quiet'' by the simplistic standard adopted for the RLQ/RQQ break in quasar studies. 
%The remaining 3CR sources with faint ($M_H>-23$) host galaxies are low-redshift FR~I sources with low radio-powers,
These sources would not be selected as part of the 3CR if they lay just a little more distant and cannot properly be compared with RLQs. 
Just two of the sources with faint host galaxies, 3C~61.1 (an FR~II) and 3C~314.1 (an FR~I) lie at higher redshifts ($z=0.186$ and $z=0.119$, respectively). In addition, seven of the lower redshift sources are worth considering: 3C~105, 3C~198, 3C~227, 3C~326, 3C~353, 3C~402 and 3C~445 are all FR~II radio sources in abnormally dim galaxies. These objects are discussed in greater detail in the following section.
Finally, we remark that just two of these dim FR~II host galaxies (3C~61.1 and 3C~445) have simple estimated masses below the $\sim3\times10^{10}~M_\odot$ mass cutoff inferred for a transition in galaxy formation mode~\citep{kauffmann+03a,khochfarsilk06,khochfarsilk09}.

It is interesting to explore how this finding compares to statistical surveys of nearby radio galaxies. 
The largest sample is that of~\citet{mauchsadler07} who examined 7824 radio sources from the 1.4~GHz NRAO VLA Sky Survey (NVSS) with galaxies brighter than $K=12.75$~mag from the 6 degree Field Galaxy Survey (6dFGS). The galaxies in the 6dFGS span the redshift range $0.003<z<0.3$, making an ideal comparison to the samples presented here. Typical colors for early-type galaxies are $H-K\approx 0.2-0.5$. 
Galaxies with luminosities $M_K>-23$ are extremely rare -- rarer than we find for $M_H>-23$. However, we note that the 6dFGS cutoff will lose $M_K>-23$ galaxies at $z\approx 0.033$ and even $M_K>-23.5$ galaxies at just $z\approx 0.041$. 
Thus, it appears that this rare population was essentially invisible to earlier surveys and have only been discovered by virtue of deep imaging of a purely radio-selected sample.

\subsection{FR-type and Unification}
\label{sec-disc-FR}
\begin{figure*}%[htbp]
\centering
%{\includegraphics[width=8.0cm]{f16.eps}}%{/Users/floyd/3C_Work/plots/host_radio.eps}}
\plotone{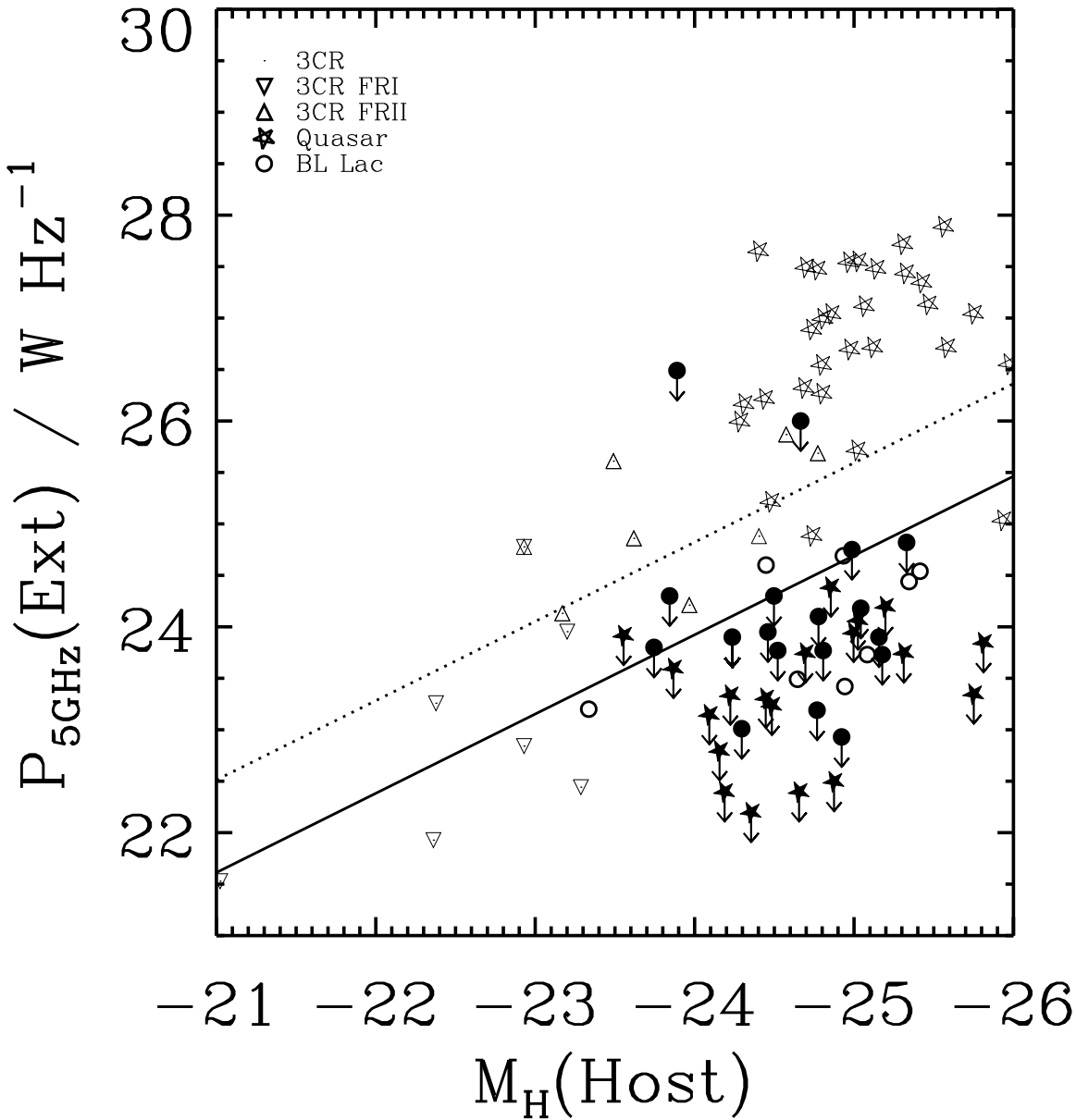}%{/Users/floyd/3C_Work/plots/host_radio.eps}}
\caption{\label{fig-host-radio} FR type diagnostic can be made according to optical luminosity vs. radio power, according to~\citet{owenledlow94}. Using only extended radio power and host galaxy luminosity avoids the effects of beaming~\citep{urry+00}. Models from~\citet{bicknell95} are overplotted 
(solid line: $\gamma_\mathrm{max}/\gamma_\mathrm{min}=10^{4}$, $\nu_c=10^{10}$~Hz, $f=1$; 
dotted line: $\gamma_\mathrm{max}/\gamma_\mathrm{min}=10^{4}$, $\nu_c=10^{11}$~Hz, $f=0.5$). 
For comparison we have plotted the RQQs (filled stars) from Dunlop et al. (2003) and Floyd et al. (2004), assuming that all their radio flux is extended. Symbols used are the same as in Figure~\ref{fig-radio}, with filled symbols indicating radio-quiet objects and open symbols radio-loud objects.}
\end{figure*}

~\citet{fanaroffriley74} showed that radio galaxies exhibit a change in morphology at a monochromatic power corresponding to roughly $10^{24.5}$~W~Hz$^{-1}$ at a rest-frame frequency of 1.4~GHz (converted from the original cutoff at 178~MHz -- \citealt{bicknell95}).
It is understood that the two classes have different bulk energy transport mechanisms (\citealt{bicknell95} and references therein). 
~\citet{owenledlow94} showed that on the optical-radio plane, the division between the two classes, which occurs over a full decade in radio power, becomes much cleaner. \citet{urry+00} took this a stage further by looking at just the extended radio power (subtracting off the core from the total) and the optical luminosity of the host galaxy alone. 
To do this, they used the 5~GHz radio flux which is sensitive to core emission and provides a core/extended distinction for the largest possible sample. This avoids the effects of beaming and cleans up the division between FR~I / BL~Lac object and FR~II / quasar even further. The dividing line was found to follow that predicted by the models of~\citet{bicknell95}, with the FR~I sources containing turbulent, transonic jets, and FR~II sources jets that are super or hypersonic (see~\citealt{bicknell95} and references therein). 
This line was later found to be consistent with a constant Eddington accretion ratio~\citep{ghisellinicelotti01}. 
In Figure~\ref{fig-host-radio} we plot the available 5~GHz data for the present sample~\citep{jenkins+77,giovannini+88}, along with RLQs from~\citet{dunlop+03} and~\citet{floyd+04}, and BL~Lac objects from~\citet{urry+00}. The division confirms once again that FR~I's are unified with BL~Lac objects, while FR~II's are analogous to RLQs.

We also note that a number of ``radio-quiet'' quasars studied by Dunlop et al. and Floyd et al. may actually occupy a similar region of the host luminosity -- extended radio power plane to the FR~I's and BL~Lac objects, if their radio luminosities are dominated by an extended component. This is uncertain since RQQs have not been studied in such great depth in the radio as RLQs, and nothing is known of their radio structure.
% -- whether they are core or lobe-dominated.
%The existence of a true radio-loudness ``dichotomy'' has been thrown into doubt by~\cite{hooper+95,white+00,cirasuolo+03}. It is clear from Figure~\ref{fighost-radio} that more observations of RQQs are required, since present flux limits cannot even distinguish ostensibly radio quiet objects from radio sources as powerful as some of the 3CR!
% and this finding would indeed agree that some RQQs may actually be as radio-loud as the FR~I sources in the 3CR.
%CAN WE SAY ANYTHING MORE ABOUT THIS??? Look at any RQQ radio detections as a case study...
%Type 1 \& 2 unification schemes (e.g.,~\citealt{antonucci93} and references therein).

% 3CR targets with low-lum hosts:::
%3c61.1      -22.176637       2     0.186000
%3c314.1      -22.959768       1     0.119000

%3c76.1      -22.275918       1    0.0320000
%3c84      -22.199066       1    0.0170000
%3c98      -22.929552       1    0.0300000
%3c264      -22.377382       1    0.0200000
%3c270      -22.632350       1   0.00700000
%3c272.1      -21.023162       1   0.00300000
%3c274      -22.548812       1   0.00400000
%3c310      -22.929902       1    0.0540000
%3c386      -22.734066       1    0.0170000
%3c449      -22.360966       1    0.0170000

%3c105      -22.875035       2    0.0890000
%3c198      -22.549904       2    0.0810000
%3c227      -22.659435       2    0.0860000
%3c326      -22.986119       2    0.0880000
%3c353      -22.994753       2    0.0300000
%3c402      -22.938752       2    0.0230000
%3c445      -22.207837       2    0.0560000

%3c71      0.85894369       0   0.00379300 	-- NGC1068 -- not in 3CRR
%3c403.1      -22.732030       0    0.0550000 	-- not in 3CRR
%3c405      -22.983337       0    0.0560000 	-- not in 3CRR
%3c442      -22.264128       0    0.0260000 	-- not in 3CRR

\section{Sources with faint host galaxies}
\label{sec-faint}
In this Section we discuss the eight FR~II radio sources with anomalously faint ($M_H>-23$) host galaxies and the unusually radio-powerful FR~I, 3C~314.1. Extinction maps were generated for these nine sources by dividing the optical WFPC2 images~\citep{dekoff+96,martel+99} by the NICMOS images, after registering, scaling and smoothing the former to the plate scale and resolution of the latter. In several cases, the optical-IR colors suggest strong extinction, sufficient to close some of the gap to $L^\star$ in $H$ band, but not all of it.
Note there are an additional four with intermediate luminosities $-23<M_H<L^\star$: 3C~111, 3C~234, 3C~371 and 3C~390.3 which we do not discuss here.\\

%\begin{itemize}
{\bf 3C~61.1} is an FR~II radio source in an under-luminous host galaxy ($M_H=-22.2$), with apparent spiral arm structure (see paper II and references therein) in spite of its high radio power ($\log_{10} L_{178}=26.27$). However, this source is the brightest member of a small group of galaxies visible on the NICMOS image, and is probably associated with a larger cluster that includes a radio-quiet AGN at $z=0.184$~\citep{kristian+78} some 30\arcsec~to the east. It is clearly very dusty from the optical morphology~\citep{dekoff+96}. The extinction map is complicated by the star formation in the apparent spiral arms, but the central region appears to be heavily obscured, and we estimate $\sim 0.5~H$~mag of extinction in the central region of this source (assuming a diffuse dust model after~\citealt{mathis90}), giving an unextincted $H$ band luminosity of $M_H=-22.7$. It is likely to be the brightest cluster member (though this is uncertain without a proper study of the host galaxy of the nearby AGN).\\
{\bf 3C~105} is an FR~II radio source at $z=0.089$, hosted by a highly flattened and faintly disturbed (in our $H$ band image) elliptical with $M_H(\mathrm{Host}) = -22.88$. The host galaxy of 3C~105 is known to be exceptionally red, with strong stellar absorption lines~\citep{tadhunter+93,cohen+99}, and both existing optical {\em HST} images provide insufficient detection to produce a color-map. We estimate an $R-H$ color for the host galaxy of $\approx8\pm2$ (very approximate due to the barely $3\sigma$ detection in $R$ band) indicating a very heavily obscured host galaxy. Cohen et al. commented that the lack of polarized flux could be due to the obscuration of the scatterers, but discount this hypothesis on the basis that the obscuring material would have to cover a large fraction of the ISM. The {\em HST} images, although shallow in the optical, suggest that a large quantity of absorbing material could indeed be the cause. \\
{\bf 3C~198} is an FR~II radio source at $z=0.081$ in a slightly elongated elliptical host galaxy with $M_H(\mathrm{Host})=-22.55$ and a detectable nuclear point source. The IR nucleus and lack of common reference points on both the $H$ band and optical images mean that we were unable to produce a reasonable color-map. This source appears to be the brightest member of a small group (see~\citealt{mchardy74}).\\
{\bf 3C~227} is a broad line FR~II radio galaxy at $z=0.086$. We can clearly see the host galaxy ($M_H=-22.66$) and the bright active nucleus on this NIR image. Both the optical and UV images show a very bright unresolved nucleus~\citep{martel+99,allen+02}. The strength of the nucleus destroys any information on the color-map. However,~\citet{cohen+99} found a strong Balmer decrement, and a likely $V$ band extinction of $A_V\approx1.7$~mag. using Keck spectropolarimetry. They and~\citet{prieto+93} conclude that the intrinsic brightness of the object is likely to put it in the quasar class. We would conservatively estimate $A_H\approx0.3$ based on these earlier findings, assuming a diffuse ISM dust model~\citep{mathis90}.\\
%{\bf 3C~310} is an FR~I radio source that appears to be the brightest member ($M_H=-22.9$) of a small group or cluster that is visible on the WFPC2 mosaic, and has $R-H=2.8$. However, it shows no major extinction features in the optical-IR color-map. If we assume this is a normal elliptical ($R-H=2.25$) reddened by dust we infer $\sim0.1$~H mag of extinction across this source for a diffuse ISM dust model~\citep{mathis90}. This would give the galaxy an unextincted luminosity of $M_H=-23$, once again low in comparison with RLQ host galaxies.
{\bf 3C~314.1} is an abnormally radio-powerful FR~I radio source ($\log_{10} L_{178}=25.42$) in a faint host galaxy ($M_H=-22.96$). It is apparently not in any cluster~\citep{allington-smith+93}, exhibiting an under-density of nearby galaxies relative to an average elliptical. It shows no major extinction features in the optical-IR color-map, and is unusually blue for an elliptical galaxy ($R-H\approx1.5$). We note that this object is flagged by~\citet{buttiglione+09} as a relic radio galaxy which is no longer being actively powered by the jet.\\
% It is difficult to reconcile 3C~314.1 with the picture of radio galaxies as being drawn from the most massive elliptical galaxies.
{\bf 3C~326} is an FR~II radio source and exhibits a LINER spectrum~\citep{simpson+96}, and the host galaxy is quite dim ($M_H=-22.99$) and red ($R-H = 2.4$). From the optical image~\citep{dekoff+96}, it is extremely dusty, and appears to be an edge-on S0. The color-map shows up to 0.3 $H$ mag of extinction in the central regions of 3C~326 (assuming a diffuse dust model after~\citealt{mathis90}), suggesting an unextincted galaxy luminosity of $M_H\approx-23.3$.\\
{\bf 3C~353} is an FR~II radio source with a LINER spectrum~\citep{simpson+96} hosted by a round giant elliptical with $M_H(\mathrm{Host})=-22.99$ at $z=0.03$. There are no signs of disturbance, with very circular ($e\approx0.04$) isophotes on the optical image~\citep{martel+99}. The color-map shows a smooth gradient in extinction, with the highest values up to $V=H=2.7$ at the center, where a small dust lane is visible near the core~\citep{dekoff+00}. We estimate $A_H\approx0.1$ mag for this source in the central region, assuming a diffuse dust model after~\citealt{mathis90}.\\
{\bf 3C~402} is an FR~II radio source hosted by a smooth, elongated elliptical galaxy with $M_H(\mathrm{Host})=-22.94$ at $z=0.023$.
The galaxy is found to be quite red, with $V-H=3.0$ on average, rising to around 3.5 in the central regions.\\
{\bf 3C~445} is an FR~II radio source hosted by a $M_H(\mathrm{Host})=-22.21$ round elliptical at $z=0.056$. In both the optical and IR images, the morphology is dominated by the bright nuclear point source. This point source dominates the color-map which thus provides no useful information on extinction. \citet{cohen+99} deduce between about 1.1 and 2.5 mag of $V$ band extinction from their spectropolarimetric study. This is enough provide up to 0.45 $H$ mag of extinction assuming a diffuse dust model after~\citealt{mathis90}.\\
%\end{itemize}

All of the sources discussed above merit closer attention to determine their stellar and black hole masses.
We note that all except two (3C~61.1, 3C~445)) of these low-mass objects are above the~\citet{kauffmann+03a} $3\times10^{10}$~M$_\odot$ cutoff (given our mass-to-light ratio assumptions) but all are significantly less luminous than typical quasar host galaxies. 
%Of course, significant scatter exists in the mass-to-light ratio estimate used here, but insufficient to bring these objects into the same league as the RLQ host galaxies
%, but the luminosity cutoff reported for RLQ host galaxies is higher than the mass cutoff reported for galaxy formation ???
Two of the abnormally faint FR~II host galaxies (3C~61.1, 3C~198) are found to be the brightest galaxies in a small cluster or group. The remaining six FR~II's show anomalously red colors, but insufficient evidence of extinction to bring them up to $L^\star$. The FR~I source 3C~314.1 is the only unusually blue galaxy. None of the faint objects correspond to the five disky (low-$n$) FR~II galaxies discussed in Paper~II and mentioned in Section~\ref{sec-disc-morph}. Those objects were all found to have close companions and either to be undergoing merger or to be likely recent post-merger systems.

Other published exceptions to the general rule include PKS~0131--36, 0313--192 and PKS~1413+135. PKS~0131--36 is a nearby FR~II radio source hosted by a typical S0 galaxy, NGC~612 (see~\citealt{emonts+08} and references therein). 0313--192 is a powerful ($\sim10^{24} h^{-2}_{75}$~W~Hz$^{-1}$ at 20 cm) FR~I radio source in a highly flattened disk-dominated galaxy~\citep{ledlow+98,ledlow+01}. PKS~1413+135 is a ``red quasar'' with BL~Lac object properties in the NIR, but no noticeable optical nucleus, located in a disky galaxy with large amounts of dust obscuration (\citealt{perlman+02} and references therein).

\section{Quasars and radio galaxies in context}
\label{sec-cont}

% ?????? ADD IN \citet{merloniheinz08}: 2 modes above a certain accretion rate???
% ?????? ADD IN \citet{merloniheinz08}: 2 modes above a certain accretion rate???
% ?????? ADD IN \citet{merloniheinz08}: 2 modes above a certain accretion rate???
% ?????? ADD IN \citet{merloniheinz08}: 2 modes above a certain accretion rate???
% ?????? ADD IN \citet{merloniheinz08}: 2 modes above a certain accretion rate???
% ?????? ADD IN \citet{merloniheinz08}: 2 modes above a certain accretion rate???
% ?????? ADD IN \citet{merloniheinz08}: 2 modes above a certain accretion rate???
% ?????? ADD IN \citet{merloniheinz08}: 2 modes above a certain accretion rate???

It remains curious that the low-luminosity and low-$n$ sources do not appear in existing samples of RLQ host galaxies.
This may be simply explained through the different methods of selection: in the 3CR, we observe all the brightest radio-objects in the northern sky, and these may be bright due to any combination of intrinsic luminosity, proximity and beaming. In the RLQ samples of~\citet{dunlop+03,floyd+04}, care has been taken to remove artificially ``radio-loud'' sources by avoiding sources that are likely to be beamed -- objects that {\em would} appear in the 3CR are dropped from the RLQ samples. However, we also note that~\citet{best+05b} find no correlation between radio power and black hole mass for radio-loud objects. Thus, it should not necessarily be expected that beamed sources should have fainter host galaxies.

Much of the drive in quasar host galaxy studies has been to confront the technical challenge of observing the host galaxies most luminous quasars. In selecting for optically luminous quasars (radio-loud or radio-quiet) there is a dual selection effect for larger, more luminous host galaxies: A priori we do not know whether the luminosity is host galaxy or AGN-dominated; and if optical quasars push the Eddington limit~\citep{floyd+04} we are likely select for more massive black holes and thus  (through the well-known black hole to bulge mass correlation) more massive host galaxy bulges. Radio selection does not necessarily select for high black hole mass since the radio luminosity is 2--3 orders of magnitude lower than the Eddington limit~\citep{ghisellinicelotti01}.

It is possible that samples of RLQs are simply not sufficiently large to have detected such comparatively rare faint objects, of which we only find eight in the present 3CR sample, or that techniques are insufficiently sophisticated to detect them with current technology (i.e., ``naked quasars''). 
Additionally, it is possible that quasar host galaxy luminosities can be overestimated due to scattered light from the nucleus~\citep{young+09}, or due to PSF contamination issues~\citep{kim+08}. We feel that this latter effect is unlikely to be a problem in the Dunlop, Taylor and Floyd samples, given the detailed statistical analysis and testing of the method provided by~\citet{mclure+99, floyd+04, floyd+08}. Finally we note that any strong star formation in the quasar host galaxies will have the effect of boosting the optical luminosities relative to the simple model assumed here. This will in effect bring the bulk luminosities of the quasar host galaxies down toward the mean luminosity of our sample.

While there is a noticeable disagreement with the results of quasar host galaxy studies, we find that the 3CR radio galaxies are not unusual with respect to the normal elliptical galaxy population. 
Indeed, only two FR~II sources -- 3C~61.1 and 3C~445 -- have galaxies dim enough to fall below $3\times10^{10}~M_{\odot}$~\citep{kauffmann+03a,khochfarsilk06,khochfarsilk09} assuming the simplistic conversion from $H$ band luminosity of~\citet{zibetti+02}. These two objects likely have sufficient ($\sim0.5$~mag) $H$ band extinction to bring them above the mass cutoff without assuming a very different mass-to-light ratio.
Thus, we have a dilemma: powerful radio sources appear to be able to form in even modestly massive elliptical galaxies (albeit in low numbers), whereas RLQs require a much more luminous host galaxy. 
%This supports the notion that the optical and radio AGN are separate phenomena~\citep{best+05b}.
Moreover, studies of RLQs and RQQs have generally found a requirement for RLQ host galaxies to be more luminous than their RQQ counterparts~\citep{dunlop+03,floyd+04}. Thus, selecting for both optical and radio luminosity results in more luminous host galaxies than {\em both} selecting for optically luminous objects {\em and} for radio-luminous ones. 
%But RLQ host galaxies are more homogeneously luminous than both RG's and RQQs! 
There is no obvious reason why either the scattered light~\citep{young+09} or PSF contamination~\citep{kim+08} effects should be more significant in RLQs than in RQQs. We therefore echo the conclusion of~\citet{best+05b} that radio and optical AGNs are separate phenomena. 
%At the highest luminosities it appears these phenomena are correlated, but we can still have somewhat powerful radio sources that are not optical quasars, and vice versa. 

%However, the discrepancies discussed above demonstrate the difficulty associated with separating quasar host galaxies from their nuclei and underlines the importance of obtaining a sample of type 2 (hidden) quasars such as radio galaxies for detailed morphological and dynamical studies.

%However, the existence of some faint 3CR FR~II host galaxies with bright nuclei (Figure~\ref{fig-nuchost}) remains at odds with the picture of RLAGN as the exclusive domain of the most giant elliptical galaxies, and supports the idea that optical and radio AGN are separate phenomena. In Figure~\ref{fig-nuchost} we see that the luminosity of a galaxy constrains the optical luminosity of any AGN contained, following a simple relationship that is consistent with the Eddington limit for a fixed black hole/spheroid mass ratio of 0.0012~\citep{merrittferrarese01}, and a constant mass-to-light ratio for the host galaxies of 1.2~\citep{zibetti+02}. All FR~II sources plotted are radio-loud by the definition used by Dunlop et al. (2003). It would appear to be the selection of optically luminous quasars, rather than radio-loud optically luminous quasars that gives the cutoff in host galaxy luminosity described in that paper. 
%At high radio energies, the correlation between radio and emission line luminosity strongly supports unification~\citep{kauffmann+08} and provides the reason for this bias in the optically luminous RLQ sample.

\section{Conclusions}
\label{sec-conc}
%\item To determine what subset of galaxies, drawn from the nearby universe, are capable of hosting a radio-loud AGN.
%\item To examine whether RLAGN type is affected by host galaxy.
%\item To determine whether there are any systematic biases introduced in the study of quasar host galaxies.
%\item To produce a sample of type 2 RLQs that can act as a proxy for the type 1 equivalents for detailed dynamical and population studies.
Here, we again summarize the main findings. 
\begin{itemize}
\item{\bf In terms of ellipticity,} the low-$z$ 3CR are found to be in excellent agreement with~\citet{pahre99} sample of ellipticals drawn from across a range of environments, and with Floyd and Dunlop quasar samples. The mergers within the 3CR (identified as such in Paper~II) fit in well with the general merger population of~\citet{rothberg+04}. 
\item {\bf In terms of \sersic index,} there is good agreement with the Floyd et al. quasar and Urry et al. BL~Lac object samples,  and very poor agreement with the merger population, even when considering just the 3CR mergers in isolation from the remainder of the 3CR sample. 
%The 3CR FR~I's follow a very similar morphology distribution to the general cluster early type galaxy population of~\citet{ferrarese+06}.
\item {\bf In terms of host galaxy luminosity,} the 3CR are generally well matched to the~\citet{BBF92,pahre99} and ~\citet{mobasher+99} elliptical galaxy samples, but exhibit far more faint objects than the quasar samples and the BCG's.  
\item The Virgo cluster early-type galaxy sample of Ferrarese et al. offers an interesting contrast to the 3CR, clearly illustrating that the latter is drawn from only the most luminous section of a cluster's population. 
\item The 3CR and its merging subsample have a similar luminosity range to the mergers of~\citet{rothberg+04}, but the merging radio galaxies are generally found to be larger, with the Rothberg mergers on the low side of the Kormendy relation.
\item In terms of radio--optical properties, the FR~I's in the 3CR unify well with the properties expected of BL~Lac objects, and FR~II's with RLQs. However, a larger spread is seen in both the morphology and the host galaxy luminosity of the 3CR sources than would be expected from samples of BL~Lac objects and RLQs studied so far. 
\end{itemize}
%We find that the FR~I sources are similar to radio-quiet AGN~\citep{ledlowowen96,sikora+07} in terms of their host environments. Some RQQs may actually have extended radio luminosities consistent with those of some BL~Lac objects, given current observational data. A concerted effort at obtaining radio detections, and if possible morphologies for RQQs is essential to test this possibility. 

We confirm findings from earlier work (e.g.,~\citealt{ledlowowen96,dunlop+03}) that an elliptical host galaxy is a prerequisite for radio-loud AGN activity. However, we identify several radio galaxies ($\sim20$\%) that fall below the observed RLQ host galaxy luminosity cutoff at $\sim L^\star$. These objects have a luminosity distribution closer to that of the normal elliptical galaxy population, and were missed in the NVSS-6dFGS survey due to the $K$ band flux limit.
% than is observed among quasar host galaxies which are preferentially drawn from the high end of the luminosity function. 
The same finding is echoed in the morphological comparison -- the 3CR host galaxies exhibit a range of \sersic index that is entirely consistent with the general giant elliptical galaxy population, whereas quasar host galaxies show a narrower range of morphology.

%While the vast majority of the FR~II members of the low-$z$ 3CR are hosted by a luminous ($\gtsim L^\star$) relaxed elliptical galaxy, there are of the order ofwhich do not fit into this category. These objects are all found to be ``unusual'' in some way: being brightest members of a (small) group, being very gas rich, or showing some sign of recent or ongoing merger. These latter sources are massive by comparison with typical mergers, and typically are in poor groups or in the field.

We conclude that the morphological difference is due to quasar selection effects and / or contamination of quasar host galaxy flux by scattered nuclear flux. It would seem highly probable that given accurate coronographic observations, or a higher resolution image of the central regions of quasars, we would observe the same spread in morphologies in the host  galaxies of these objects as well. 
However, we do not believe that the luminosity discrepancy is so easily explained. The fainter FR~II host galaxies identified here clearly merit more detailed study and form an interesting subset of the 3CR, containing dusty sources and brightest apparent members small groups.
%It would be interesting to explore the detailed kinematic differences between FR~II 3CR objects that are involved in mergers with those that are not. It appears that this dichotomy exists also in quasar host galaxies (e.g.,~\citep{bahcall+97}) and also within the putative parent population of elliptical galaxies themselves~\citep{dunlop+03}.

%The radio selection of objects and optical selection of quasars both result in different populations to the combined selection of RLQs and we conclude that the radio and optical AGN phenomena are separate. 
%A better multi-wavelength data set for a statistical sample is important.

While we now have a reasonable view of the quasars of host galaxies, (and an impressively deep view in a small number of cases -- see~\citealt{bennert+08}), several problems remain. One is the technical issue of separating any scattered nuclear flux from the host galaxy flux. Note that this is distinct from the problem of simply separating out the host galaxy from the PSF. The problem is reviewed by~\citet{young+09} who argue that the effect is likely to affect the measured luminosities of quasar host galaxies significantly. The extent of the problem can be easily tested using space-based optical polarimetry of a bright quasar. The second issue is one of sample bias. Existing quasar host galaxy studies focus on small samples with likely selection biases toward bright host galaxies. We need a statistical survey of low-redshift quasar host galaxies in at least two bands in order to be able to place meaningful constraints on the masses of these objects to provide a baseline against which to compare higher redshift samples.

\acknowledgements
D.F. acknowledges the support of a Magellan Fellowship from AAL, and the hospitality of OCIW and Las Campanas Observatory during the writing of this paper. We thank Alister Graham for helpful comments on the manuscript and Barry Rothberg and Marc Postman for providing their reduced images and other data. This paper is based on observations with the NASA/ESA {\em Hubble Space Telescope}, obtained at the Space Telescope Science Institute, which is operated by the Association of Universities for Research in Astronomy, Inc. (AURA), under NASA contract NAS5-26555. 
This research has made use of the NASA/IPAC Extragalactic Database (NED) which is operated by the Jet Propulsion Laboratory, California Institute of Technology, under contract with the National Aeronautics and Space Administration. We gratefully acknowledge the insightful input of an anonymous referee who made a number of constructive criticisms that significantly improved this paper.

%%%%%%%%%%%%%%%%%%%%%%%%%%%%%%%%%%%%%%
%\newpage
\bibliographystyle{astron}
\bibliography{mss,full_lib}
\newpage

%%%%%%%%%%%%%%%%%%%%%%%%%%%%%%%%%%%%%%%%%%%%%%%%%%%
\begin{deluxetable*}{l|rrrrr|r|rrrr|r}
%\begin{table*}{l|rrrrr|r|rrrr|r}
  \tabletypesize{\tiny}
  \tablecolumns{12}
  \tablewidth{0pc}
  \tablecaption{\label{tab-stats} Full sample statistics: $D$ and $p$ from the two-sided K-S test comparing the full 3CR sample with each other sample.}
  \tablehead{
    \colhead{Property} &
    \colhead{U+00$^{a}$} & \colhead{MM01$^{b}$} & \colhead{D+03$^{c}$} & 
    \colhead{F+04$^{d}$} & \colhead{T+96$^{e}$} & \colhead{RJ04$^{f}$} & 
    \colhead{BBF92$^{g}$} & \colhead{P99$^{h}$} & \colhead{M+99$^{i}$} & 
    \colhead{F+06$^{j}$} & \colhead{G+96$^{k}$}}
\startdata
$M_H({\mathrm Host})$&0.47&0.34&0.62&0.57&0.44&0.29&0.23&0.15&0.11&0.68&0.81\\
  &0.00&0.06&0.00&0.00&0.00&0.01&0.00&0.11&0.82&0.00&0.00\\
\hline
$R_{1/2}$&0.33&0.32&0.52&0.19&0.62&0.30&0.49&0.26&0.32&0.65&0.60\\
  &0.00&0.09&0.00&0.65&0.00&0.00&0.00&0.00&0.00&0.00&0.00\\
\hline
$\epsilon=1-b/a$&0.36&0.15&0.14&0.22&0.26&0.34&0.25&0.06&...&0.20&...\\
  &0.00&0.89&0.67&0.42&0.03&0.00&0.00&0.99&...&0.04&...\\
\hline
$n$&0.27&...&0.42&0.17&...&0.34&...&...&...&0.29&0.57\\
  &0.12&...&0.00&0.73&...&0.00&...&...&...&0.00&0.00\\
\hline
$d$&...&...&...&...&...&0.42&0.28&...&...&...&...\\
  &...&...&...&...&...&0.00&0.00&...&...&...&...
%  \hline
  \enddata
  %  \tablecomments{}
  \tablerefs{$^a$ \citet{urry+00}; 
    $^b$\citet{mcleod01}; 
    $^c$\citet{dunlop+03}; 
    $^d$\citet{floyd+04}; 
    $^e$\citet{taylor+96}; 
    $^f$\citet{rothberg+04}; 
    $^g$\citet{BBF92}; 
    $^h$\citet{pahre99}; 
    $^i$\citet{mobasher+99}; 
    $^j$\citet{ferrarese+06}; 
    $^k$\citet{graham+96}}
\end{deluxetable*}
%\end{table*}

%%%%%%%%%%%%%%%%%%%%%%%%%%%%%%%%%%%%%%%%%%%%%%%%%%%
\begin{deluxetable*}{l|rrrrr|r|rrrr|r}
%\begin{table*}{l|rrrrr|r|rrrr|r}
  \tabletypesize{\tiny}
  \tablecolumns{12}
  \tablewidth{0pc}
  \tablecaption{\label{tab-stats_FRi} FR~I statistics: $D$ and $p$ of the two-sided K-S test comparing the 3CR FR~I sample with each other sample.}
    \tablehead{
    \colhead{Property} &
    \colhead{U+00$^{a}$} & \colhead{MM01$^{b}$} & \colhead{D+03$^{c}$} & 
    \colhead{F+04$^{d}$} & \colhead{T+96$^{e}$} & \colhead{RJ04$^{f}$} & 
    \colhead{BBF92$^{g}$} & \colhead{P99$^{h}$} & \colhead{M+99$^{i}$} & 
    \colhead{F+06$^{j}$} & \colhead{G+96$^{k}$}}
\startdata
$M_H({\mathrm Host})$&0.66&0.55&0.82&0.75&0.62&0.47&0.42&0.30&0.12&0.68&0.87\\
  &0.00&0.00&0.00&0.00&0.00&0.00&0.00&0.03&0.83&0.00&0.00\\
\hline
$R_{1/2}$&0.52&0.13&0.68&0.34&0.76&0.18&0.67&0.13&0.14&0.50&0.73\\
  &0.00&0.99&0.00&0.15&0.00&0.58&0.00&0.80&0.86&0.00&0.00\\
\hline
$\epsilon=1-b/a$&0.33&0.22&0.22&0.23&0.38&0.46&0.37&0.18&...&0.34&...\\
  &0.05&0.69&0.44&0.58&0.01&0.00&0.00&0.41&...&0.01&...\\
\hline
$n$&0.42&...&0.57&0.34&...&0.36&...&...&...&0.10&0.70\\
  &0.02&...&0.00&0.14&...&0.02&...&...&...&0.99&0.00\\
\hline
$d$&...&...&...&...&...&0.46&0.30&...&...&...&...\\
  &...&...&...&...&...&0.00&0.03&...&...&...&...
%  \hline
  \enddata
  %  \tablecomments{}
  \tablerefs{$^a$ \citet{urry+00}; 
    $^b$\citet{mcleod01}; 
    $^c$\citet{dunlop+03}; 
    $^d$\citet{floyd+04}; 
    $^e$\citet{taylor+96}; 
    $^f$\citet{rothberg+04}; 
    $^g$\citet{BBF92}; 
    $^h$\citet{pahre99}; 
    $^i$\citet{mobasher+99}; 
    $^j$\citet{ferrarese+06}; 
    $^k$\citet{graham+96}}
\end{deluxetable*}
%\end{table*}

%%%%%%%%%%%%%%%%%%%%%%%%%%%%%%%%%%%%%%%%%%%%%%%%%%%
\begin{deluxetable*}{l|rrrrr|r|rrrr|r}
%\begin{table*}{l|rrrrr|r|rrrr|r}
  \tabletypesize{\tiny}
  \tablecolumns{12}
  \tablewidth{0pc}
  \tablecaption{\label{tab-stats_FRii} FR~II statistics: $D$ and $p$ of the two-sided K-S test comparing the 3CR FR~II sample with each other sample.}
    \tablehead{
    \colhead{Property} &
    \colhead{U+00$^{a}$} & \colhead{MM01$^{b}$} & \colhead{D+03$^{c}$} & 
    \colhead{F+04$^{d}$} & \colhead{T+96$^{e}$} & \colhead{RJ04$^{f}$} & 
    \colhead{BBF92$^{g}$} & \colhead{P99$^{h}$} & \colhead{M+99$^{i}$} & 
    \colhead{F+06$^{j}$} & \colhead{G+96$^{k}$}}
\startdata
$M_H({\mathrm Host})$&0.47&0.29&0.61&0.54&0.43&0.25&0.23&0.15&0.28&0.74&0.82\\
  &0.00&0.20&0.00&0.00&0.00&0.06&0.03&0.26&0.11&0.00&0.00\\
\hline
$R_{1/2}$&0.28&0.40&0.48&0.18&0.64&0.37&0.46&0.30&0.37&0.74&0.62\\
  &0.03&0.03&0.00&0.78&0.00&0.00&0.00&0.00&0.00&0.00&0.00\\
\hline
$\epsilon=1-b/a$&0.39&0.14&0.17&0.26&0.20&0.29&0.20&0.10&...&0.15&...\\
  &0.00&0.95&0.55&0.30&0.23&0.02&0.07&0.77&...&0.40&...\\
\hline
$n$&0.28&...&0.41&0.20&...&0.43&...&...&...&0.41&0.60\\
  &0.14&...&0.00&0.63&...&0.00&...&...&...&0.00&0.00\\
\hline
$d$&...&...&...&...&...&0.41&0.28&...&...&...&...\\
  &...&...&...&...&...&0.00&0.00&...&...&...&...
%  \hline
  \enddata
  %  \tablecomments{}
  \tablerefs{$^a$ \citet{urry+00}; 
    $^b$\citet{mcleod01}; 
    $^c$\citet{dunlop+03}; 
    $^d$\citet{floyd+04}; 
    $^e$\citet{taylor+96}; 
    $^f$\citet{rothberg+04}; 
    $^g$\citet{BBF92}; 
    $^h$\citet{pahre99}; 
    $^i$\citet{mobasher+99}; 
    $^j$\citet{ferrarese+06}; 
    $^k$\citet{graham+96}}
\end{deluxetable*}
%\end{table*}
%%%%%%%%%%%%%%%%%%%%%%%%%%%%%%%%%%%%%%%%%%%%%%%%%%%

\end{document}